\documentclass[preprint,12pt]{elsarticle}

\usepackage{natbib}
\usepackage[labelfont=bf]{caption}
\usepackage{subcaption}
\usepackage[utf8]{inputenc}  
\usepackage[T1]{fontenc}     
\usepackage{lmodern}         
\usepackage[scaled]{helvet} 
\usepackage[english]{babel}  
\usepackage{graphicx}        
\usepackage{comment}         
\usepackage[dvipsnames]{xcolor}
\usepackage{xfrac}
\biboptions{sort&compress}   

\usepackage{hyperref}

\usepackage{amsmath,amsfonts,amssymb}

\usepackage{xcolor}
\usepackage{gensymb}

\usepackage{siunitx}
\usepackage{textgreek}

\renewcommand{\v}[1]{\ensuremath{\mathbf{#1}}} 
\let\cross=\times
\renewcommand{\times}{\cdot} 
\newcommand{\gv}[1]{\ensuremath{\mbox{\boldmath$ #1 $}}} 
\newcommand{\diff}{\mathrm{d}}
\newcommand{\pd}[2]{\frac{\partial #1}{\partial #2}}

\newcommand{\grad}[1]{\gv{\nabla} #1} 
\renewcommand{\div}[1]{\gv{\nabla} \cdot #1} 

\newcommand{\laplacian}[1]{\grad^2 #1}

\newcommand{\ifrac}[2]{#1 / #2}

\usepackage{lineno}

\journal{arXiv}

\begin{document}

\begin{frontmatter}



\title{Interaction between corner and bulk flows during drainage in granular porous media}

\author[inst1]{Paula Reis}
\author[inst1,ify]{Gaute Linga}
\author[inst1]{Marcel Moura}
\author[inst1]{Per Arne Rikvold}
\author[inst1,inst4]{Renaud Toussaint}
\author[inst1,inst2]{Eirik Grude Flekkøy}
\author[inst1,inst3]{Knut Jørgen Måløy}

\affiliation[inst1]{organization={PoreLab, The Njord Centre, Department of Physics, University of Oslo},
            city={Oslo},
            country={Norway}}

\affiliation[ify]{organization={PoreLab, Department of Physics, Norwegian University of Science and Technology},
            city={Trondheim},
            country={Norway}}

\affiliation[inst2]{organization={PoreLab, Department of Chemistry, Norwegian University of Science and Technology},
            city={Trondheim},
            country={Norway}}

\affiliation[inst3]{organization={PoreLab, Department of Geoscience and Petroleum, Norwegian University of Science and Technology},
            city={Trondheim},
            country={Norway}}

\affiliation[inst4]{organization={ITES/Institut Terre et Environnement de Strasbourg, CNRS UMR7063 - Université de Strasbourg},
            city={Strasbourg},
            country={France}}               

\begin{abstract}

Drainage in porous media can be broken down into two main mechanisms: a primary piston-like displacement of the interfaces through the bulk of pore bodies and throats, and a secondary slow flow through corners and films in the wake of the invasion front. In granular porous media, this secondary drainage mechanism unfolds in connected pathways of pendular structures, such as capillary bridges and liquid rings, formed between liquid clusters. To represent both mechanisms, we proposed a dynamic dual-network model for drainage, considering that a gas displaces a wetting liquid from quasi-2D granular porous media. For this model, dedicated analyses of the capillary bridge shapes and hydraulic conductivity were conducted so that the secondary drainage mechanism could be properly quantified at finite speeds. With the model, an investigation of the wetting-phase connectivity and flow during drainage was carried out, covering a broad range of flow conditions. Results indicate that the span of liquid-connected structures in the unsaturated region, as well as their ability to contribute to flow, varies significantly with Capillary and Bond numbers.

\end{abstract}

\begin{graphicalabstract}
\includegraphics[width=0.95\textwidth]{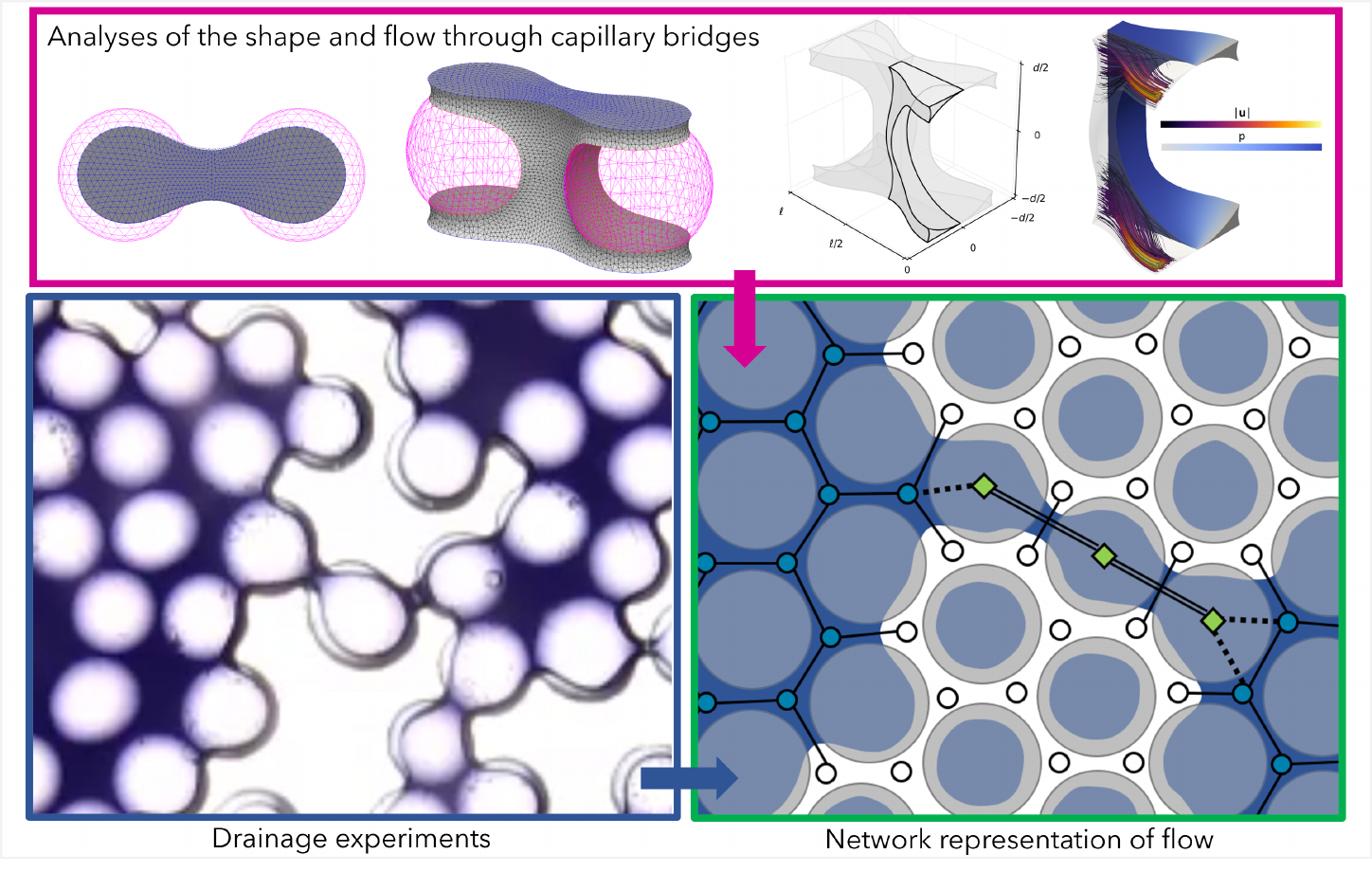}
\end{graphicalabstract}

\begin{highlights}
\item We present a modeling approach for integrating corner flow in a dynamic pore-network model for drainage in granular porous media.
\item The model relies on a dual-network formulation of the flow through pores and bridges, where the conductance through individual bridges is computed by direct simulations.
\item Results indicate that wetting-phase connectivity and flow in unsaturated granular media varies significantly and nontrivially with $Ca$ and $Bo$.
\end{highlights}


\end{frontmatter}


\section{Introduction}
\label{sec:intro}

The displacement of a liquid by a gaseous phase in granular porous media is a central component of diverse natural and engineered subsurface processes, such as CO$_2$ geological sequestration \cite{juanes2010footprint,aryana2012experiments,datta2013drainage,zacharoudiou2018impact}, enhanced oil recovery \cite{alvarado2010enhanced,afzali2018comprehensive}, and water retention in soils \cite{or2013foam,hoogland2016drainage,hoogland2016drainageb,hoogland2017capillary}. Understanding and properly designing these processes, therefore, depend on how sensibly the pore-scale dynamics of liquid mobilization by gas can be characterized and upscaled to the geological length scale of interest \cite{toussaint2012two,moebius2014pore}. In this work, we focus on the pore-scale description of such flows under the assumption that the liquid phase preferentially wets the grains constituting the porous medium and that the viscosity and density of the gas can be neglected. In particular, we investigate how the wetting-phase connectivity and flow through corners during drainage can be effectively represented in pore-scale models, leading to more realistic predictions of transport properties and liquid residual saturations under a broad range of flow conditions.  

Drainage displacements are commonly sorted into different categories according to the balance between capillary, viscous, and gravitational forces taking place during the flow \cite{lenormand1990liquids}. In the absence of significant viscous or gravitational forces, the motion of gas as it invades a liquid-filled random porous medium delineates a pattern known as capillary fingering \cite{lenormand1989capillary,maaloy1992dynamics}. Under this regime, the largest pore throats at the fluid's interface are drained one at a time, since they exhibit the lowest values of capillary pressure thresholds for invasion, $P_t$, given by the Young-Laplace Equation \eqref{eq:Pt}. Following this criterion only, the interface advancement traps clusters of the wetting phase and forms fractal fingers that can be well described by invasion percolation theory \cite{wilkinson1983invasion,lenormand1989capillary}.

\begin{equation}
    P_t = \gamma \left( \frac{1}{r_1} + \frac{1}{r_2} \right) 
    \label{eq:Pt}
\end{equation}
where $\gamma$ is the interfacial tension between the phases and $r_1$ and $r_2$ are the principal radii of curvature of the interface moving through a pore throat. 

Departing from capillary-dominated to faster flows, the pattern outlined by the gas motion is modified by the onset and evolution of viscous instabilities, leading to viscous fingering \cite{homsy1987viscous}. In this case, 
the progression of the interface is not governed solely by the size of the available pore throats but also by the pressure drop across them \cite{maaloy1985viscous}. This altered criterion for pore invasion is a consequence of the viscous pressure gradient developed in the liquid, which descends along the direction of the flow and favors drainage at the most advanced points of the front. As a result, the efficiency of liquid mobilization is low, and the gas tends to flow through thin fingers aligned with the flow direction \cite{lovoll2004growth,lovoll2011influence}. 

In contrast to fingered flows, the invasion front between gas and liquid during drainage may be stable when gravitational effects come into play, with the denser liquid phase lying initially under the gaseous phase \cite{lovoll2005competition}. As proposed by \citet{meheust2002interface}, the crossover for viscous instability in the presence of gravity occurs when the generalized bond number, $Bo^*=Bo-Ca$, falls below zero. Thus, when the bond number -- given by  $Bo=\Delta \rho g a^2 \mathbin{/} \gamma $, where $\Delta \rho $ is the density difference between the phases, $g$ is the gravitational component in the direction of the flow, and $a$ is the typical pore size -- exceeds the capillary number -- given by $Ca= v \mu a^2 \mathbin{/} \gamma k$, where $v$ and $\mu$ are the Darcy velocity and the viscosity of the wetting phase, and $k$ is the porous medium permeability -- the 
the hydrostatic pressure gradient in the liquid overcomes the destabilizing viscous pressure gradient, suppressing the unbounded growth of fingers and favoring the invasion of pores far from the tip of the front.

The preceding description of drainage flow regimes -- capillary fingering, viscous fingering, and stable front displacement -- focuses on the evolution of the main invasion front as it percolates through a porous medium. This front defines the border between a saturated region, labeled here as the main defending cluster, and an unsaturated region where liquid can be found in clusters, corners, and films \cite{hoogland2016drainage,moura2019connectivity}. While the role of corners in wetting-phase transport has been extensively investigated in imbibition \cite{zhao2016wettability,golmohammadi2021impact,primkulov2021wettability,primkulov2022avalanches,liu2022systematic,zhao2022dynamic,pavuluri2023interplay,suo2024spontaneous}, few studies have been dedicated to adequately representing drainage via corners in the wake of the main invasion front progression. Next, we present a brief review of studies concerning the role of corner flow in drainage. Then, we establish how the present work fits in with the current literature on this topic.

In a study concerning water retention in soils, \citet{hoogland2016drainage} proposed a distinction between two main drainage mechanisms: a rapid piston-like displacement at the front, and a slow gravity-driven corner flow behind the front. In that work, the authors used invasion-percolation theory and experiments on small columns filled with sand and glass beads to predict the saturation at which the network of water-filled pores is disrupted and drainage is dominated by corner flow. In addition, they suggested a star-shaped pore geometry to estimate water retention and transport in natural pores. In an accompanying study \cite{hoogland2016drainageb}, the foam drainage equation (FDE) \cite{or2013foam} was employed to predict macroscopic drainage dynamics in the slow corner-flow zone behind the front. The alternative use of the FDE was based on the idea that the flow through channels formed between bubbles in foams, known as Plateau borders, could represent the flow in crevices between soil particles. As a result, the wetting phase is assumed to be fully connected by liquid-filled corners behind the front.

The hypothesis of overall wetting-phase connectivity in unsaturated granular porous media can, however, lead to overestimated water removal rates and inaccurate predictions of liquid residual saturations \cite{bryant2003wetting}. For this reason, conceptualizing flow in a network of angular pores and throats with interconnected liquid-filled corners may be inadequate. In fact, when the flow through the bulk of pores and throats is disrupted in granular porous media, pathways of corner flow may remain, where liquid rings formed at grain-grain contact points and capillary bridges join disconnected elements of the network of pores and throats. These corner-flow pathways evolve with time and do not coincide with the original network of pores and throats \cite{moura2019connectivity}. In both quasi-2D \cite{aursjo2014film, chen_2018,moura2019connectivity}  and 3D \cite{Scheel_2008,kharaghani2021three,dong2022characterization} grain packings, it has been verified in visualization experiments that liquid located in rings around contact points between smooth solid surfaces and capillary bridges can form large locally connected structures behind the front. Still, these do not span the entire unsaturated zone, allowing the drainage of only a fraction of liquid clusters in this region. 

Such description of wetting-phase connectivity has been represented in a few pore-scale models for both drying \cite{vorhauer2015drying,chen_2018,kharaghani2021three} and drainage \cite{bryant2003wetting,reis2023simplified} in granular media. \citet{vorhauer2015drying} proposed a pore-network model (PNM) of drying in quasi-2D regular porous media in which liquid clusters could be formed not only by interconnected pores and throats in the network but also by adjacent liquid rings. This approach was later expanded to 3D porous media consisting of simple cubic lattices of spheres by \citet{kharaghani2021three}, leading to better drying rate estimates than PNM with clusters of pores and throats only. Similarly, \citet{chen_2018} proposed a numerical model to reproduce drying in a circular quasi-2D porous medium composed of cylinders arranged as neighboring spirals between parallel plates. In this particular geometry, no clusters of liquid-filled pores are formed and long chains of capillary bridges connect the central part of the circle to its open outer rim, enhancing the drying rate. Based on this simplified corner-flow geometry, their model was the first to simulate viscous flows through chains of capillary bridges, by employing viscous resistance values obtained with direct numerical simulation (DNS) of Stokes flow on single bridge shapes. This, combined with realistic bridge snap-off pressure estimates, enabled their model to reasonably predict the maximum capillary bridge chain length for different imposed evaporation rates.

As for models representing drainage, \citet{bryant2003wetting} developed a quasi-static PNM based on a 3D dense packing of spheres in which pendular rings could act as connecting points for liquid in neighboring pores/throats. Although their model only included short-range connectivity rules that could not account for the whole unsaturated region, estimates of residual liquid saturation matched experimental results significantly better than models assuming unrestricted liquid connectivity. With a similar goal, \citet{reis2023simplified} proposed a modified invasion-percolation (IP) model for slow drainage under the effects of capillary and gravitational forces. Their model could identify long-range extra connectivity provided by liquid rings and bridges, qualitatively reproducing experiments presented by \citet{moura2019connectivity} in quasi-2D porous media consisting of spherical beads trapped between parallel planes.

In the present work, the wetting-phase connectivity concept presented in \citet{reis2023simplified} was incorporated in a dynamic PNM for drainage, in which a dual-lattice network approach was used to represent both networks of connected pores and throats, and of connected rings and capillary bridges. The interest in moving from an IP model to a dynamic PNM is twofold: investigating the effect of moderate/high $Ca$ on corner flow, and more adequately representing corner flow during slow drainage. Using a quasi-static model, we could not differentiate the time scales required to drain a pore belonging to the front from a pore belonging to a cluster connected by corner flow. To further improve this point, viscous resistance in bridges was calculated using DNS of Stokes flow through various single bridge shapes, similarly to the work of \citet{chen_2018}, and implemented in the network of corner flow. Besides the new model, we present a comprehensive evaluation of how the balance between capillary, viscous, and gravitational forces influence the wetting-phase connectivity and flow in the unsaturated region trailing the main drainage front in granular porous media.

\section{Pore-Scale Model}
\label{sec:met}

The pore-scale model proposed in this work to represent drainage with corner flow belongs to the category of Dynamic Pore-Network Models \cite{joekar2012analysis}. In these models, the pore space is conceptualized as networks of geometrically-simplified elements, in which fluid configurations are presumed according to flow conditions and wetting properties. Based on these simplifications, multiphase flow through the networks is normally solved by imposing mass conservation on nodes (representing pores) and assuming Poiseuille flow on the edges (representing throats). Capillary effects are incorporated into the calculations using the Young--Laplace equation (see eq.\  \eqref{eq:Pt}) on network elements containing more than one phase. Gravitational effects can also be included, by computing the individual contribution to the hydrostatic pressure of the fluids occupying each network element. While Dynamic PNMs are not able to resolve geometric details related to the fluids' interfaces and pore shapes, as achieved with DNS models, the possibility of simulating bigger domains can offer insights into flow phenomena stretched over several thousands of pores ($10^3-10^4$), as explored in this study.

Within the dynamic PNM category, the one presented here was specifically designed to reproduce drainage in a quasi-2D porous medium consisting of a Hele-Shaw cell filled with a monolayer of glass beads. This design of porous medium has been employed in several experimental studies of drainage \cite{maaloy1985viscous,maaloy1992dynamics,meheust2002interface,lovoll2004growth,lovoll2005competition,lovoll2011influence,toussaint2012two,moura2015impact,moura2019connectivity,ayaz2020gravitational} and represented in a quasi-static PNM in \citet{reis2023simplified}. Such structure is particularly beneficial to our study, as it allows us to clearly distinguish between flow phenomena in saturated and unsaturated regions while retaining fundamental 3D features of corner flow in granular materials. Behind the main invasion front, liquid rings remain between each bead and the Hele-Shaw cell planes, linking otherwise disconnected liquid-filled pores and throats. Notably, draining such a porous medium evidences how the pathways of corner flow do not coincide with the main network of pores and throats \cite{moura2019connectivity}, a key aspect explored in this work and detailed in section \ref{sec:duallat}.

Another particular aspect of the proposed model comes from the choice of represented wetting and non-wetting fluids, namely a glycerol-water mixture and air. Given their significant difference in viscosity and density, pressure variations within the non-wetting phase are assumed to be negligible. Further details of how this assumption affects the flow calculation in each network element, as well as the model's governing equations, are discussed in sections \ref{sec:q_nets} and \ref{sec:goveq}. As a final noteworthy feature of the presented PNM, estimates of the viscous resistance through capillary bridges were obtained with DNS of Stokes flow through a wide range of bridge shapes, as discussed in section \ref{sec:g_bridge}.

\subsection{Dual-lattice network representing pores, throats, and corner-flow pathways}
\label{sec:duallat}

The representation of Hele-Shaw cells filled with glass beads as networks of pores and throats was carried out here in the same way as proposed by \citet{reis2023simplified}. The disposition of the beads between the planes was assumed to form a regular triangular lattice (TL), leading to a hexagonal lattice (HL) of pores connected by throats. During drainage simulations, starting from a saturated pore space, the original pore network with HL topology evolved according to its fluid occupation. Nodes representing pores in this lattice could be labeled as filled by gas or liquid, while edges representing throats were deleted from the original grid when completely filled by gas. In this way, the remaining connected parts of the hexagonal lattice network correspond directly to the liquid clusters in the porous medium. More details on this network approach to represent fluids in pores and throats, as well as its relation to graph theory concepts, can be found in \citet{reis2023simplified}.

Using the hexagonal-lattice network as a scaffold, the flow through the bulk of pores and throats was represented in the model. As proposed in \citet{hoogland2016drainage} and \citet{moura2019connectivity}, the fast piston-like flow of gas as it invades the bulk of liquid-filled pores and throats is responsible for mobilizing most of the drained volume, and is therefore termed as the primary drainage mechanism. In opposition, the slow corner-flow dominated drainage in the unsaturated region \cite{hoogland2016drainage} is termed the secondary drainage mechanism \cite{moura2019connectivity}. Besides the difference in time scales required for drainage via both mechanisms, it has been verified experimentally that they also do not happen in coinciding paths in granular media. For this reason, we chose to represent corner flow through an evolving secondary network built on a triangular lattice -- the dual lattice of the hexagonal pore network.

\begin{figure}[ht!]
     \centering
     
     \begin{subfigure}[t]{0.32\textwidth}
         \centering
         \includegraphics[width=0.95\textwidth]{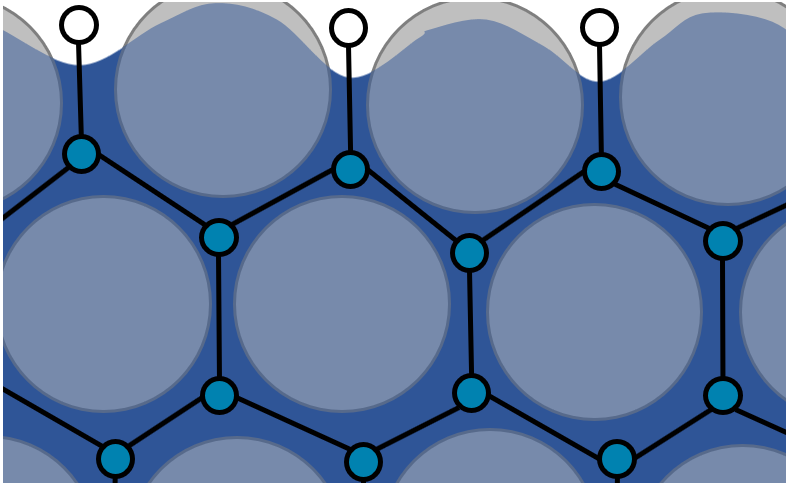}
         \caption{}
         \label{fig:dl_1}
     \end{subfigure}
     \hfill
     \begin{subfigure}[t]{0.32\textwidth}
         \centering
         \includegraphics[width=0.95\textwidth]{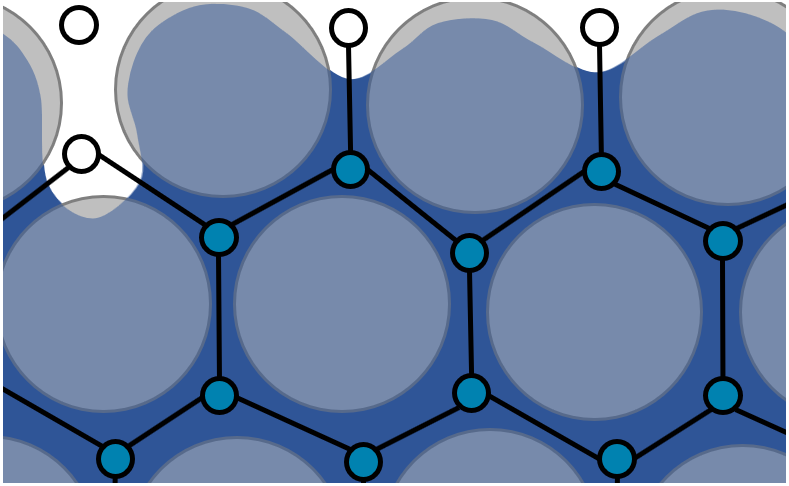}
         \caption{}
         \label{fig:dl_2}
     \end{subfigure}
     \hfill
     \begin{subfigure}[t]{0.32\textwidth}
         \centering
         \includegraphics[width=0.95\textwidth]{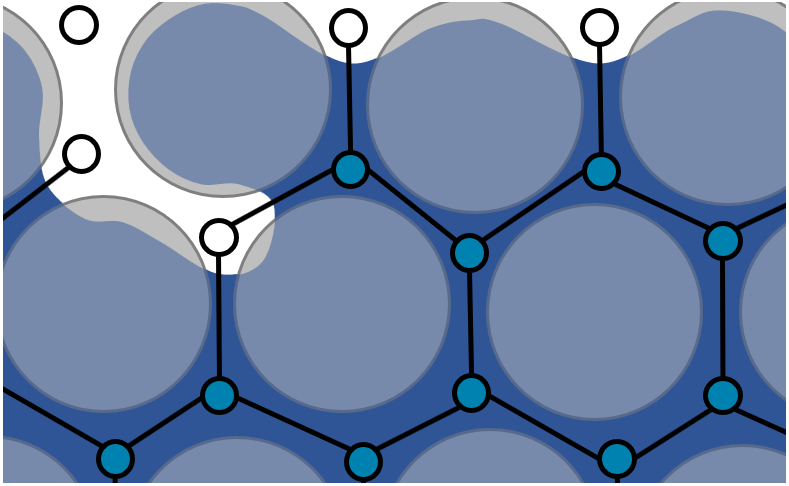}
         \caption{}
         \label{fig:dl_3}
     \end{subfigure}
     \hfill
     \begin{subfigure}[t]{0.32\textwidth}
         \centering
         \includegraphics[width=0.95\textwidth]{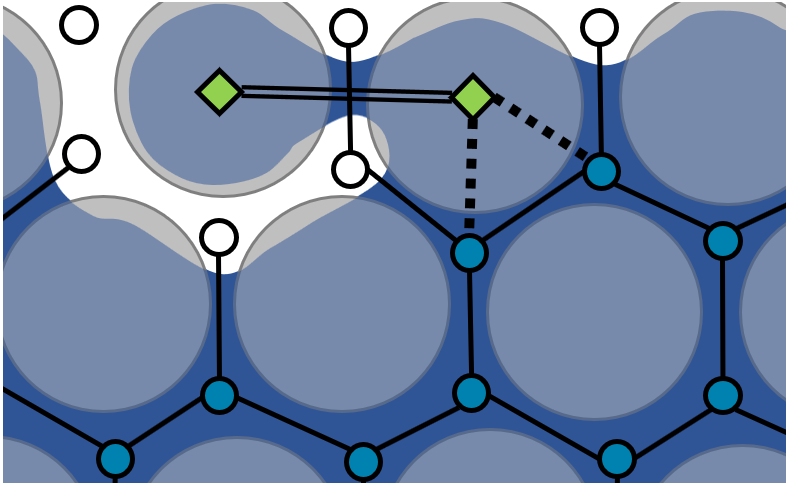}
         \caption{}
         \label{fig:dl_4}
     \end{subfigure}
     \hfill
     \begin{subfigure}[t]{0.32\textwidth}
         \centering
         \includegraphics[width=0.95\textwidth]{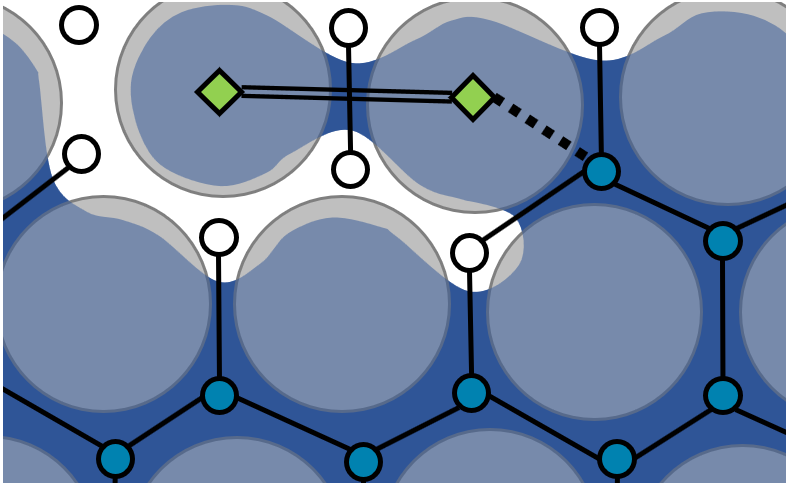}
         \caption{}
         \label{fig:dl_5}
     \end{subfigure}
     \hfill
     \begin{subfigure}[t]{0.32\textwidth}
         \centering
         \includegraphics[width=0.95\textwidth]{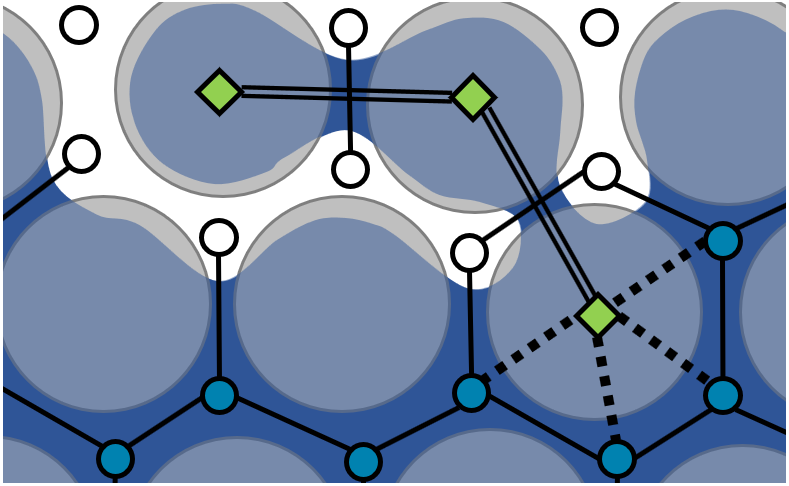}
         \caption{}
         \label{fig:dl_6}
     \end{subfigure}
     \hfill
     \begin{subfigure}[t]{0.32\textwidth}
         \centering
         \includegraphics[width=0.95\textwidth]{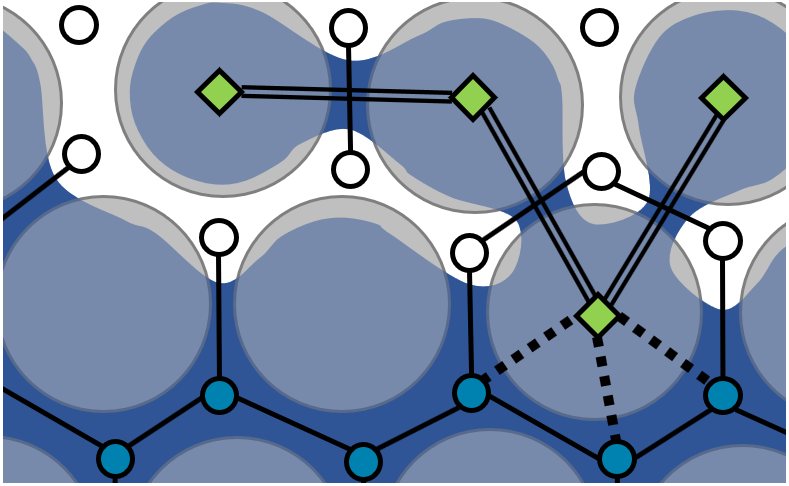}
         \caption{}
         \label{fig:dl_7}
     \end{subfigure}
     \hfill
     \begin{subfigure}[t]{0.32\textwidth}
         \centering
         \includegraphics[width=0.95\textwidth]{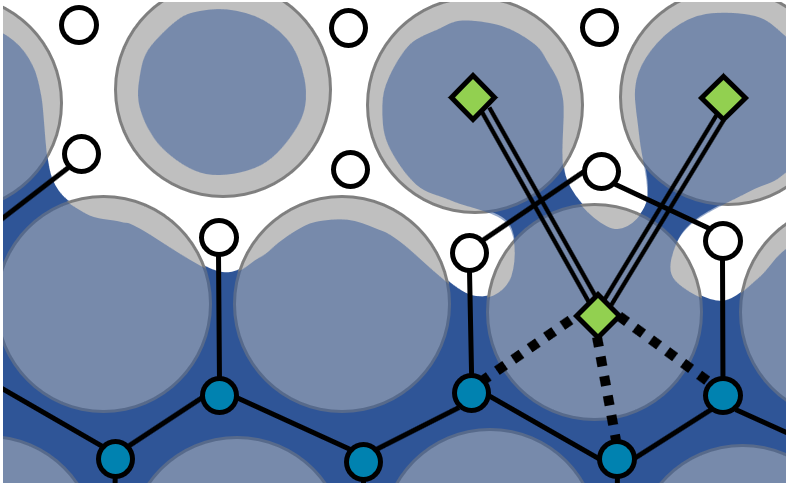}
         \caption{}
         \label{fig:dl_8}
     \end{subfigure}
     \hfill
     \begin{subfigure}[t]{0.32\textwidth}
         \centering
         \includegraphics[width=0.95\textwidth]{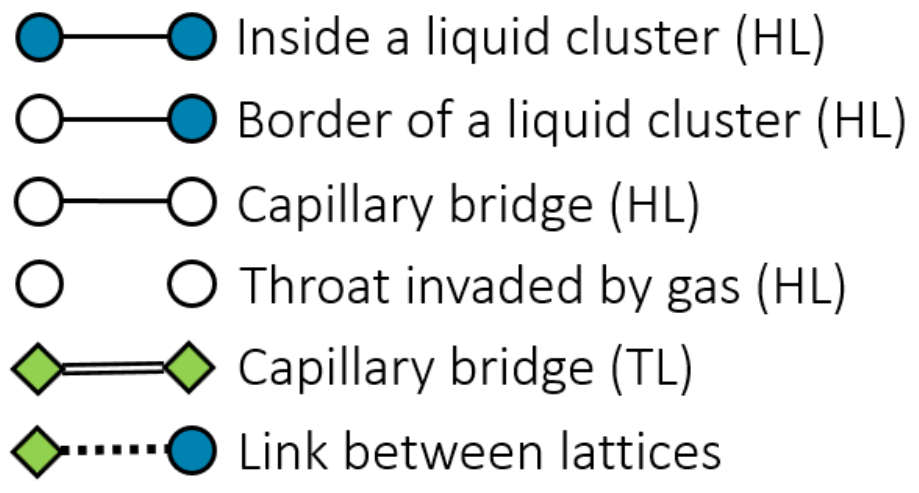}
         \label{fig:legend}
     \end{subfigure}
     
        \caption{Illustration of the dual-network representation of flow paths, as the quasi-2D granular porous medium is drained. Network elements are defined in the bottom right insert. Small circle and diamond shapes represent the network nodes in the hexagonal and triangular lattices, respectively, and correspond to the endpoints of pore throats and capillary bridges. Simple, double, and dashed lines represent the network edges, representing pore throats, capillary bridges, and the connections between them, in that order. From (a) to (c), gas fully invades two consecutive pore throats, leading to the deletion of their representative edges from the hexagonal-lattice network. From (d) to (g), the movement of the gas-liquid interface leads to the formation of capillary bridges, which are then represented with new nodes and edges on the dual-lattice of the network of pores and throats. From (g) to (h) the transformation of the network due to the snap-off of a bridge is illustrated.}
        \label{fig:duallat_seq}
\end{figure}

The schematic representation of a Hele-Shaw cell filled with beads undergoing drainage in Figure \ref{fig:duallat_seq} helps illustrate the dual-lattice network approach. In the sketched image, depicting a top view of the porous medium, large gray circles represent the beads and blue areas represent the liquid occupation of the porous medium, at the frontier between saturated and unsaturated zones. Along with the sketch, the corresponding dual-lattice network representation of the wetting-phase connectivity is shown.  From Fig. \ref{fig:dl_1} to Fig. \ref{fig:dl_3}, we see the non-wetting phase invading two new pores and throats, and the resulting alterations in the network representing pores connected by throats. At this point, the hexagonal-lattice (HL) network still corresponds to the region connected by liquid. From Fig. \ref{fig:dl_4} to Fig. \ref{fig:dl_7}, further advancement of the gas into the porous medium leaves behind capillary bridges between adjacent beads, which can establish permeable enough paths for the liquid to flow \cite{chen2017control,chen_2018,moura2019connectivity}. These flow paths are perpendicular to the ones in the hexagonal lattice, and could not be represented in the previously established network of pores connected by throats. Therefore, to account for the flow through the capillary bridges, we created new nodes and edges in the dual lattice of the original hexagonal pore network. In the new lattice, edges (double lines in Fig. \ref{fig:duallat_seq}) represent the capillary bridges, while the nodes (green diamond shapes in Fig. \ref{fig:duallat_seq}) are the bridges' endpoints, coinciding with the glass beads' positions. As dual lattices do not share nodes or edges, connecting elements were also created, so that the liquid flow could be represented in a single network. In this way, the new links, illustrated in Fig. \ref{fig:duallat_seq} as dashed lines, represent the flow between liquid clusters and the capillary bridges. As a final remark, the flow through capillary bridges may also be interrupted, if the capillary pressure for the snap-off of a bridge is reached, as illustrated in Fig. \ref{fig:dl_8}. In this case, an isolated liquid ring can be formed between the bead and each Hele-Shaw cell plane, as represented by the blue-filled circle on top of the bead. 

\begin{figure}[ht!]
    \centering
    \includegraphics[width=1\textwidth]{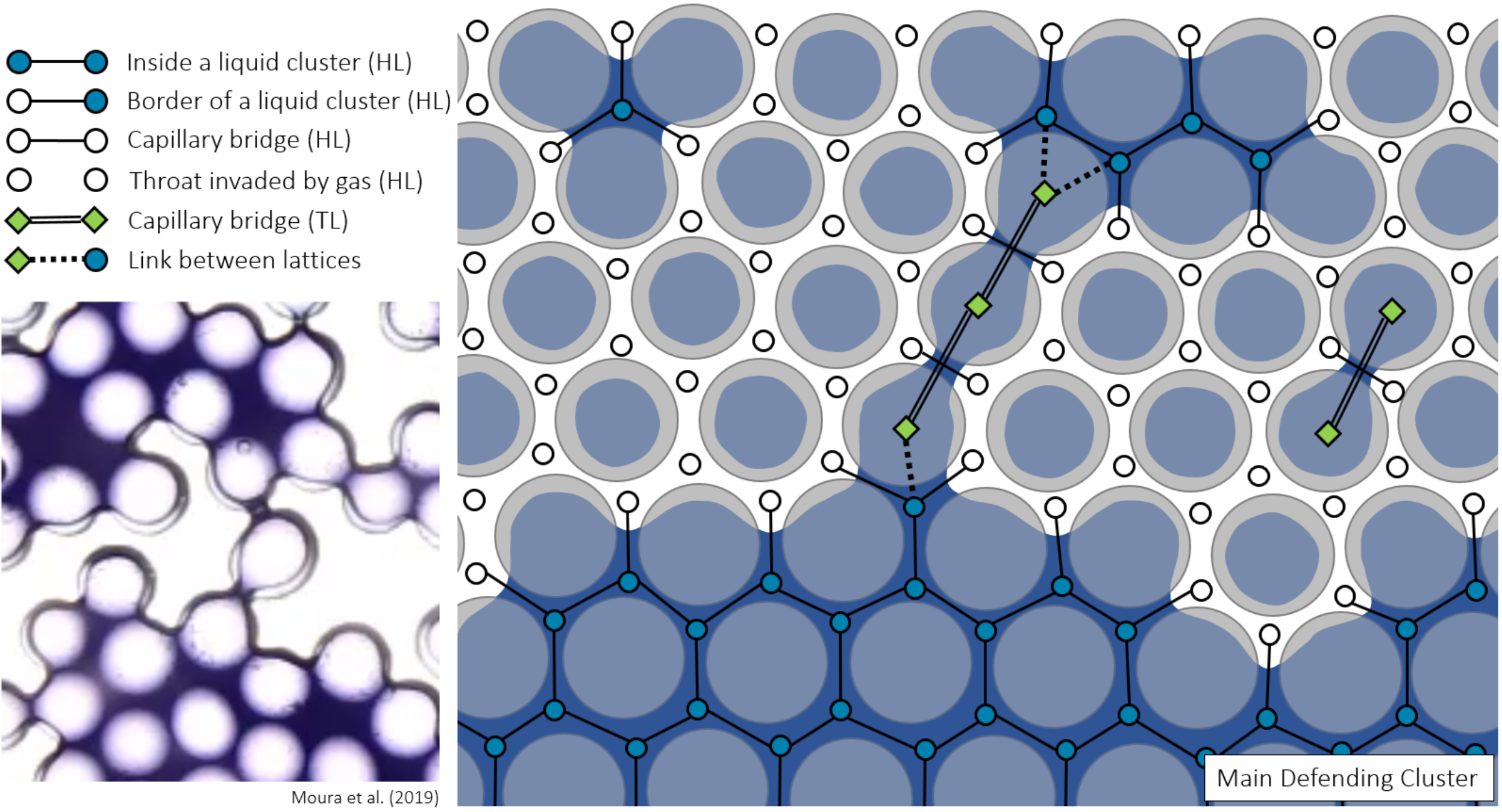}
    \caption{Schematic representation of saturated and unsaturated zones during drainage in a Hele-Shaw cell filled with glass beads and the corresponding flow networks. Large grey circles represent glass beads, and blue areas correspond to the liquid occupation of the pore space. As for the network representation: on the hexagonal lattice, small blue-colored circles represent liquid-filled pores, while white ones are filled with gas. Edges representing throats only exist if filled with liquid. As seen in the bottom-left insert, clusters connected by capillary bridges are disconnected in this network. On the triangular lattice, edges representing capillary bridges establish pathways for corner flow.}
    \label{fig:duallat}
\end{figure}

Figure \ref{fig:duallat} represents a more complex liquid configuration on a larger domain, which includes the main defending cluster, a cluster connected to the front by a chain of two capillary bridges, a small disconnected cluster, a disconnected bridge, and several isolated liquid rings.
In this figure, an experimental photo \cite{moura2019connectivity} from a similar liquid configuration can also be seen in the bottom-left insert.

\subsection{Flow through different network elements}
\label{sec:q_nets}

\begin{figure}[ht!]
    \centering
    \includegraphics[width=0.5\textwidth]{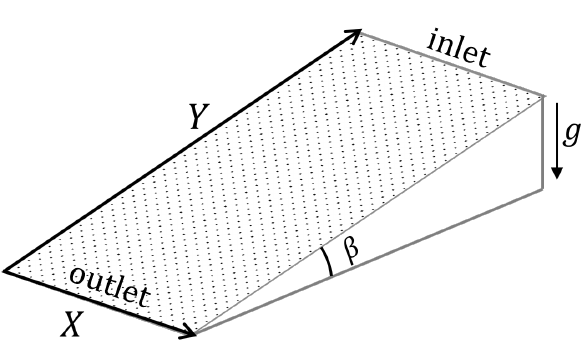}
    \caption{Schematic representation of a tilted rectangular Hele-Shaw cell filled with a monolayer of spherical glass beads. $X$ and $Y$ are the global coordinates inside the cell, and $\beta$ is the inclination angle of the cell with respect to the horizontal plane. The air inlet is positioned at the top of the cell, while the liquid outlet is positioned at the bottom of the cell. In this way, gravitational pressure fields stabilize the flow.}
    \label{fig:hele-shaw}
\end{figure}

The dual-lattice network described in section \ref{sec:duallat} represents the wetting-phase occupation in the modified Hele-Shaw cell porous medium, schematically illustrated in \ref{fig:hele-shaw}. During drainage, liquid can only flow through the portions of this network connected to the liquid outlet, with the rest being permanently trapped. During this process, the pressure in the air-invaded portions of the porous medium is assumed to remain constant, due to this phase's relatively low density and viscosity. 

A converging-diverging shape was adopted in this work for the pore throats (see sec. \ref{sec:g_throat}) \cite{aker1998two}. In this way, we do not specify a particular shape or flow resistance to the pores, and nodes in the networks act as mere throat connecting points. Therefore, to calculate the liquid flow out of the porous medium we only need to solve the flow through edges, which can represent four different scenarios: $(i)$ liquid-filled throats located inside clusters, $(ii)$ throats containing air and liquid separated by a meniscus at the clusters' borders with the non-wetting phase, $(iii)$ capillary bridges, and $(iv)$ the junctions between capillary bridges and clusters. Altogether, these liquid structures compose a single network that connects to a liquid outlet, where the flow is locally governed by Darcy's law.

In section \ref{sec:g_throat}, we describe the flow calculation through pore throats, represented by edges in the hexagonal lattice. Then, in sections \ref{sec:g_bridge} and \ref{sec:so_bridge}, we detail the procedure used to calculate the flow through capillary bridges, represented by edges in the triangular lattice, and the criterion for capillary bridge snap-off. Finally, in section \ref{sec:g_conn_elements}, the flow between capillary bridges and clusters is detailed.

\subsubsection{Flow Through Pore Throats}
\label{sec:g_throat}

In the proposed model, pore throats formed between adjacent glass beads in the Hele-Shaw cells were idealized as converging-diverging cylinders with parabolic profiles given by Eq.\ \eqref{eq:rx_complete}.

 \begin{subequations}

  \begin{equation}
    r(x)=a+b\left(x-\frac{l}{2}\right)^2
    \label{eq:rx}
\end{equation}

   \begin{equation}
    a= r_{t}
    \label{eq:a}
\end{equation}
   \begin{equation}
    b= r_{b}\left(\frac{2}{l}\right)^2
    \label{eq:b}
\end{equation}

\label{eq:rx_complete}
\end{subequations}

\noindent{where $l$ is the length of the throat, $r_t$ is the radius at the throat's constriction, corresponding to half the distance between the surfaces of adjacent beads, and $r_b$ is the radius of the beads. Values of $l$, $r_b$, and the distribution of $r_t$ were based on the experimental work of \citet{moura2019connectivity} and are the same as used in \citet{reis2023simplified}.}

\begin{figure}[ht!]
    \centering
    \includegraphics[width=0.6\textwidth]{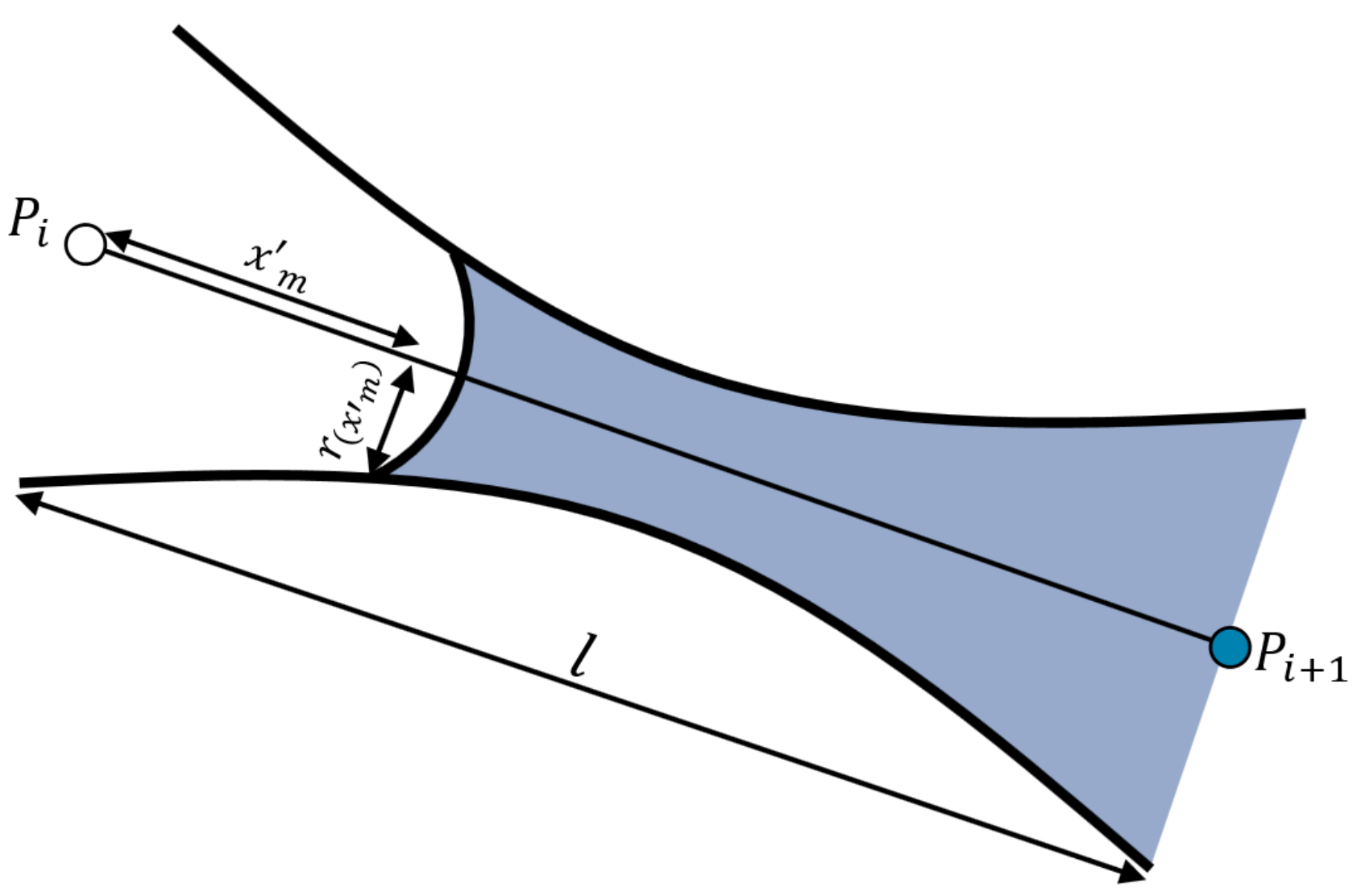}
    \caption{Schematic representation of a converging-diverging pore throat containing a meniscus between the gas and liquid phases.}
    \label{fig:poreel}
\end{figure}

With the profile given by Eq. \eqref{eq:rx_complete}, the flow through pore throats is calculated assuming Poiseuille flow and varies according to the fluid occupation status. At liquid-filled throats, the contributions of viscous and gravitational effects to the pressure are considered. At throats containing both phases, as illustrated in Fig. \ref{fig:poreel}, capillary pressure effects are also taken into account. Next, we present the procedure to calculate the flow through throats containing a meniscus separating two phases. From that, single-phase flow can be obtained as a particular case of the presented equations, in which the meniscus position $x'_m=0$ and the capillary pressure $\Delta P_{cap}=0$.

Using Fig. \ref{fig:poreel} as a guide, we assume that the pressure drop across the pore throat, $\Delta P = P_i-P_{i+1}$, can be decomposed into three components, $\Delta P = \Delta P_{visc} + \Delta P_{grav} + \Delta P_{cap}$, in which $\Delta P_{visc}$ represents the pressure variation due to viscous dissipation during flow, $\Delta P_{grav}$ represents the hydrostatic pressure and $\Delta P_{cap}$ represents the capillary pressure. Each of these components of the pressure drop is calculated according to Eqs.\ \eqref{eq:deltaPs}:
\begin{subequations}
  \begin{equation}
    \Delta P_{visc}=\frac{8q}{\pi}\left( \int_{0}^{x'_m}\frac{\mu _{g}}{r(x'_m)^4}dx' +  \int_{x'_m}^{l}\frac{\mu _{l}}{r(x'_m)^4}dx' \right)
    \label{eq:dP_visc}
    \end{equation}
   \begin{equation}
    \Delta P_{grav}=\rho _{l} g \left( h_{i+1}-h_i \right) \left(\frac{l-x'_m}{l}\right)
    \label{eq:dP_grav}
    \end{equation}
   \begin{equation}
    \Delta P_{cap}=\gamma\left( \frac{\cos{\theta_b}}{r(x'_m)} + \frac{\cos{\theta_p}}{r_b}\right)
    \label{eq:dP_cap}
    \end{equation}\label{eq:deltaPs}\end{subequations}
where $q$ is the flow rate through the throat, $x'_m$ is the meniscus position, $\mu _g$ and $\mu _l$ are the viscosities of gas and liquid, respectively, $\rho _l$ is the density of liquid, $g$ is the gravitational acceleration, $h_{i+1}$ and $h_i$ are elevations at the throat's endpoints, calculated as $(Y_{i+1}-Y_i)\sin{\beta}$, with $(X,Y)$ being the global coordinates of the Hele-Shaw cell and $\beta$ its angle of tilt to the horizontal plane, as shown in Fig. \ref{fig:hele-shaw}, $\theta _b$ is the contact angle between the air-liquid interface and the beads' surface, and $\theta _p$ is the contact angle between the air-liquid interface and the Hele-Shaw cell planes.

Based on the decomposition of $\Delta P$ into viscous, gravitational, and capillary components presented in Eqs.\ \eqref{eq:deltaPs}, we define the hydraulic conductance of each throat as $\Gamma =\ifrac{q}{\Delta P_{visc}}$, and calculate the throat's flow rate as $q=\Gamma(\Delta P - \Delta P_{grav} - \Delta P_{cap})$. This equation for flow through individual throats is later incorporated into the model's governing equations for flow through the networks, as described in Sec.\ \ref{sec:goveq}.

\subsubsection{Flow Through Capillary Bridges}
\label{sec:g_bridge}

To model flow through the network of capillary bridges, we follow a similar approach to \citet{chen_2018} and use DNS to compute the conductance of individual liquid bridges.
By inspection of the experimental images, we find that capillary bridges often connect two adjacent beads, as seen in the bottom left insert in Fig. \ref{fig:duallat}, establishing liquid flow paths distinct from those between pore bodies and throats. We thus seek an effective expression for the conductance $\Gamma$ of a link in the dual network representing two beads connected by a capillary bridge, as a function of the bead-bead separation $\ell$ and the capillary pressure $p_c$ in the bridge (defined here as the pressure difference between the outside and the inside of the bridge, hence typically $p_c > 0$). 

We assume quasi-static conditions, such that the shape of a bridge is solely given by its equilibrium shape, i.e.\ the shape that minimizes the free energy of the air-liquid interface (constrained by liquid volume conservation).
As such, any flow through the bridge does not alter its shape. 
This approximation is good in the range of low capillary numbers considered here, which coincides with the range where corner flow can play an important role.
We note that the liquid volume of the bridge must be a function of the capillary pressure and the distance $\ell$ between the beads, hence $\Gamma = \Gamma(p_c, \ell)$.
By dimensional analysis (see \ref{sec:scaling_stokes}), we find
\begin{align}
    \Gamma = \frac{d^4}{\mu \ell} k_0 \left( \frac{p_c d}{2 \gamma}, \frac{\ell}{d} \right),
    \label{eq:g_bridge_gen}
\end{align}
where $k_0(\tilde \kappa, \tilde \ell)$ is a dimensionless function that only depends on the shape of the bridge, encoded in the dimensionless curvature $\tilde \kappa = \kappa d$ and bead separation $\tilde \ell = \ell/d$, both normalized by the bead diameter $d$.
The capillary pressure $p_c$ is related to the mean curvature $\kappa$ by the Young--Laplace law, $p_c = 2 \kappa \gamma$.

\begin{figure}[htb]
    \centering
    \begin{subfigure}[t]{0.58\textwidth}
        \centering
        \includegraphics[width=0.95\textwidth,trim=4cm 0.5cm 2.2cm 1.8cm,clip]{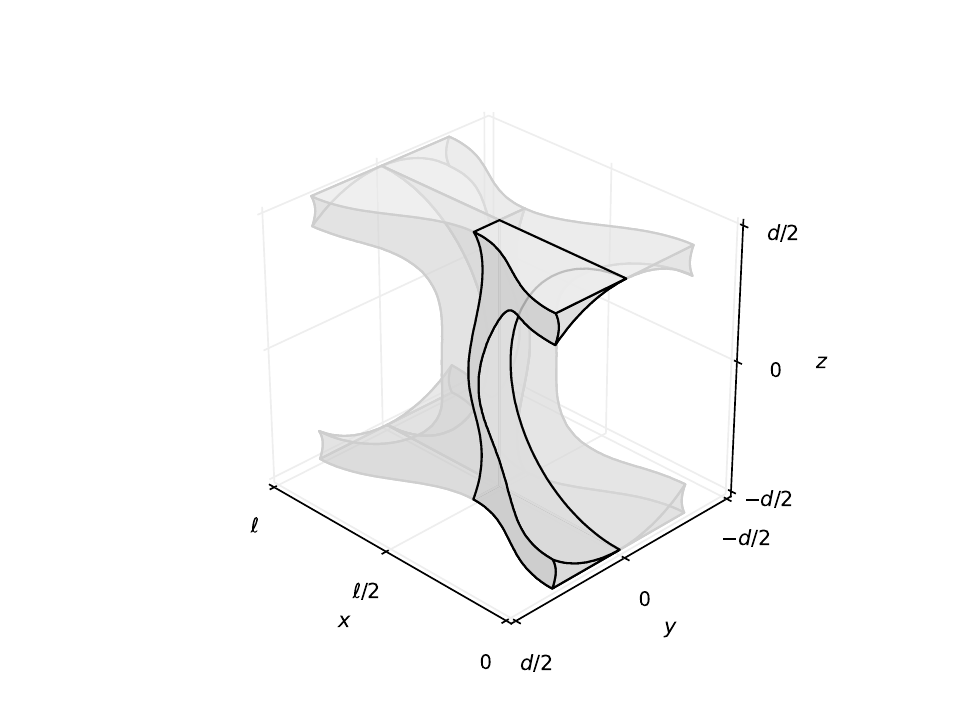}
        \caption{}
        \label{fig:bridge_schematic}
    \end{subfigure}
    \begin{subfigure}[t]{0.4\textwidth}
        \centering
        \includegraphics[width=0.95\textwidth]{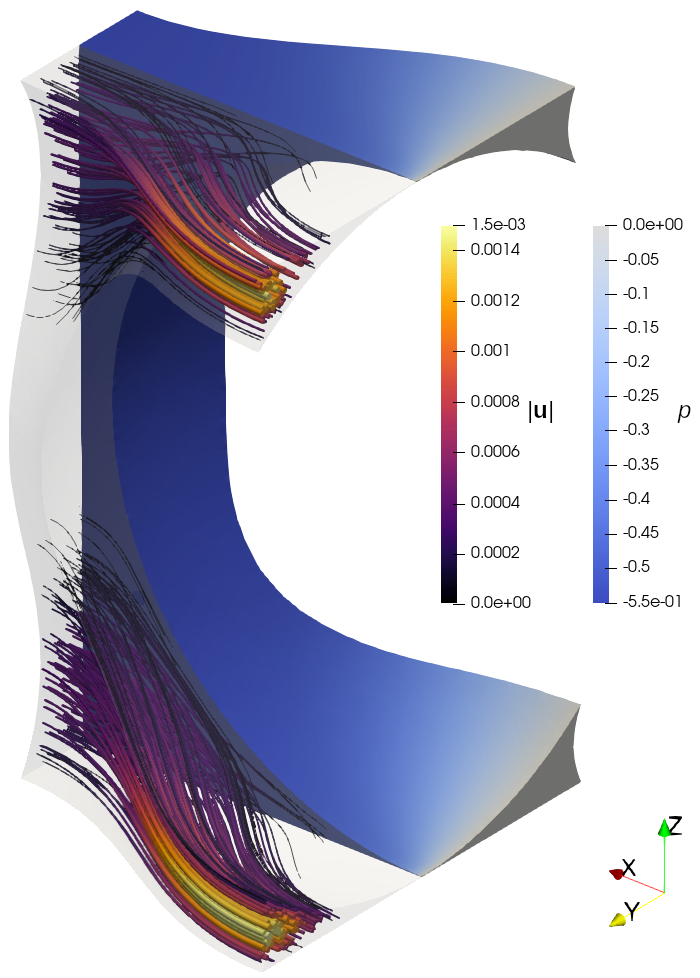}
        \caption{}
        \label{fig:bridge_flow}
    \end{subfigure}
    \caption{(a) Liquid bridge geometry. The quarter of the domain drawn with solid lines is used for CFD calculations by symmetry considerations. The solid beads have centers at $(x,y,z) = (0, 0, 0)$ and $(\ell, 0, 0)$ and are not shown directly here. (b) Visualization of the simulated flow field in the computational domain showing streamlines with velocity magnitude $|\v u|$ and pressure $p$.
    }
    \label{fig:bridge_cfd}
\end{figure}

We consider $z$ as the out-of-plane coordinate and a periodic array of beads along $x$ with centers at $y=0$, $z=d/2$. 
Thus the connected bridge and net flow is along the $x$ direction.
The liquid bridge geometry, as well as the coordinate system, is shown schematically in Fig.\ \ref{fig:bridge_schematic}.
In the $y$ direction, the surface will be is symmetric about the plane $y=0$. Moreover, in the $x$ direction, the surface is symmetric about the centers of the beads and the midplanes between two adjacent beads. These symmetry considerations limit our domain to a quarter of the original domain (cf.\ the part of Fig.\ \ref{fig:bridge_schematic} drawn with black solid lines), allowing for increased numerical resolution in the DNS. 
Note that because of a small but finite gravity, the domain is not symmetric in the $z$ direction. This symmetry breaking generally has a minute effect on the conductance, but significant effects on the connectivity and snap-off bridges, cf.\ Sec.\ \ref{sec:so_bridge}.

In order to obtain liquid bridge geometries in which to calculate the conductance by DNS, we used the software Surface Evolver \cite{brakke1992surface} to find the equilibrium air-liquid interface by surface energy minimization in the described solid geometry.
We considered a wide range of different scaled curvatures $\tilde \kappa$ and bead-bead distances $\tilde \ell$ around the values encountered in the experiments by \citet{moura2019connectivity}.
The symmetry planes were implemented as neutrally wetting boundaries (90\degree contact angle, which is mathematically equivalent).
We then smoothed the curved liquid-air surface using isotropic explicit remeshing \cite{hoppe1993mesh} via \texttt{PyMeshLab} \cite{pymeshlab}. This smoothing step was key to achieving well-defined surface normals required for imposing stress-free boundary conditions (see Eqs.\ \eqref{eq:stokes_bcs} below).
We verified that the resulting meshes were not significantly altered and still physically valid, e.g.\ compatible with the imposed contact angles, as reported in \ref{sec:validation_bridge_shape}.
The interior of the resulting closed surface meshes were tetrahedralized using \texttt{TetGen} \cite{hang2015tetgen}.

Assuming creeping flow conditions, the fluid velocity $\v u$ and pressure field $p$ inside the liquid bridge are governed by the Stokes equations:
\begin{align}
    \mu \laplacian \v u = \grad p, 
    \quad \div \v u = 0,
    \label{eq:stokes}
\end{align}
where $\mu$ is the dynamic liquid viscosity.
We decompose the pressure into a linear and a fluctuating part, $p = F x + p'$,
where $F = \Delta p / \ell $ is the imposed macroscopic pressure gradient driving a net flow through the bridge, and $p'$ represents local pressure variations.
As for the surface, the velocity field solution is symmetric about $y=0$.
Because of the linearity of the Stokes equations \eqref{eq:stokes} the resulting flow field is antisymmetric about the planes $x=0$ and $x=\ell/2$.
We further assume no-slip conditions at the solid boundaries.
This amounts to the following boundary conditions:
\begin{subequations}
\begin{align}
    u_y = \pd {u_x}{y} = \pd {u_z} {y} = 0 \quad & \textrm{at} \quad y = 0, \\
    \v u = \v 0 \quad & \textrm{at} \quad z = \pm d/2, \ \text{and} \ \sqrt{x^2 + y^2 + z^2} = d/2, \\
    \pd {u_x} x = u_y = u_z = p' = 0 \quad & \textrm{at} \quad x = 0, \ell/2.
\end{align}
On the air-water interface, we impose a free-slip condition:
\begin{align}
    \v u \cdot \hat{\v n} &= 0, \label{eq:un0}
    \\
    (\v D \v u \cdot \hat{\v n} ) \cross \hat{\v n} &= \v 0, \label{eq:freeslip}
\end{align}\label{eq:stokes_bcs}\end{subequations}
where $\v D\v u = (\grad \v u + \grad \v u^T)/2$ is the symmetrized strain rate, and $\hat{\v n}$ is the normal vector to the air-water interface.
Note that Eq.\ \eqref{eq:un0} reflects the quasi-static approximation, i.e.\ the interface shape is fixed regardless of flow.
Eq.\ \eqref{eq:freeslip} encodes that no shear stress acts along the interface, which follows from neglecting the viscosity in the air phase.

We solve numerically Eqs.\ \eqref{eq:stokes} and \eqref{eq:stokes_bcs} to obtain the shape factor $k_0(\tilde \kappa, \tilde \ell)$ as the total flux $\Phi$ in the case of $\mu = F = d = 1$ (see \ref{sec:scaling_stokes}). 
The numerical implementation relying on the finite element method is described in \ref{sec:fem_bridge} in more detail.
Note that to obtain $\Phi$ (and thus $k_0$) we must multiply the numerically obtained flux by 2 because of our symmetry-reduced computational domain.
Finally, knowing $k_0$, Eq.\ \eqref{eq:g_bridge_gen} can be used to retrieve the physical conductance.

An example of numerically obtained velocity and pressure fields is shown in Fig.\ \ref{fig:bridge_cfd}.
It is worth noting that the fastest flow occurs in the corners between the beads and the confining plates, i.e.\ near the top and bottom, while the flow in the near-contact region between two adjacent beads is negligible.
This observation is also reflected in the pressure gradients, which are more prominent close to the computational inlet plane, i.e.\ $x=0$.

\begin{figure}[htb]
    \centering
    \includegraphics[width=0.9\columnwidth]{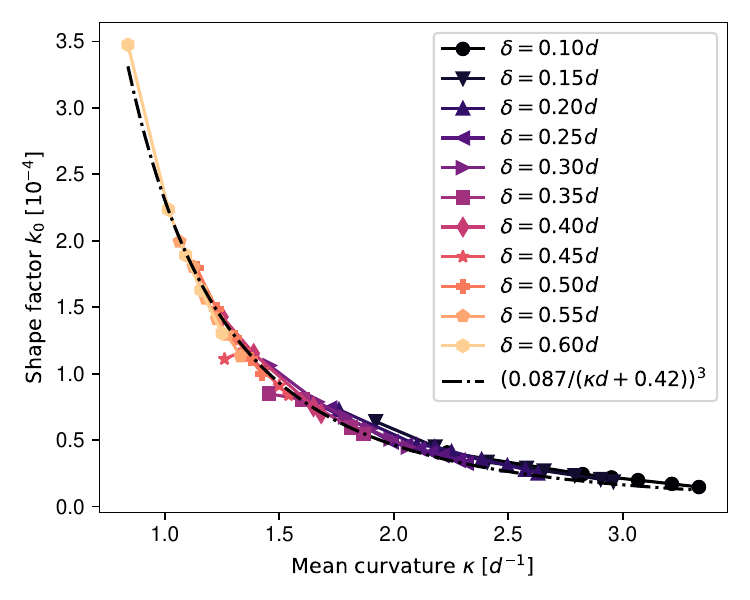}
    \caption{Shape factor $k_0$ obtained by DNS for different bead-bead separations $\ell = d + \delta$ and mean curvatures $\kappa$.}
    \label{fig:shape_factor_kappa}
\end{figure}

Figure \ref{fig:shape_factor_kappa} shows the results for the shape factor $k_0$ obtained by DNS for different bead-bead separations $\ell = d + \delta$ and mean curvatures $\kappa$.
Notably, we find that $k_0$ is independent of $\ell$, evidenced by the numerical data collapse onto a single curve, $k_0( \tilde \kappa)$.
We find that the function
\begin{align}
    k_0(\tilde\kappa) = \left( \frac{A}{\kappa + B} \right)^3,
    \label{eq:fit_cubic}
\end{align}
describes the data remarkably well, with $A = 0.0869$ and $B=0.417$ determined by a nonlinear fit of Eq.\ \eqref{eq:fit_cubic} to the data.
A physical justification for the functional form of Eq.\ \eqref{eq:fit_cubic}, as well as for the independence of $\ell$, is given in \ref{sec:theoretical_modelling_bridge_g}.
We stress that the values of $A, B$ are specific to the chosen values of the contact angles $\theta_b = 20\degree, \theta_p = 60\degree$, but in accordance with the derivation in \ref{sec:theoretical_modelling_bridge_g} we expect the approximate form of Eq.\ \eqref{eq:fit_cubic} to be fairly general.


\subsubsection{Capillary-Bridge Snap-Off}
\label{sec:so_bridge}

The liquid flow paths established by capillary bridges, as detailed in Sec. \ref{sec:g_bridge}, may be disrupted in case the capillary pressure for the snap-off of a bridge is reached during the flow. As the capillary pressure increases, the surface of a bridge becomes highly curved, making its structure progressively more slender. At a critical pressure value, the bridge becomes unstable, leading to the snap-off. The effect of capillary pressure on the bridges' shapes is illustrated in Fig. \ref{fig:bridge_SO}. In the two images on the left side (a), we see a lateral and a top view of a capillary bridge modeled with the software Surface Evolver. This bridge is formed between two adjacent spheres, which rest between two parallel planes, representing a small sample of a Hele-Shaw cell filled with glass beads. In these images, the capillary pressure corresponds to \SI{231}{\pascal}, the spheres have a diameter of \SI{1}{\milli\meter}, and are separated by a gap of \SI{0.3}{\milli\meter}.
In \ref{fig:bridge_SO}(b), we have the same sphere configuration, but the bridge is at a capillary pressure of \SI{297}{\pascal}. It's noticeable how increasing the capillary pressure made the bridge thinner, which could eventually lead to its rupture, as shown in \ref{fig:bridge_SO}(c). In this third set of images, all the liquid contained at the former capillary bridge is redistributed to the four liquid rings formed between the beads and the Hele-Shaw planes, ceasing the possibility of flow. This scenario helps illustrate how we can have liquid at the corners of granular media while not necessarily establishing an overall wetting-phase continuity.

\begin{figure}[htb]
    \centering
    \includegraphics[width=1\linewidth]{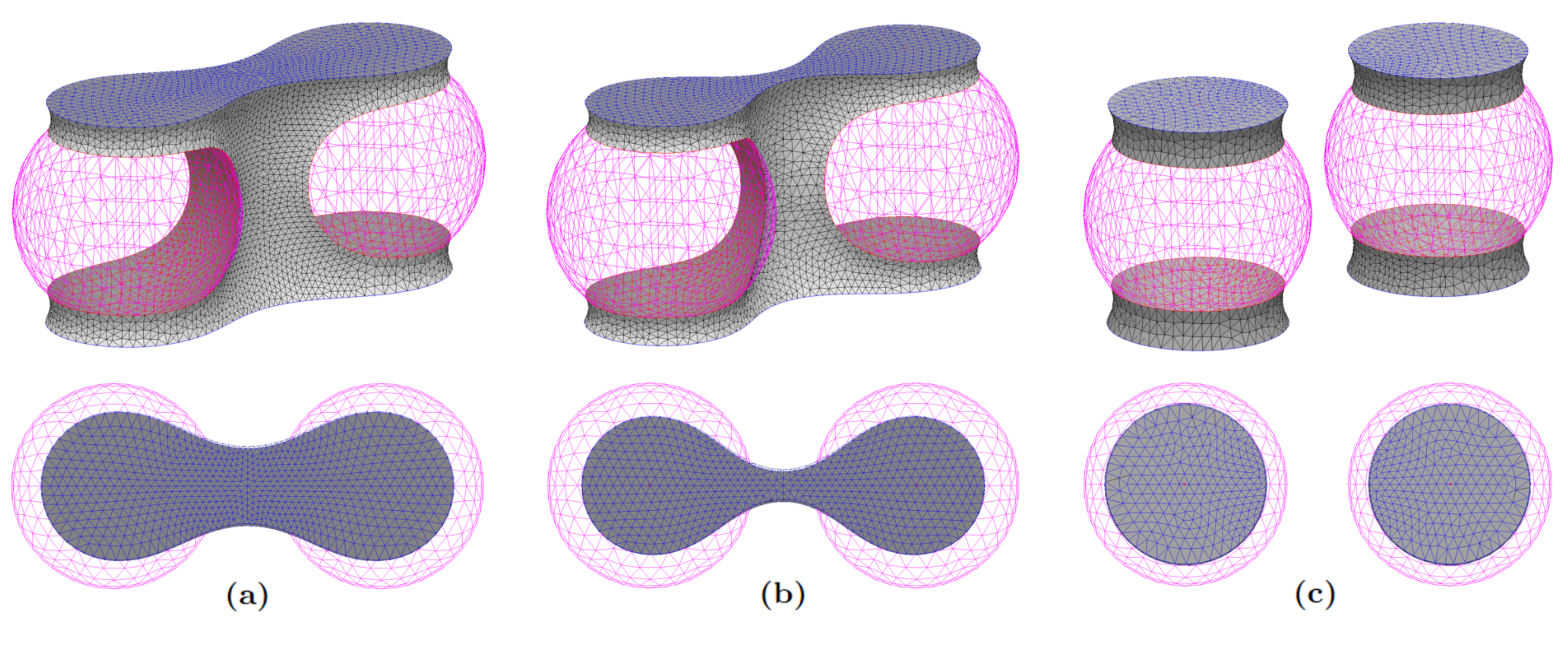}
    \caption{Examples of capillary bridges modeled with the software Surface Evolver. Here, the spherical beads placed between planes, represented by the pink mesh, have a diameter of 1 mm and are separated by a distance of 0.3 mm. The blue regions of the capillary bridge mesh represent the contact surfaces with the Hele-Shaw cell planes, while the black mesh represents the interface with air. The lateral and top views of the bridge shown in (a) correspond to a capillary pressure of 231 Pa. In (b), the same spherical bead configuration is shown, but the capillary bridge shape corresponds to a capillary pressure of 297 Pa. Higher values of capillary pressure would lead to this bridge snap-off, as shown in (c). In this case, the capillary pressure is reduced to 123 Pa, as the total liquid of the bridge is redistributed into the 4 liquid rings between the planes and beads. Solid and fluid properties described in Sec. \ref{sec:met} were used in these simulations.}
    \label{fig:bridge_SO}
\end{figure}

This phenomenon was observed in experiments conducted by \citet{moura2019connectivity} and could lead to important drainage interruptions in the wake of the invasion front. It was verified numerically \cite{reis2023simplified} that even the snap-off of single capillary bridges could disconnect and trap significant portions of liquid in quasi-2D unsaturated granular media. To incorporate these effects in our model, we adopted the same criterion for capillary bridge snap-off as proposed in \citet{reis2023simplified}. In this way, for the sake of simplicity, the capillary pressure at the snap-off of a bridge in a given throat is approximated by the maximum capillary pressure at the piston-like invasion of the same throat (Eq. \eqref{eq:dP_cap}, with $x'_m=0$), given the similar curvatures experienced by the gas-liquid interfaces in both scenarios. 

\subsubsection{Flow Between Capillary Bridges and Liquid Clusters}
\label{sec:g_conn_elements}

In sections \ref{sec:g_throat} and \ref{sec:g_bridge}, the conductances of pore throats -- represented as network elements of the hexagonal lattice -- and capillary bridges -- represented as network elements of the triangular lattice -- were defined. In order to have a single network containing elements from both lattices, we also need to define the flow through connecting links between capillary bridges and pore throats, as shown in Figs. \ref{fig:duallat_seq} and \ref{fig:duallat} by the dashed edges. 

To estimate the conductance of such links, we leveraged the resemblance of their flow paths to that of a capillary bridge. As suggested in Fig. \ref{fig:bridge_cfd}(b) and further discussed in \ref{sec:theoretical_modelling_bridge_g}, most of the viscous pressure drop over a capillary bridge occurs at the corners around the solid-solid contact points between the beads and the planes. A similar flow path through corners is followed by the liquid as it goes from a capillary bridge to a pore throat. Given that the length of these corners in a connecting link is half of that of a capillary bridge, we approximated these links' conductances as twice the conductance described in Eqs.  Eq.\ \eqref{eq:fit_cubic} and \ \eqref{eq:g_bridge_gen} for a bridge.

\subsection{Governing Equations}
\label{sec:goveq}

With the flow through pore throats, capillary bridges, and their connecting links defined in Sec. \ref{sec:q_nets}, we can write the equations for the drainage of the networks described in Sec. \ref{sec:duallat}. To do so, first, we define the incidence matrix $C$, relating the network's nodes and edges, as indicated in Eq. \eqref{eq:C}. 

\begin{equation}
c_{ij}=
\begin{cases}
        1 & \text{if edge $j$ enters node $i$}\\
       -1 & \text{if edge $j$ leaves node $i$}\\
        0 & \text{otherwise}
    \end{cases}
    \label{eq:C}
\end{equation}

Using $C$, the system of equations for flow in the networks is presented in Eqs. \eqref{eq:syseq}. Two sets of variables are defined in this system: the pressure $P_i$ in every node $i$ in the network, and the position of the meniscus $x'_{m,j}$, for every throat containing two phases. To relate these variables, volume balance (Eq. \eqref{eq:volcons}) is enforced in every node, and an equation for the motion of the menisci (Eq. \eqref{eq:xm}) is written as a function of the throat's flow rate and cross-sectional area. The choice of edge orientation in $C$ does not affect the network flow calculations for throats containing only liquid. Throats containing a meniscus are oriented leaving the gas-filled node and entering the liquid-filled node so that the sign of the capillary pressure is consistent with the flow direction. Eq. \eqref{eq:q_jfull} describes how the flow through each throat or capillary bridge is written using the matrix $C$ and concepts detailed in Sec. \ref{sec:q_nets}. 

\begin{subequations}

    \begin{equation}
        \sum_{j=1}^{n_e} c_{ij}q_j = 0
    \label{eq:volcons}    
    \end{equation}
    
    \begin{equation}
        \frac{dx'_{m,j}}{dt}=\frac{q_j}{\pi r(x'_{m,j})^2}
    \label{eq:xm}
    \end{equation}

    \begin{equation}
        q_j = -\Gamma_j\Biggl\{ \sum_{k=1}^{n_n} c_{kj} \left[P_k + h_k\rho _{l}g\left(\frac{l-x'_{m,j}\delta_m}{l}\right)\right] + \delta_m\gamma\left( \frac{\cos{\theta_b}}{r(x'_{m,j})} + \frac{\cos{\theta_p}}{r_b}\right) \Biggr\}
    \label{eq:q_jfull}
    \end{equation}
    
    \label{eq:syseq}
\end{subequations}

\noindent where $q_j$ is the volumetric flow rate through a network edge $j$, $n_e$ is the total number of edges in the network, $n_n$ is the total number of nodes in the network, $\delta_m$ is used to differentiate the calculation of flow through throats containing a meniscus, $\delta_m=1$, from the flow through liquid-filled throats, capillary bridges, and connecting links, $\delta_m=0$.

These equations are numerically solved in the proposed model by, first, approximating Eq. \eqref{eq:xm} with the Implicit Euler method, then applying the Newton-Raphson method to the resulting system of nonlinear equations, for each time step $dt$. During the drainage simulation, time steps are chosen so that the maximum displacement among the menisci does not exceed approximately $10\%$ of the throat's length. 

\begin{figure}[ht!]
     \centering
     
     \begin{subfigure}[t]{0.32\textwidth}
         \centering
         \includegraphics[width=0.95\textwidth]{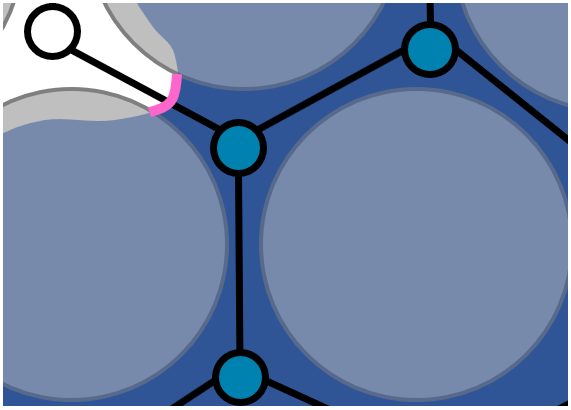}
         \caption{}
         \label{fig:xm_1}
     \end{subfigure}
     \hfill
     \begin{subfigure}[t]{0.32\textwidth}
         \centering
         \includegraphics[width=0.95\textwidth]{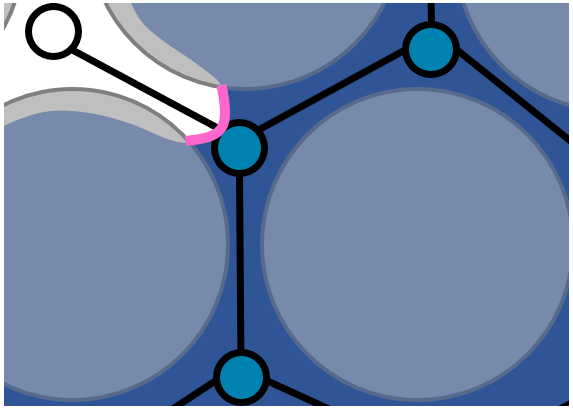}
         \caption{}
         \label{fig:xm_2}
     \end{subfigure}
     \hfill
     \begin{subfigure}[t]{0.32\textwidth}
         \centering
         \includegraphics[width=0.95\textwidth]{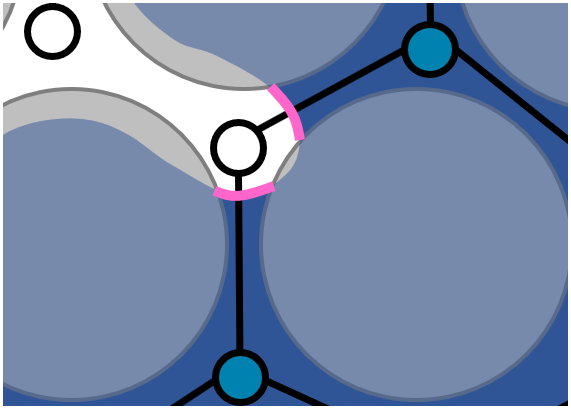}
         \caption{}
         \label{fig:xm_3}
     \end{subfigure}
     \hfill
        \caption{(a) A throat contains an advancing meniscus located at its midsection. (b) The advancing meniscus reaches the throat's full length $x'_m=l$, prompting a shift in the pore-throat network. (c) In this case, the edge representing the throat is deleted, the occupation of the node changes from liquid to gas, and new menisci are created on the edges incident on the newly gas-filled node, with initial position $x'_m=0$.}
        \label{fig:xm}
\end{figure}

In order to have networks representing only the liquid occupation of the porous medium, as introduced in Sec. \ref{sec:duallat}, whenever a meniscus reaches the full length of a throat, the network shifts as indicated in Fig. \ref{fig:xm}. In this case, the edge representing the throat is deleted, the occupation of its end node is changed from liquid to gas, and new menisci are created at the edges incident on that node. Once the pore-throat network is changed, its structure is checked for the creation of new capillary bridges. The procedure to identify the formation of bridges used here is the same as presented in \citet{reis2023simplified}. If a new bridge is formed, new nodes and edges in the triangular lattice network are created to account for the corner flow (see Fig. \ref{fig:duallat_seq}). At any time step, these bridges may also be deleted, in case the average pressure among its end nodes exceeds the bridge's snap-off pressure, defined in Sec. \ref{sec:so_bridge}. The solution of Eqs. \eqref{eq:syseq} combined with the network's updates is successively carried out until the non-wetting phase percolates through the pore space.

\subsection{Simulation Parameters}
\label{sec:nets_conds}

The networks used to generate the results presented in Sec. \ref{sec:res} represent Hele-Shaw cells with dimensions \SI{12}{\centi\meter}$ \cross $\SI{14}{\centi\meter} (width vs.\ height), filled with a monolayer of glass beads with a diameter equivalent to \SI{1}{\milli\meter}. Presuming a regular arrangement of glass beads (see Sec. \ref{sec:duallat}), this leads to HL networks with 13203 nodes and, initially, 19723 edges. Values of $r_t$ are assigned randomly to the edges, following the pore-size distribution reported in \citet{moura2019connectivity}. All edges have the same length of $l=\SI{1}{\milli\meter}$. 

As for the simulated drainage conditions, the rectangular porous medium cell is initially saturated with liquid, with the top face open to the air inflow. The bottom face of the cell is connected to a liquid outlet, to which a flow rate is prescribed, while the lateral faces are impermeable. The porous medium cells can also have a tilt angle $\beta$ with respect to the horizontal plane so that the flow is affected by gravitational forces (see Fig. \ref{fig:hele-shaw}). Under these conditions, drainage is carried out until the air percolates the cell.

Therefore, we assume that nodes at the top of the networks are initially filled with air, while all others contain liquid. At the bottom of the network, all nodes are connected by high-conductance edges to a single outlet node, to which a fixed drainage rate is assigned. In the presented results, the following set of flow rates was tested $q_{net}=[10^{-3},\SI{3.5e-2},10^{-2},10^{-1},1,10] \si{\milli\liter\per\minute}$. 
As for the network tilt angle, the following values were used: $\beta=[0,10,20,30,45,60,75,90]^{\circ}$. In this manner, a large range of capillary and Bond numbers were explored in this study. For each set of tested drainage parameters, $40$ different networks were used in the simulations, so that the results reflect the flow conditions, not a particular porous medium realization with randomly pore-throat sizes.

To compare the results with experimental data presented in \citet{moura2019connectivity}, the wetting phase represented in the model is a glycerol-water mixture, with viscosity and density equivalent to $\mu_l=\SI{5e-2}{\pascal\second}$ and $\rho_l=\SI{1.2e3}{\kilogram\per\meter\cubed}$, 
respectively. The non-wetting phase is air, within which we considered the effects of viscous and gravitational forces on the pressure field to be negligible. For the calculation of the conductance of throats containing both liquid and gas, nonetheless, a viscosity of $\mu_g=\SI{1.8e-5}{\pascal\second}$ was used to represent the air, so that numerical problems were avoided due to $g_j\to\infty$ when the meniscus reached the throat's length. Besides, different contact angles are assumed between the fluids' interface and the glass beads, $\theta_b=\SI{20}{\degree}$, and between the fluids' interface and the Hele-Shaw cell planes $\theta_p=60^{\circ}$, which were covered with a contact paper sheet for experiments presented in \citet{moura2019connectivity}.

\section{Results}
\label{sec:res}

As illustrated in Fig. \ref{fig:FlowCond}, the broad range of drainage flow conditions adopted in this study led to distinct flow regimes, including capillary fingering, stable displacement, and viscous fingering. In the axis of Fig.\ref{fig:FlowCond}, Bond numbers were calculated as 
$ Bo= \rho_l g \sin{\beta} a^2 \mathbin{/} \gamma $, where $a$ is the typical pore size, of $\SI{1}{\milli\meter}$. Capillary numbers were calculated as $Ca= q_{net} \mu_l a^2 \mathbin{/} A\gamma k_{net}$, where $A$ is the networks transversal area, of $\SI{1.2}{\centi\meter\squared}$, and $k_{net}$ is the average network permeability of $\SI{3.8e-5}{\centi\meter\squared}$.

In each image, white represents the gas occupation of a pore network at breakthrough, while black represents both the liquid-filled regions and the solid matrix. As proposed by \citet{meheust2002interface}, we can notice that a stable drainage front was only achieved when $Bo^*=Bo-Ca>0$. On the top row of flow cases, corresponding to $Bo=0$, we notice the transition from the capillary fingering type of instability to viscous fingering, as $Ca$ increases from $\SI{2.8e-5}{}$ to $\SI{2.8e-1}{}$. Unstable flows are also present in the last column, when $Ca=\SI{2.8e-1}{}$, as the gravitational pressure gradient is not sufficient to bound the growth of viscous fingers. These results suggest that the model proposed in sec. \ref{sec:met} can appropriately represent the interplay between capillary, viscous, and gravitational forces during drainage in porous media at finite speeds.

In the next sections, detailed images of the cases delineated in red, which correspond to a subset of the stable displacement cases, are shown separately for clarity. These cases correspond to tilt angles of $\beta=[20,30,60]^{\circ}$ and imposed liquid withdrawal rates of $q_{net}=[0.01, 0.1, 1]\si{\milli\liter\per\minute}$.

\begin{figure}[ht!]
    \centering
    \includegraphics[width=0.9\textwidth]{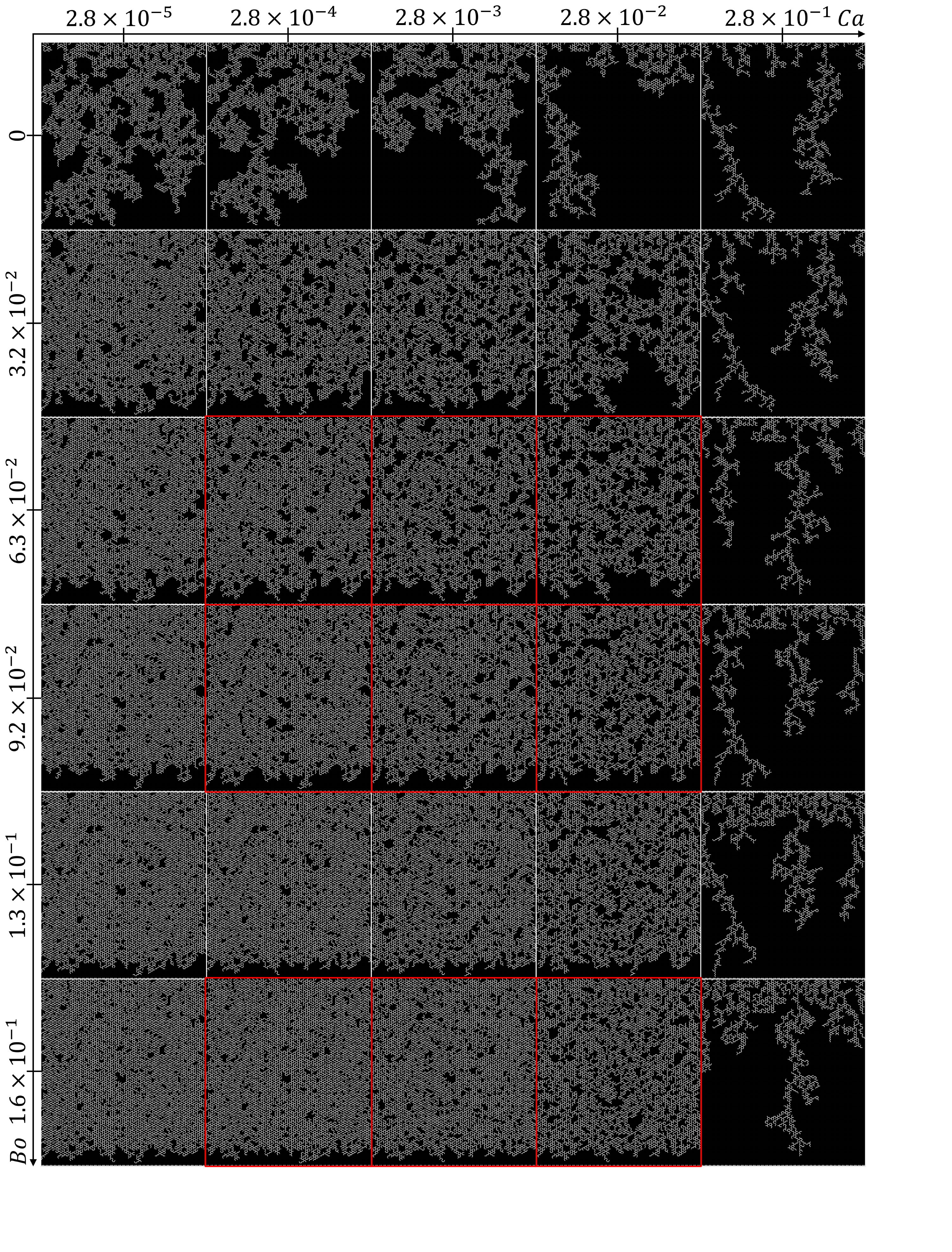}
    \caption{Gas occupation of a pore network at breakthrough, for various investigated flow conditions. As observed, stable displacement, capillary, and viscous fingering regimes were covered in our analyses.}
    \label{fig:FlowCond}
\end{figure}

\subsection{Total drainage through capillary-bridge-connected paths }
\label{sec:sat_mech}

Figure \ref{fig:InvasionMode} represents the regions of a pore network that were invaded by gas under different flow conditions, separated by drainage mechanism. Green areas correspond to the primary drainage mechanism, during which pores along the main invasion front are invaded in a fast piston-like displacement. Yellow areas correspond to the secondary drainage mechanism, indicating pores that had belonged to clusters but were drained using connected pathways formed by capillary bridges and liquid rings. All nine cases in this figure represent stable gas displacements, and, as expected, decreasing the Bond number and/or increasing the capillary number led to larger bypassed liquid clusters and reduced primary drainage efficiency. From these clusters, we can verify that a more significant fraction could be later drained via corner flow in slower displacements, when compared to fast drainage, for the same Bond numbers.

\begin{figure}[ht!]
     \centering
     
     \begin{subfigure}[t]{0.32\textwidth}
         \centering
         \includegraphics[width=0.95\textwidth]{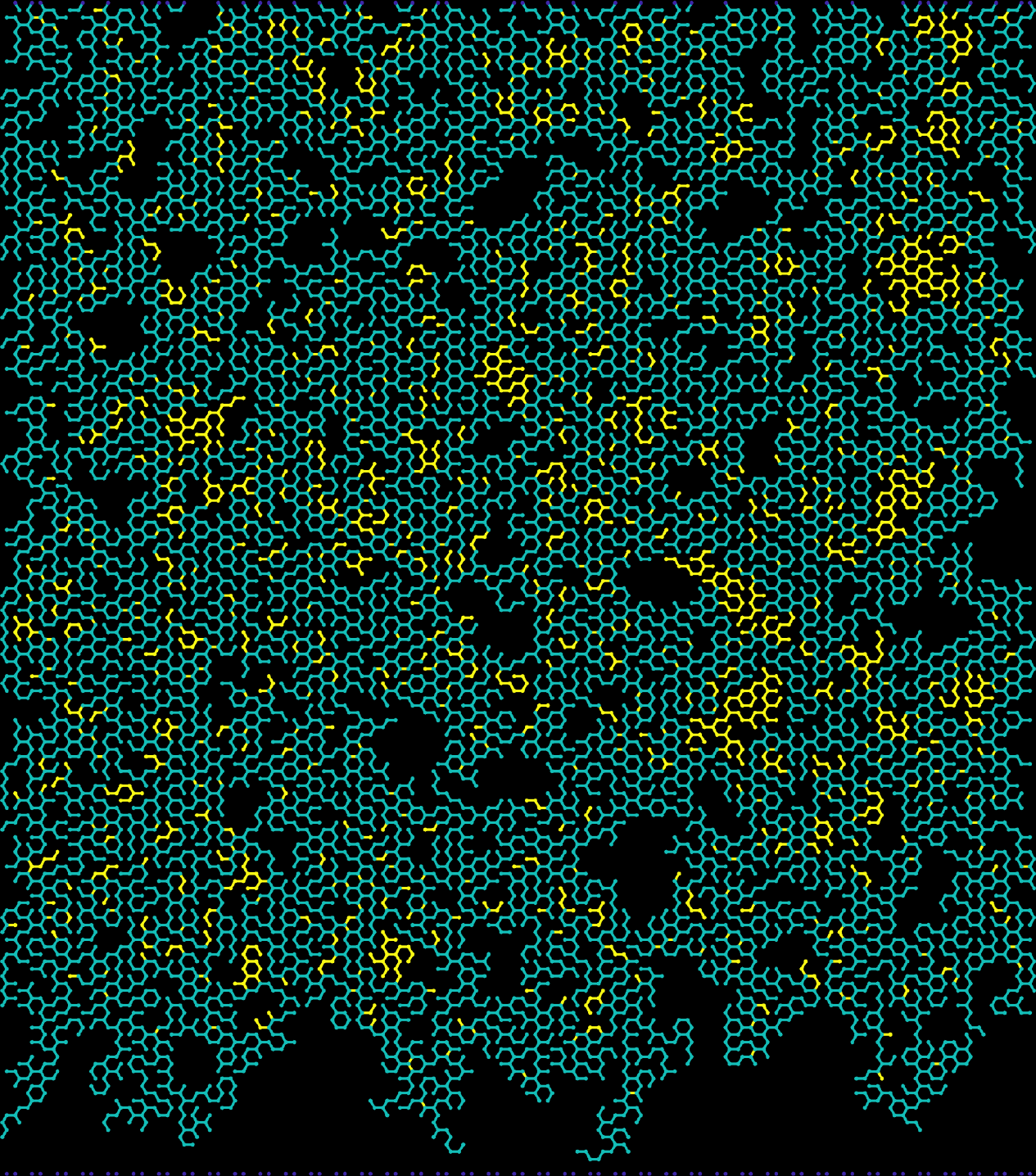}
         \caption{$Bo\approx 6 \times 10^{-2}$ $Ca\approx3\times10^{-4}$}
         \label{fig:im_20_0.01}
     \end{subfigure}
     \hfill
     \begin{subfigure}[t]{0.32\textwidth}
         \centering
         \includegraphics[width=0.95\textwidth]{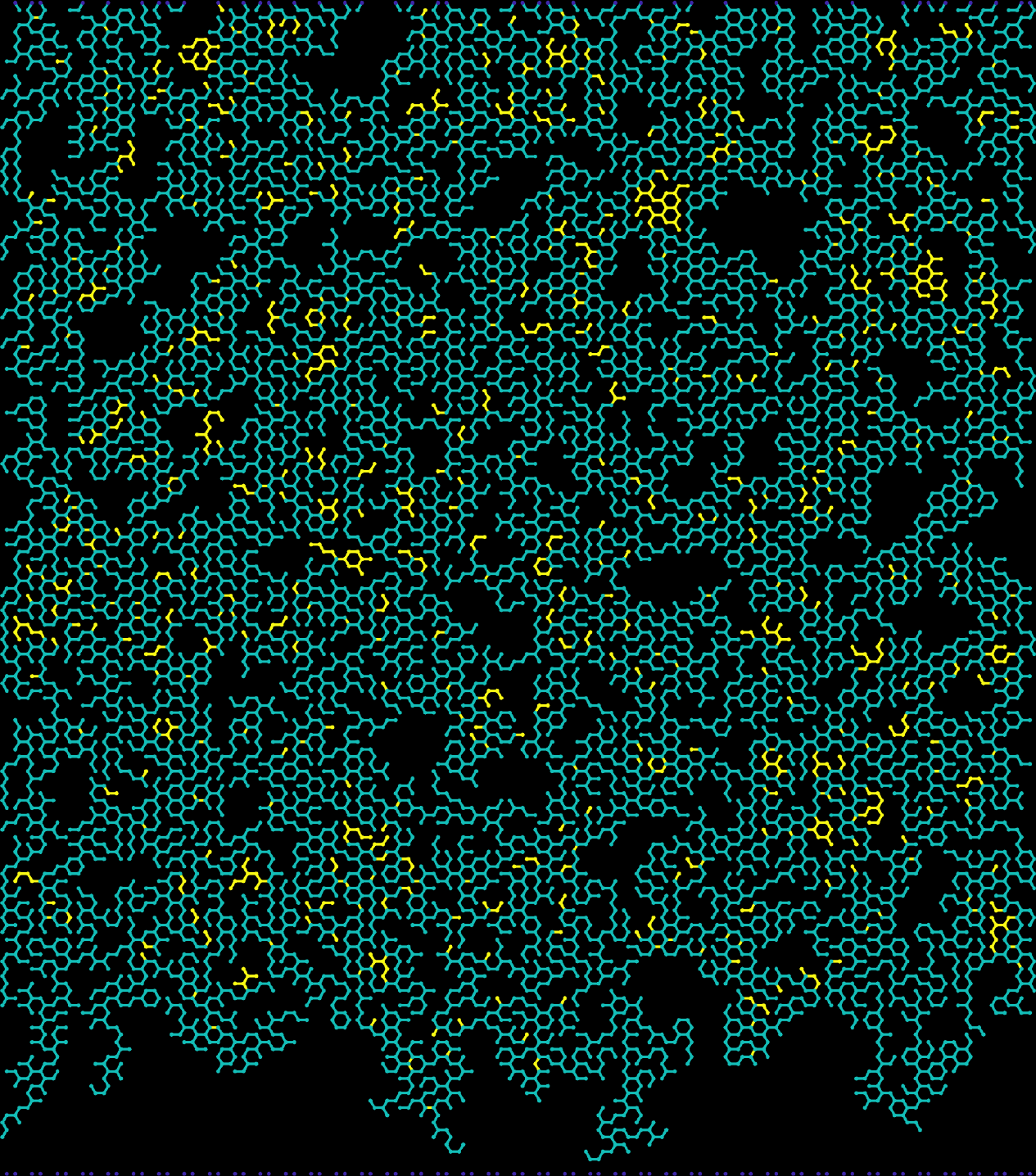}
         \caption{$Bo\approx 6 \times 10^{-2}$ $Ca\approx3\times10^{-3}$}
         \label{fig:im_20_0.1}
     \end{subfigure}
     \hfill
     \begin{subfigure}[t]{0.32\textwidth}
         \centering
         \includegraphics[width=0.95\textwidth]{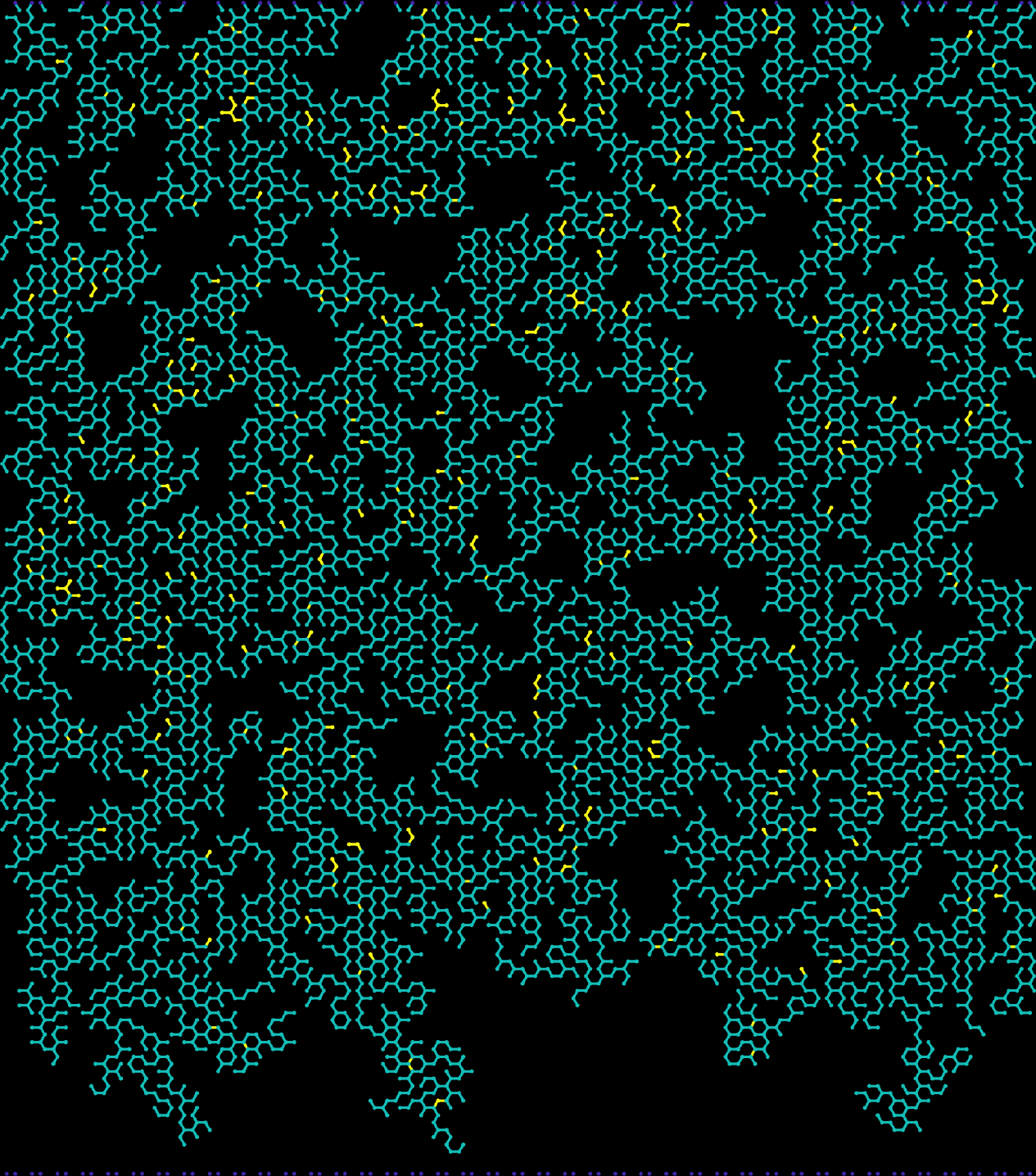}
         \caption{$Bo\approx 6 \times 10^{-2}$ $Ca\approx3\times10^{-2}$}
         \label{fig:im_20_1}
     \end{subfigure}
     \hfill
     \begin{subfigure}[t]{0.32\textwidth}
         \centering
         \includegraphics[width=0.95\textwidth]{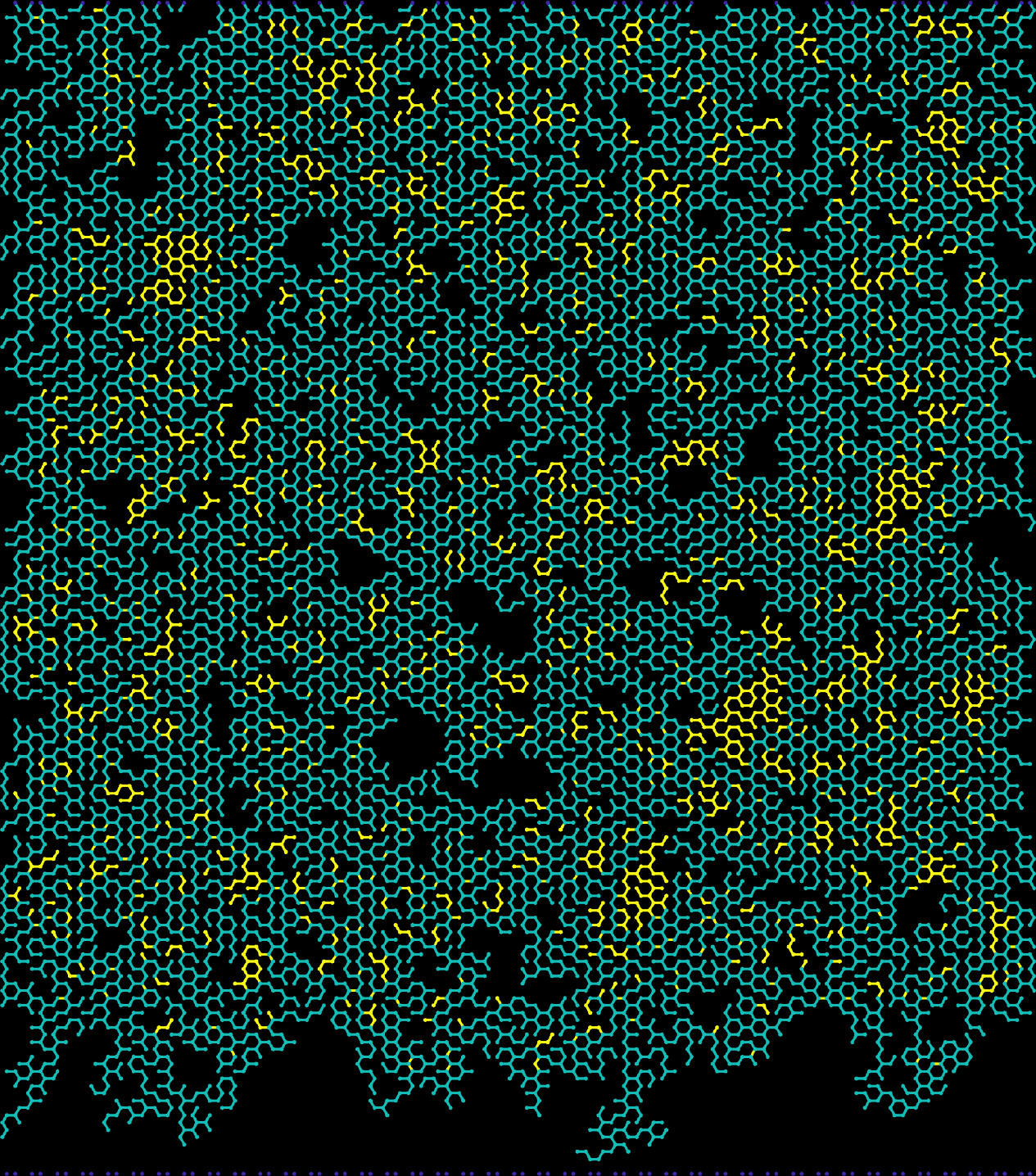}
         \caption{$Bo\approx9\times 10^{-2}$ $Ca\approx3\times10^{-4}$}
         \label{fig:im_30_0.01}
     \end{subfigure}
     \hfill
     \begin{subfigure}[t]{0.32\textwidth}
         \centering
         \includegraphics[width=0.95\textwidth]{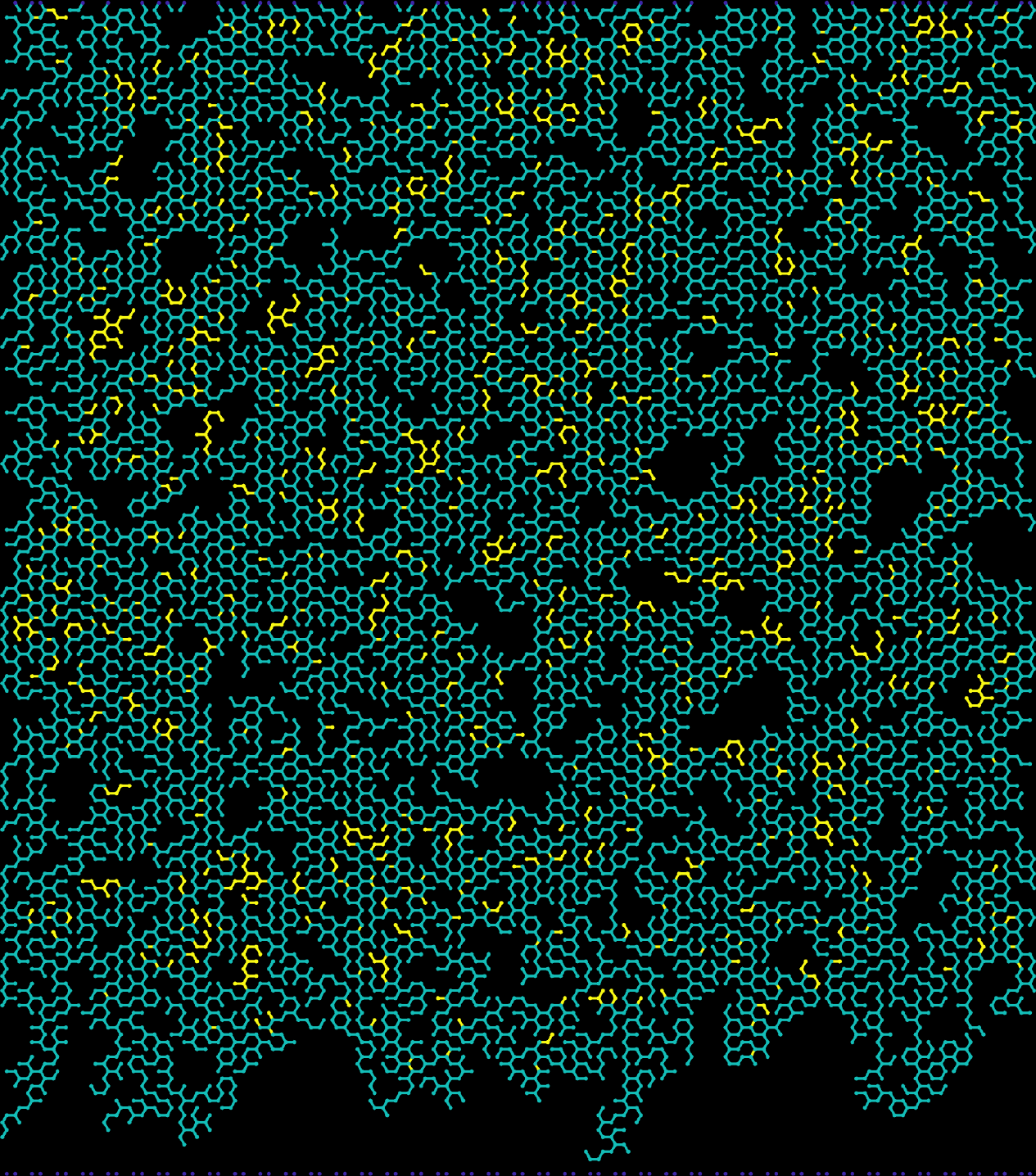}
         \caption{$Bo\approx9\times 10^{-2}$ $Ca\approx3\times10^{-3}$}
         \label{fig:im_30_0.1}
     \end{subfigure}
     \hfill
     \begin{subfigure}[t]{0.32\textwidth}
         \centering
         \includegraphics[width=0.95\textwidth]{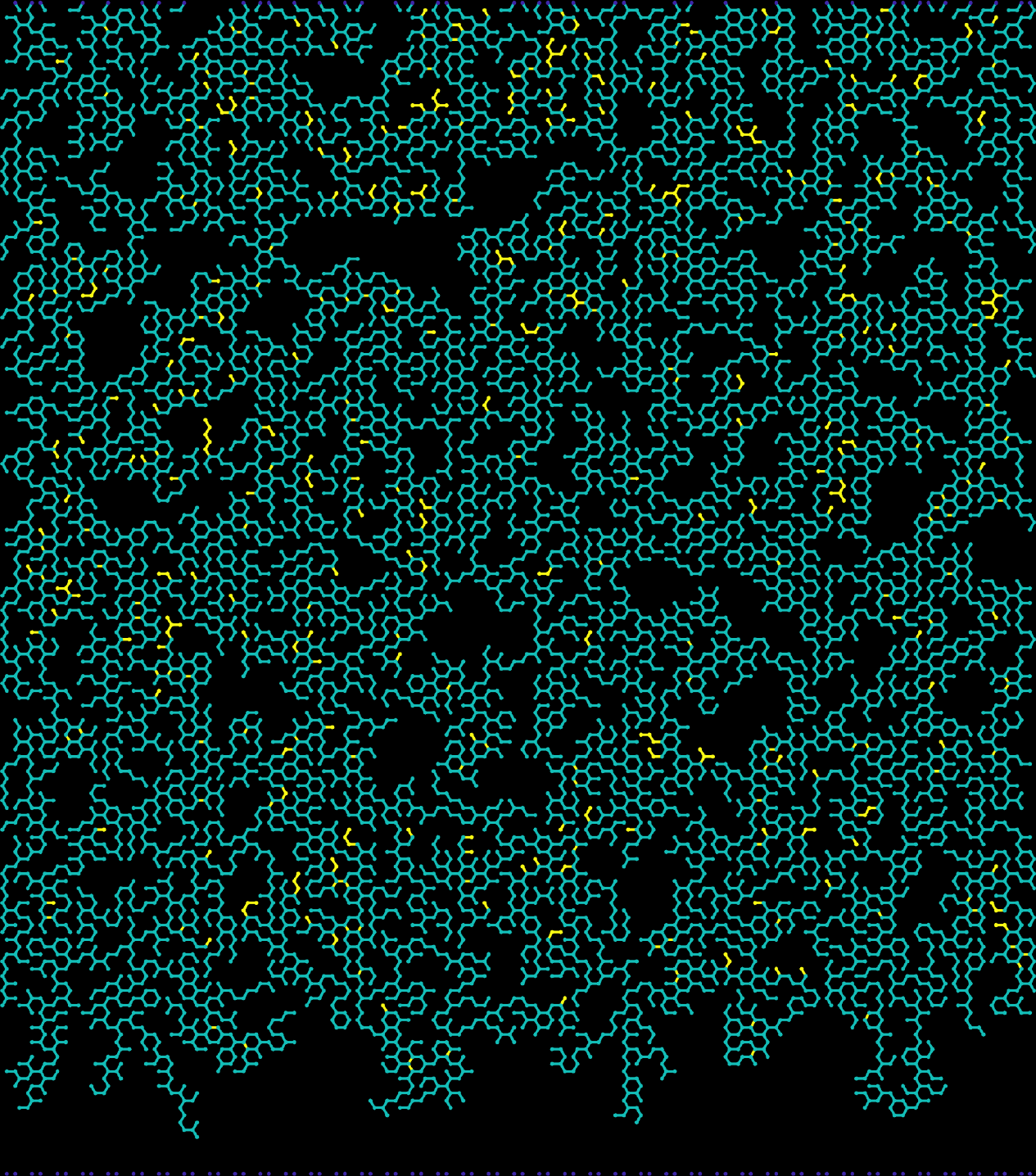}
         \caption{$Bo\approx9\times 10^{-2}$ $Ca\approx3\times10^{-2}$}
         \label{fig:im_30_1}
     \end{subfigure}
     \hfill
     \begin{subfigure}[t]{0.32\textwidth}
         \centering
         \includegraphics[width=0.95\textwidth]{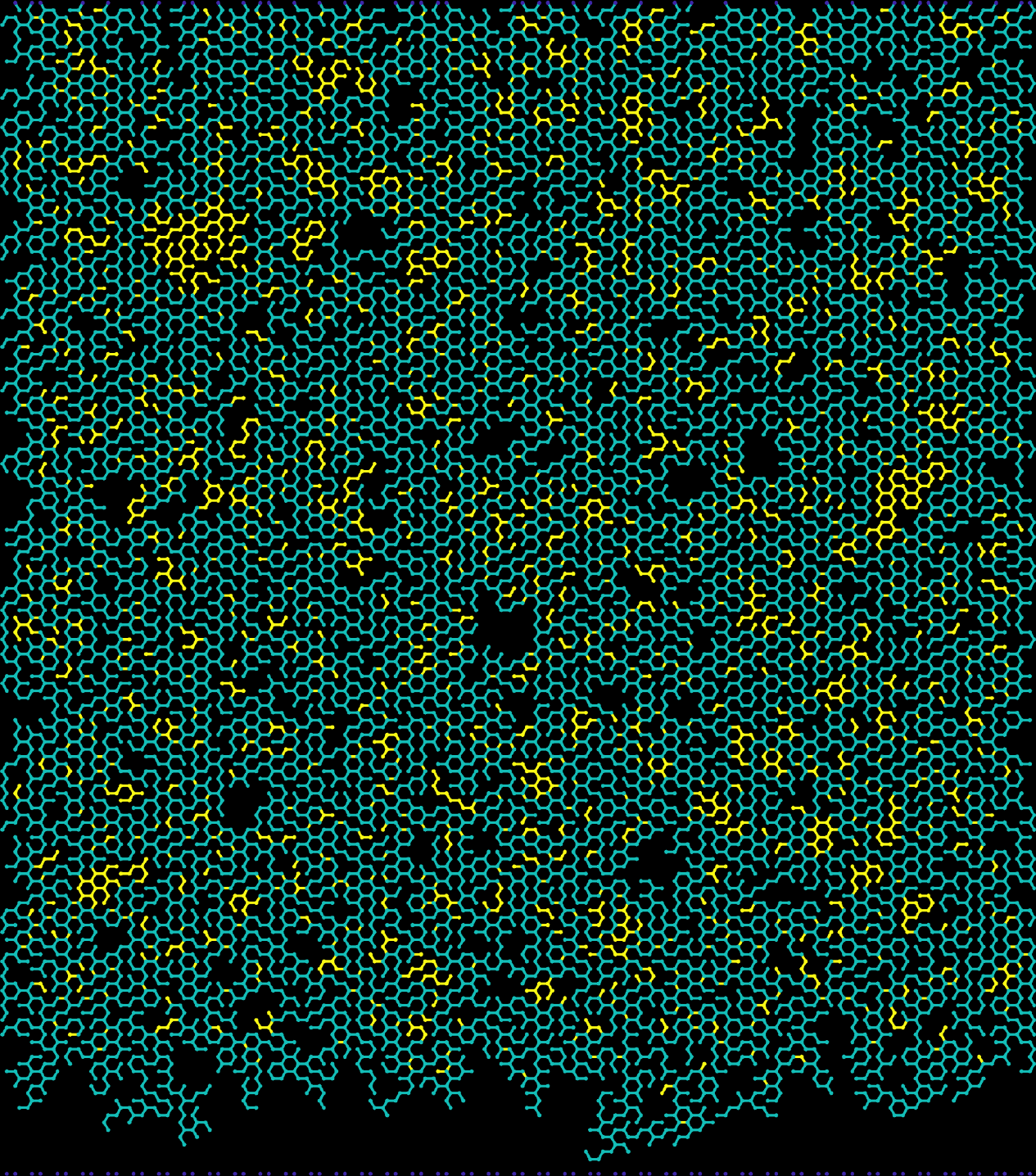}
         \caption{$Bo\approx1.6\times 10^{-1}$ $Ca\approx3\times10^{-4}$}
         \label{fig:im_60_0.01}
     \end{subfigure}
     \hfill
     \begin{subfigure}[t]{0.32\textwidth}
         \centering
         \includegraphics[width=0.95\textwidth]{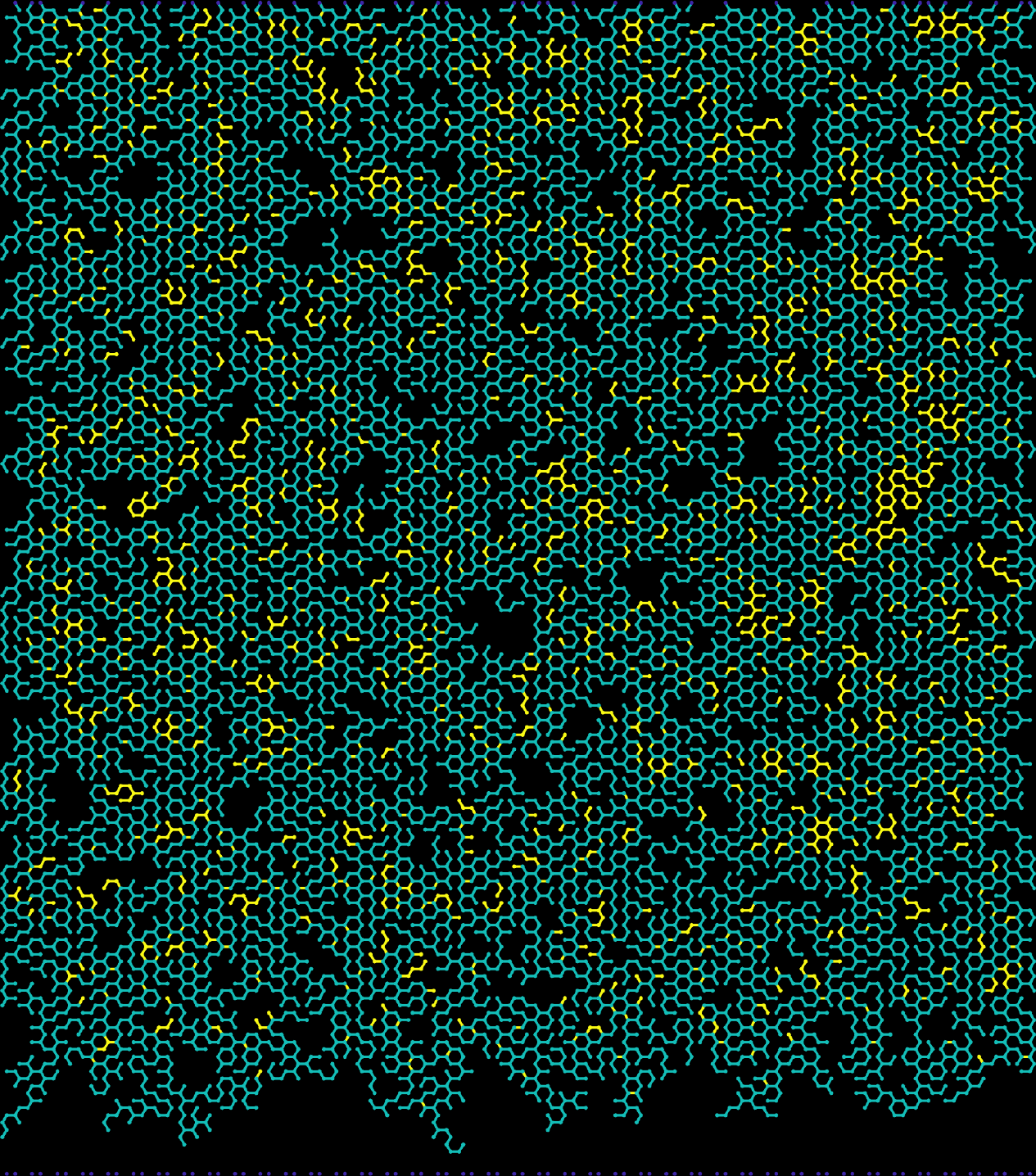}
         \caption{$Bo\approx1.6\times 10^{-1}$ $Ca\approx3\times10^{-3}$}
         \label{fig:im_60_0.1}
     \end{subfigure}
     \hfill
     \begin{subfigure}[t]{0.32\textwidth}
         \centering
         \includegraphics[width=0.95\textwidth]{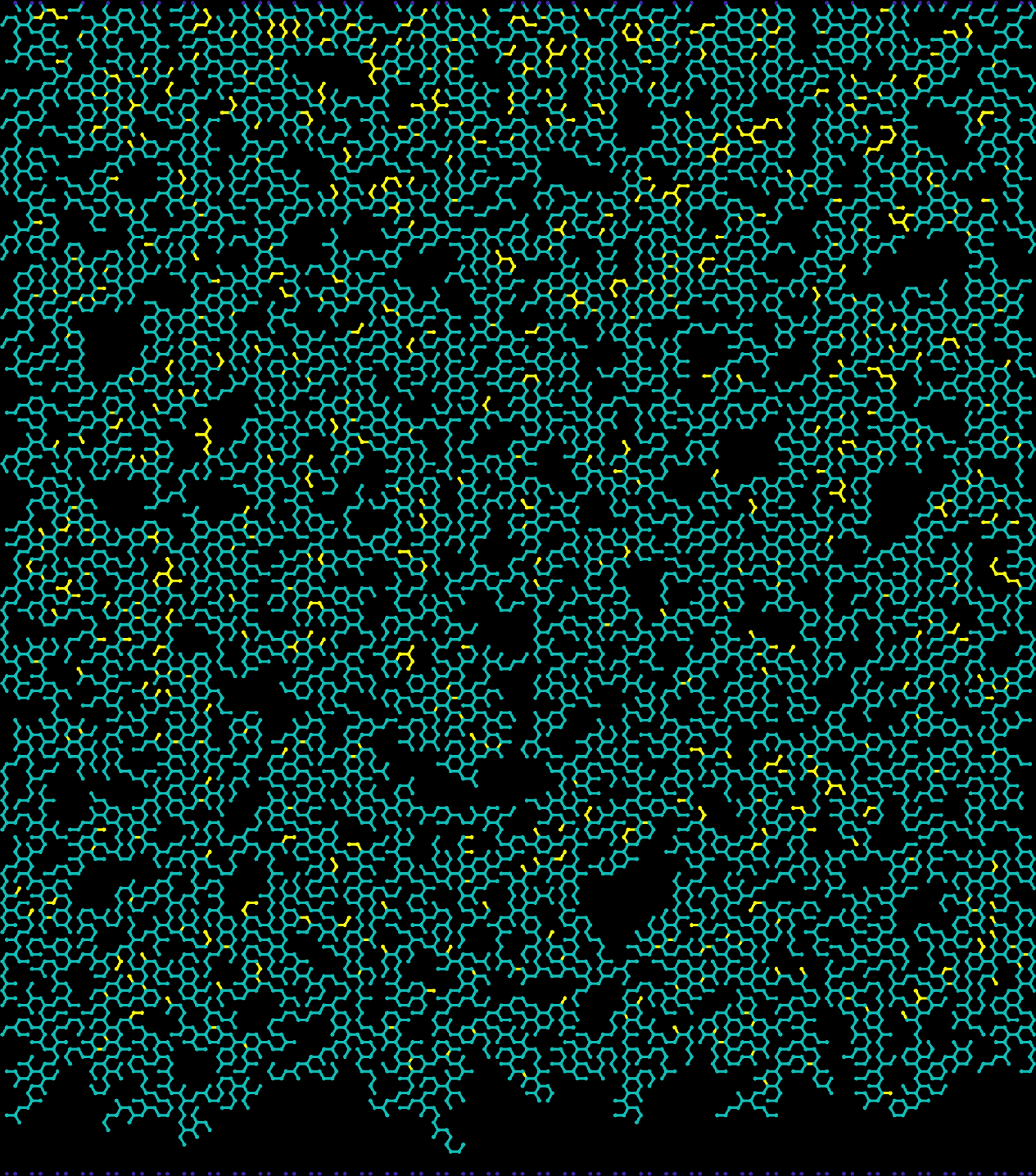}
         \caption{$Bo\approx1.6\times 10^{-1}$ $Ca\approx3\times10^{-2}$}
         \label{fig:im_60_1}
     \end{subfigure}
     
        \caption{Representation of the gas occupation of the pore networks at breakthrough, under various flow conditions. In green, we have pores and throats occupied under the primary drainage mechanism. In yellow, we have the regions of the network only drained due to the liquid connectivity established by capillary bridges and liquid rings.}
        \label{fig:InvasionMode}
\end{figure}

\begin{figure}[ht!]
     \centering
     
     \begin{subfigure}[t]{0.32\textwidth}
         \centering
         \includegraphics[width=0.95\textwidth]{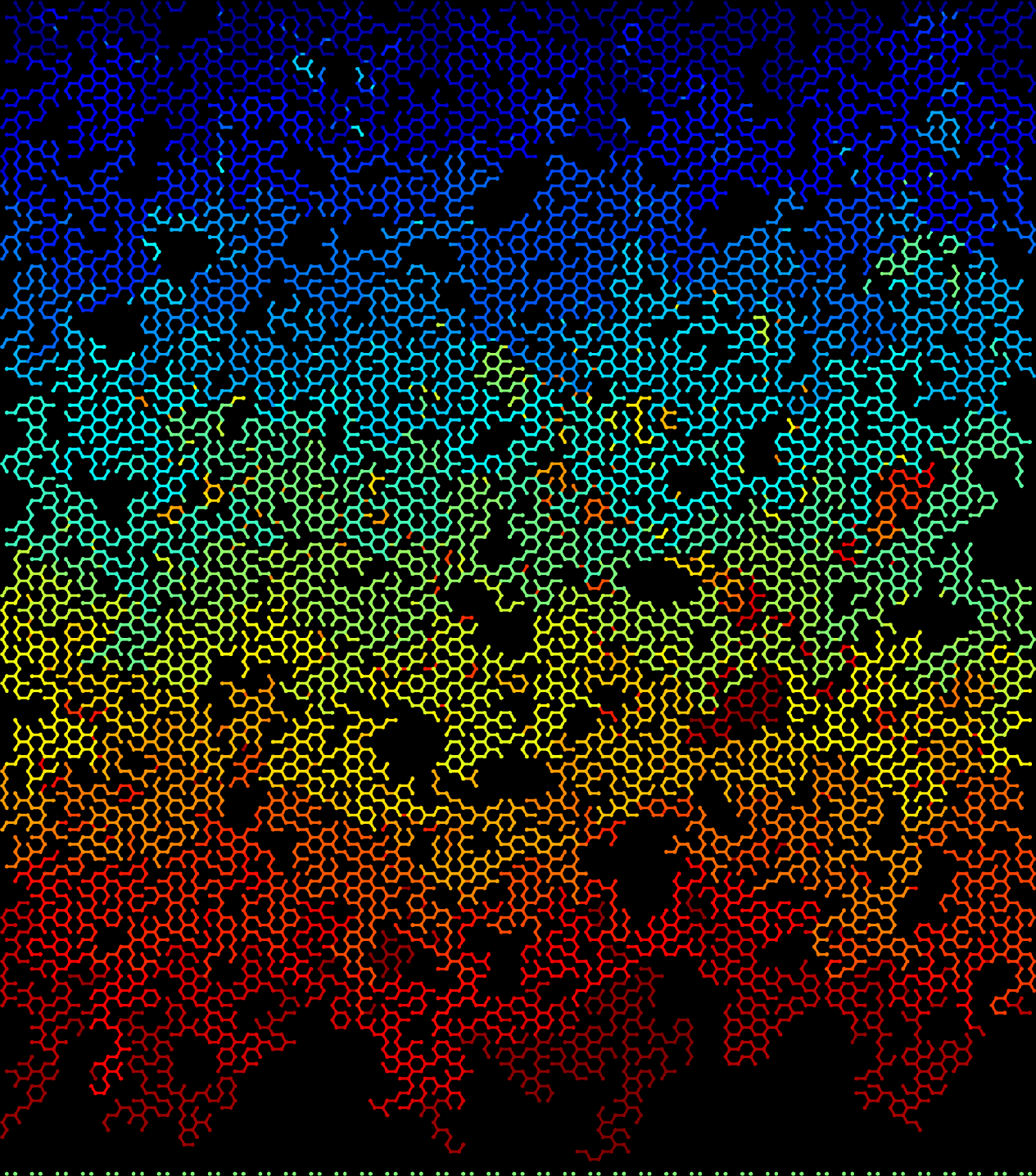}
         \caption{$Bo\approx 6 \times 10^{-2}$ $Ca\approx3\times10^{-4}$}
         \label{fig:it_20_0.01}
     \end{subfigure}
     \hfill
     \begin{subfigure}[t]{0.32\textwidth}
         \centering
         \includegraphics[width=0.95\textwidth]{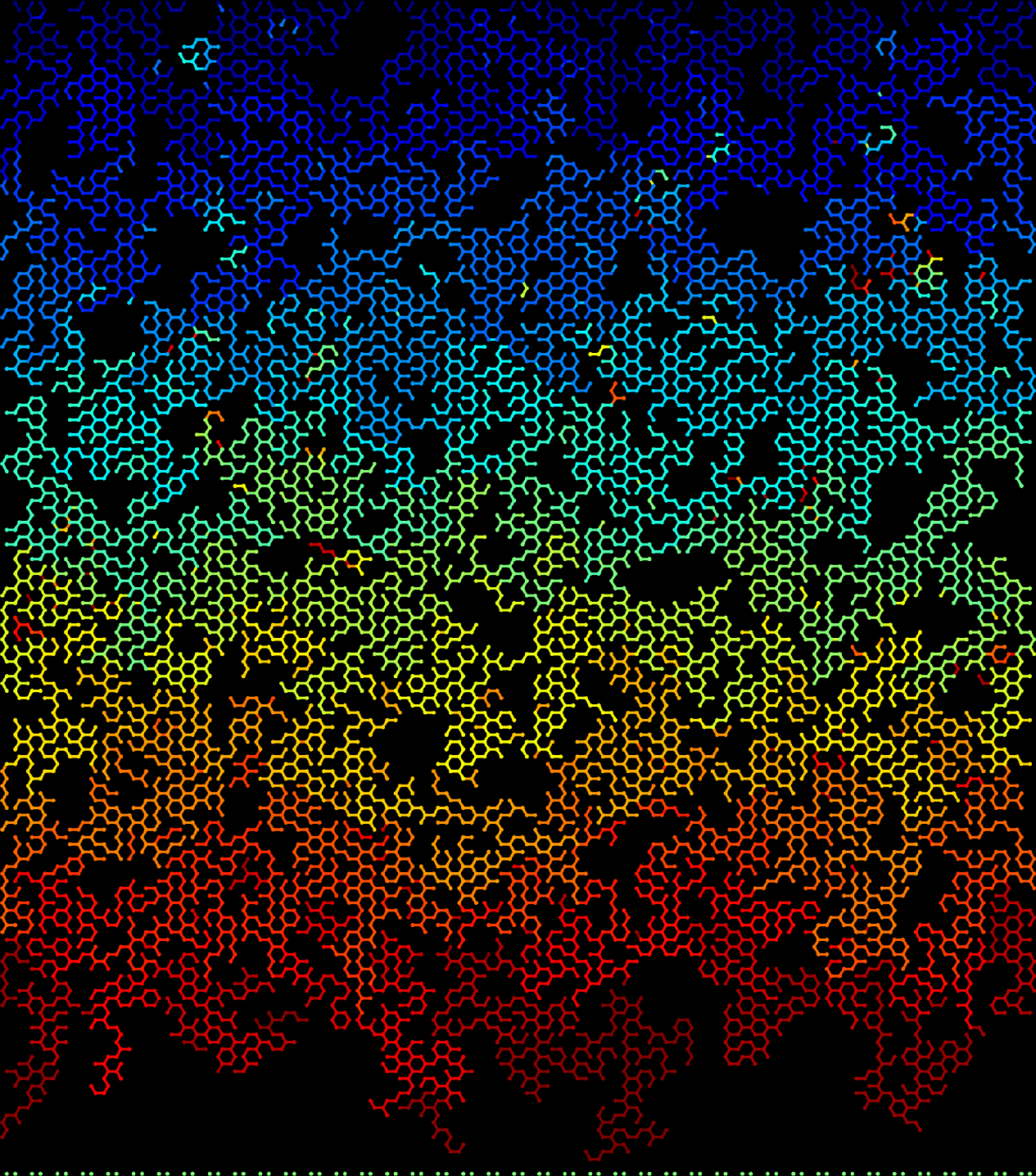}
         \caption{$Bo\approx 6 \times 10^{-2}$ $Ca\approx3\times10^{-3}$}
         \label{fig:it_20_0.1}
     \end{subfigure}
     \hfill
     \begin{subfigure}[t]{0.32\textwidth}
         \centering
         \includegraphics[width=0.95\textwidth]{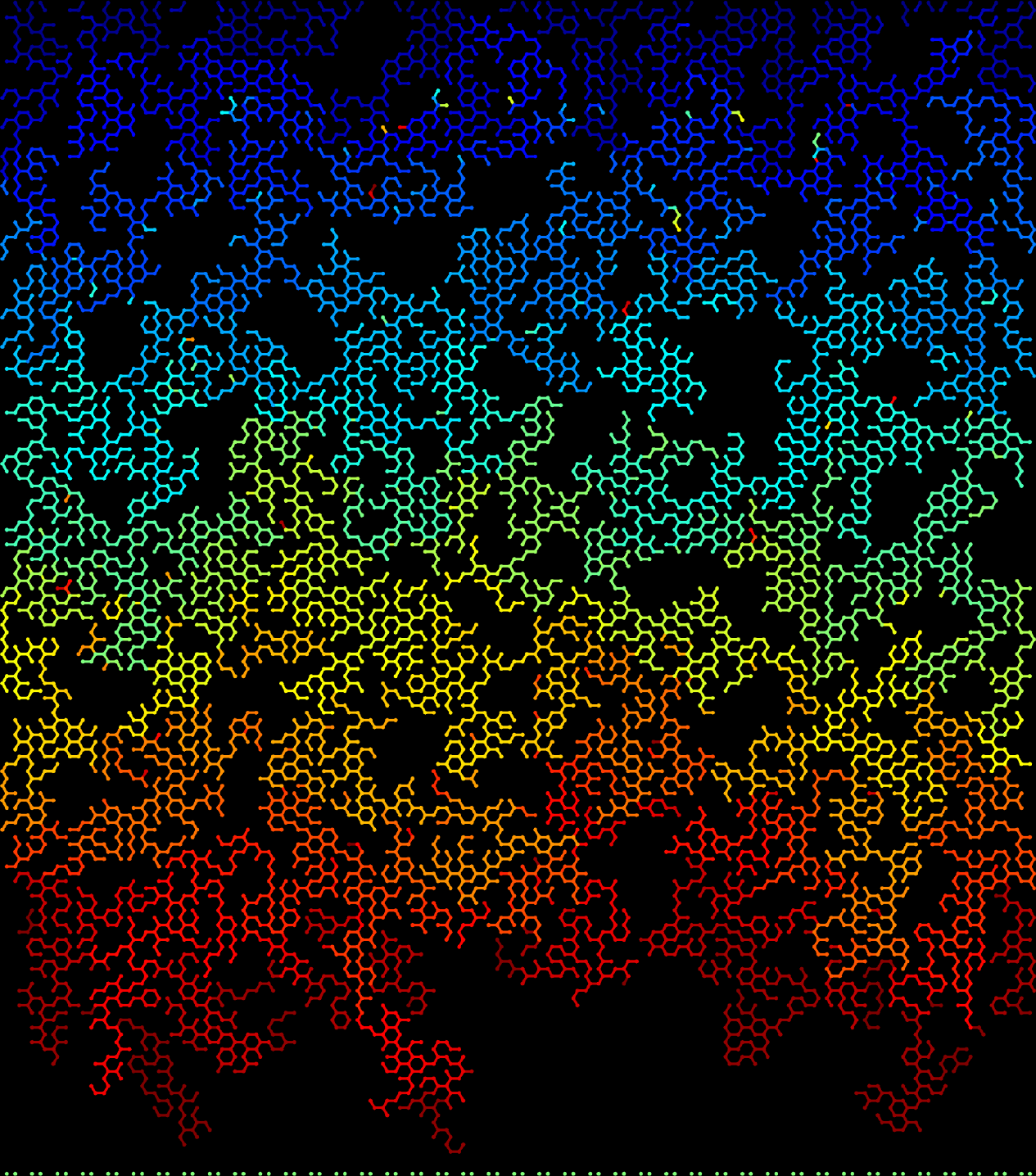}
         \caption{$Bo\approx 6 \times 10^{-2}$ $Ca\approx3\times10^{-2}$}
         \label{fig:it_20_1}
     \end{subfigure}
     \hfill
     \begin{subfigure}[t]{0.32\textwidth}
         \centering
         \includegraphics[width=0.95\textwidth]{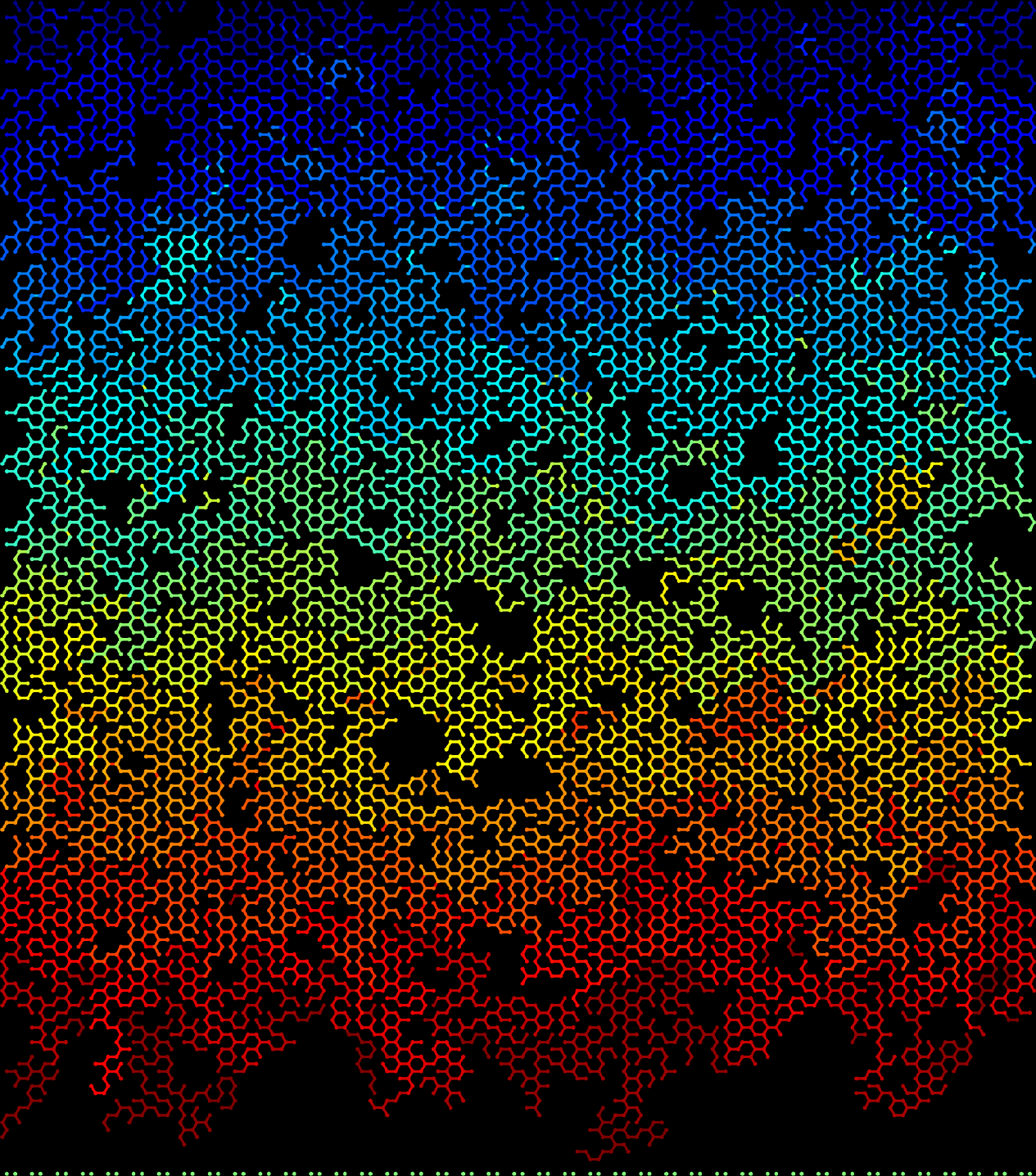}
         \caption{$Bo\approx9\times 10^{-2}$ $Ca\approx3\times10^{-4}$}
         \label{fig:it_30_0.01}
     \end{subfigure}
     \hfill
     \begin{subfigure}[t]{0.32\textwidth}
         \centering
         \includegraphics[width=0.95\textwidth]{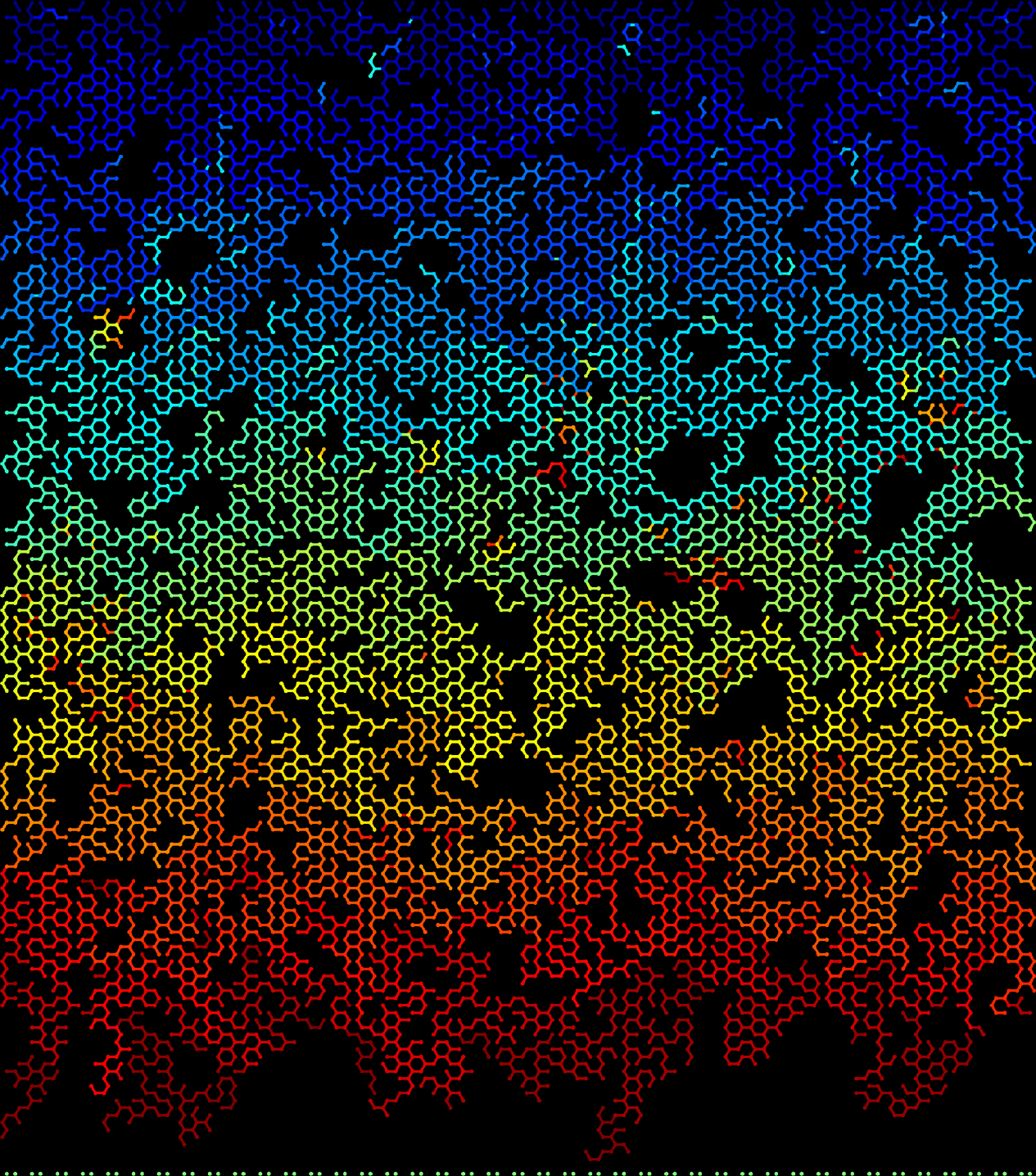}
         \caption{$Bo\approx9\times 10^{-2}$ $Ca\approx3\times10^{-3}$}
         \label{fig:it_30_0.1}
     \end{subfigure}
     \hfill
     \begin{subfigure}[t]{0.32\textwidth}
         \centering
         \includegraphics[width=0.95\textwidth]{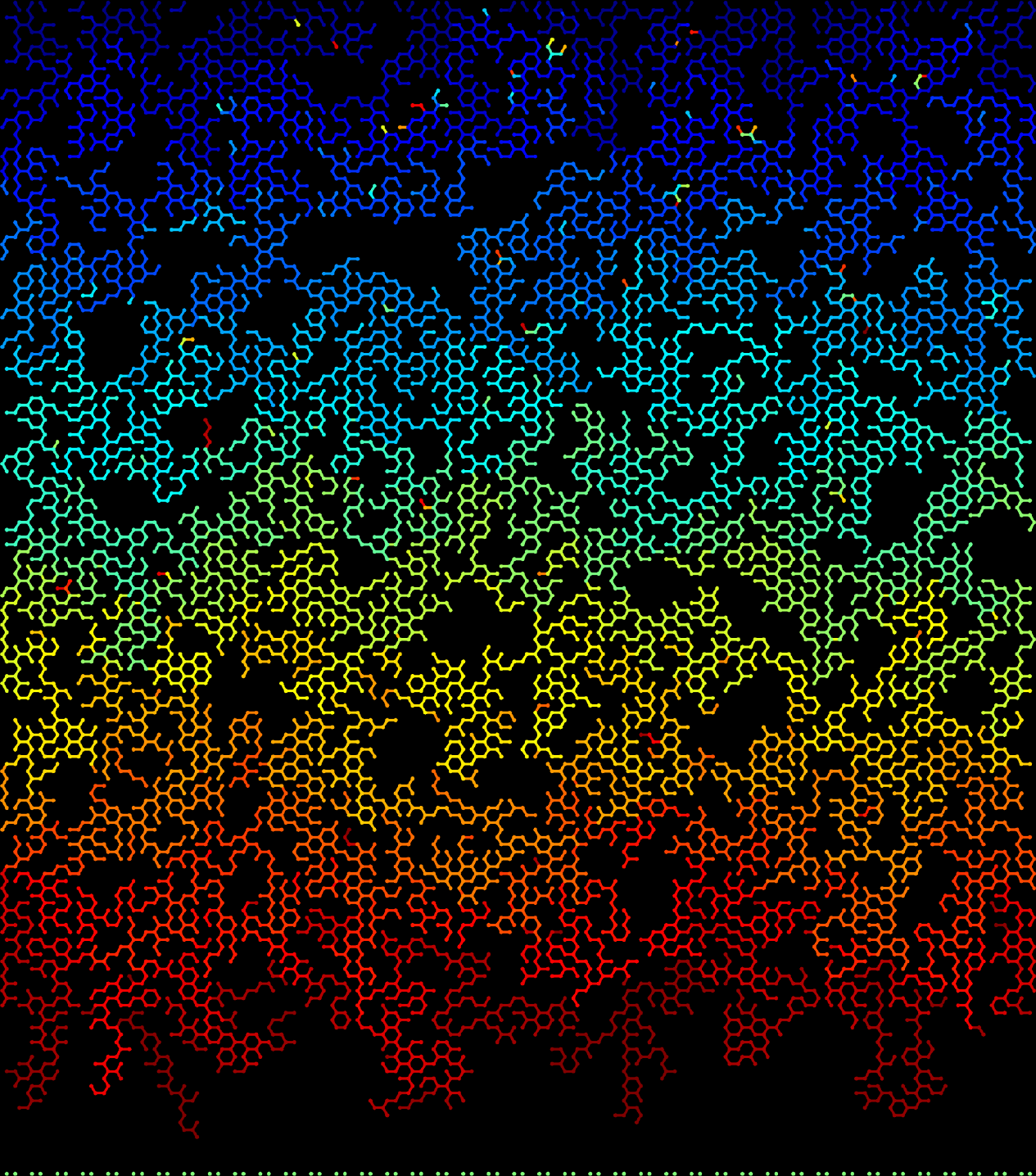}
         \caption{$Bo\approx9\times 10^{-2}$ $Ca\approx3\times10^{-2}$}
         \label{fig:it_30_1}
     \end{subfigure}
     \hfill
     \begin{subfigure}[t]{0.32\textwidth}
         \centering
         \includegraphics[width=0.95\textwidth]{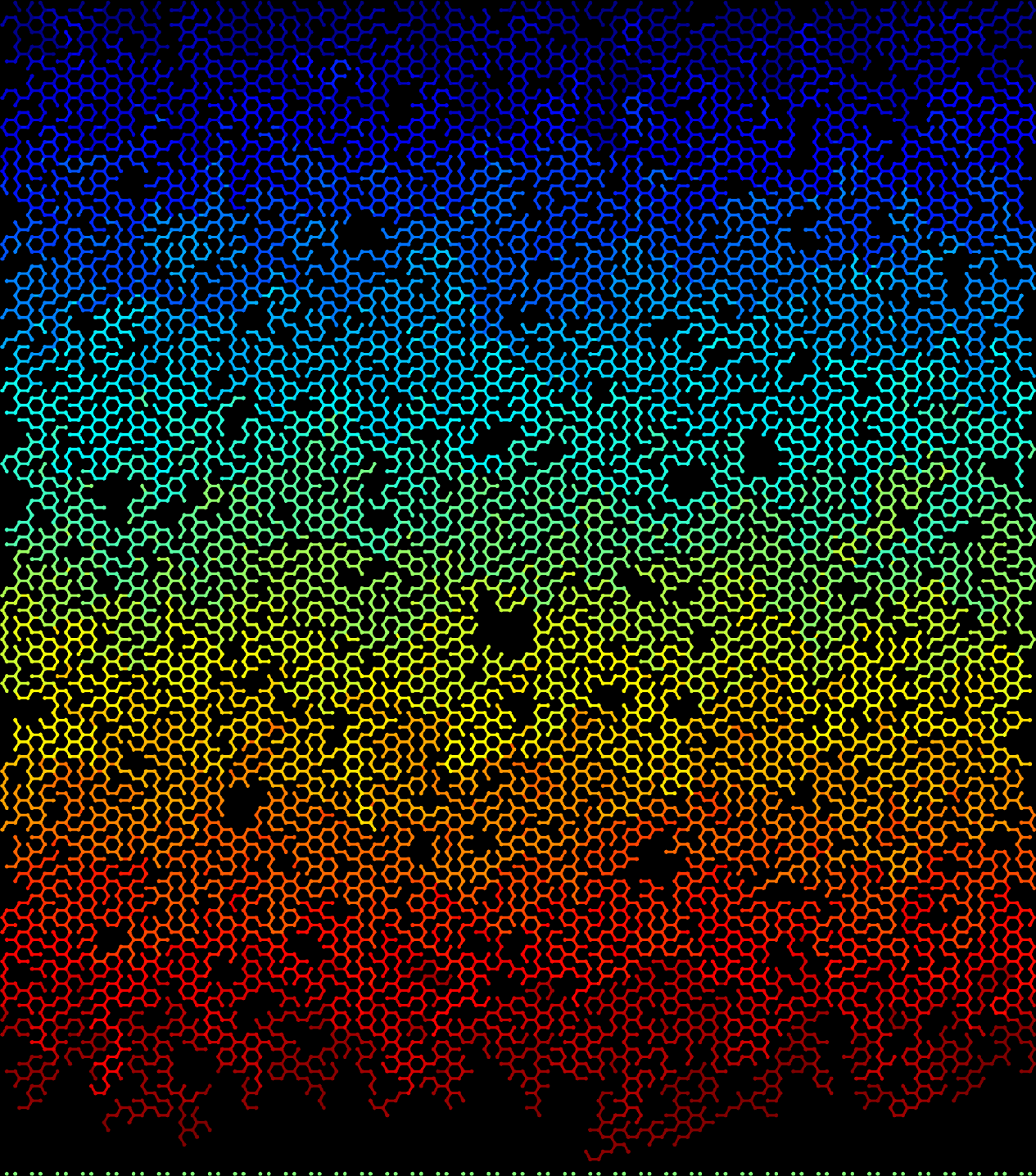}
         \caption{$Bo\approx1.6\times 10^{-1}$ $Ca\approx3\times10^{-4}$}
         \label{fig:it_60_0.01}
     \end{subfigure}
     \hfill
     \begin{subfigure}[t]{0.32\textwidth}
         \centering
         \includegraphics[width=0.95\textwidth]{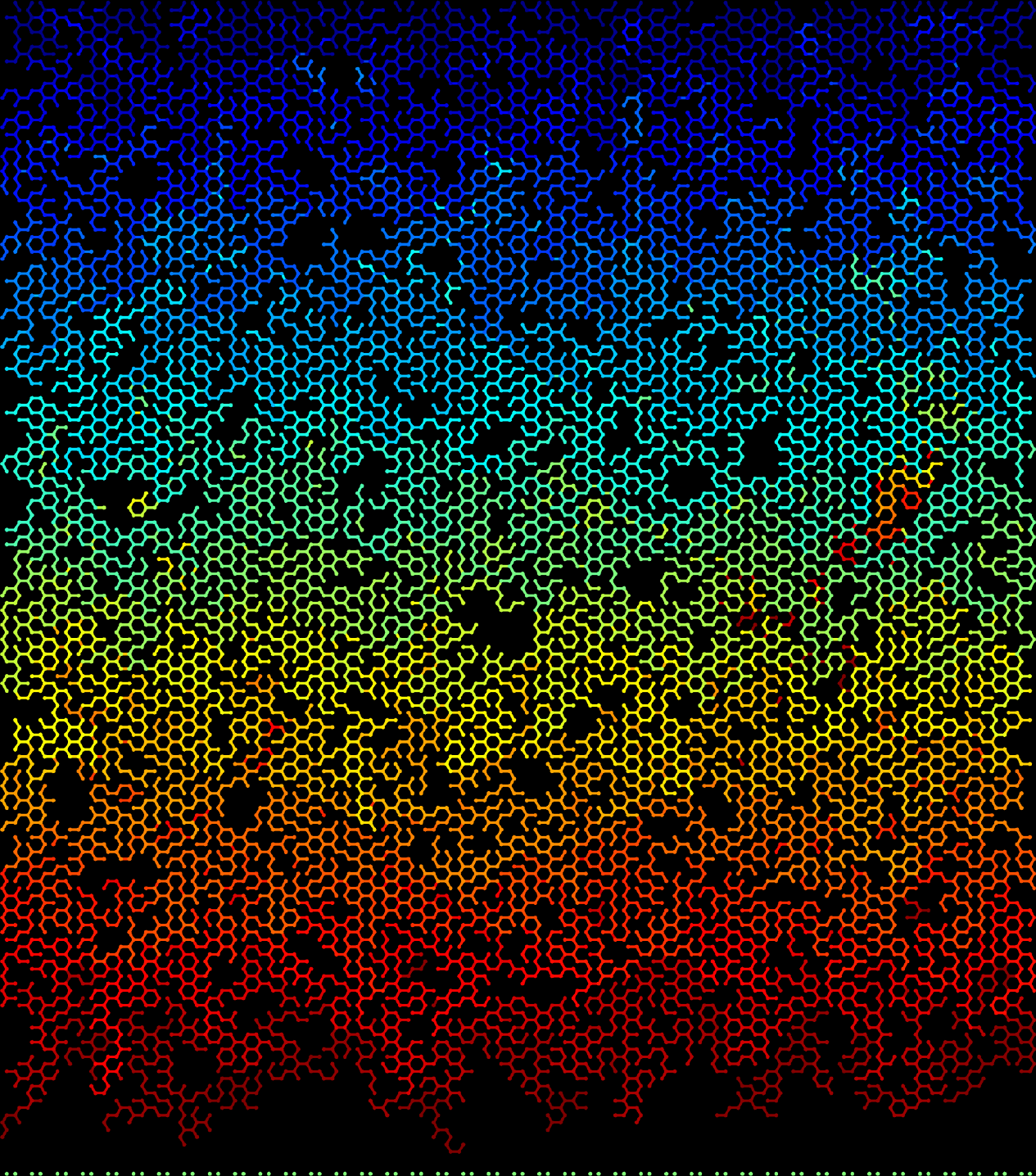}
         \caption{$Bo\approx1.6\times 10^{-1}$ $Ca\approx3\times10^{-3}$}
         \label{fig:it_60_0.1}
     \end{subfigure}
     \hfill
     \begin{subfigure}[t]{0.32\textwidth}
         \centering
         \includegraphics[width=0.95\textwidth]{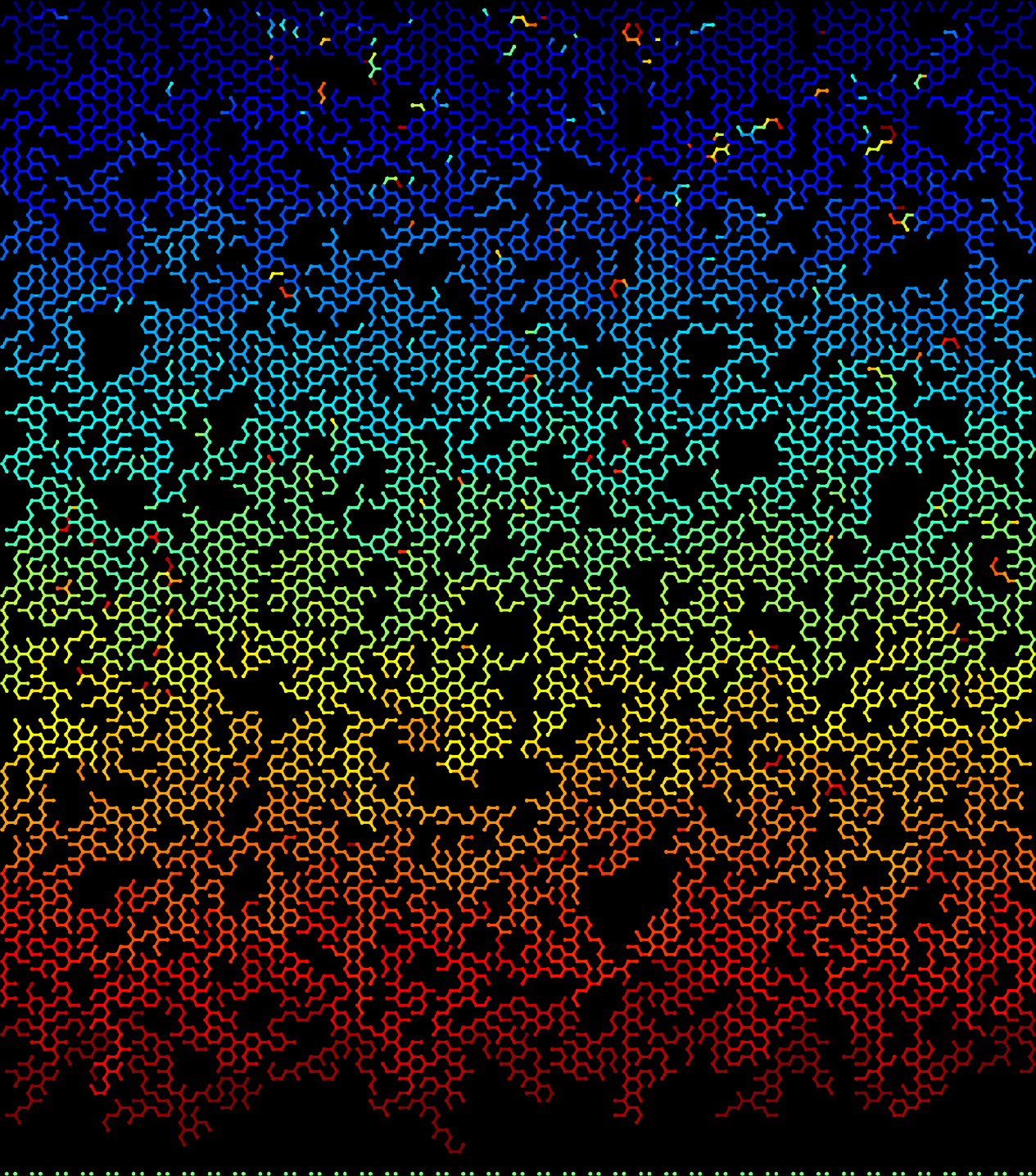}
         \caption{$Bo\approx1.6\times 10^{-1}$ $Ca\approx3\times10^{-2}$}
         \label{fig:it_60_1}
     \end{subfigure}
     
        \caption{Representation of the time of the gas invasion of each drained pore and throat at breakthrough. From dark blue to dark red, the colors indicate early to late times of invasion. Black regions in the images represent liquid clusters that were not drained.}
        \label{fig:InvasionTime}
\end{figure}

The additional time required for clusters to be drained to the front through corners can be inferred from Fig. \ref{fig:InvasionTime}. These images illustrate the same cases shown in Fig. \ref{fig:InvasionMode}, but now the colors, from dark blue to dark red, represent ascending gas invasion times. The horizontal color stratification seen in Fig. \ref{fig:InvasionTime} indicates the stable movement of the gas front from the top of the pore networks, where the gas inlet is located, to their bottom, from which liquid is withdrawn. Small isolated patches with distinctively later invasion times than their neighbors indicate that, under some flow conditions, clusters can be drained through corners far behind the invasion front. Examples of that are the red patches at the center of Fig. \ref{fig:InvasionTime}(a) and the bright spots at the top of Fig. \ref{fig:InvasionTime}(i). Conversely, secondary drainage mechanism events may also take place predominantly in the vicinity of the main invasion front, as indicated in Fig. \ref{fig:InvasionTime}(g). In this case, a significant part of the liquid removal is attributed to the secondary drainage mechanism, as seen in Fig. \ref{fig:InvasionMode}(g), but no noticeable delay is associated with these events.

\begin{figure}[ht!]
    \centering
    \includegraphics[width=0.75\textwidth]{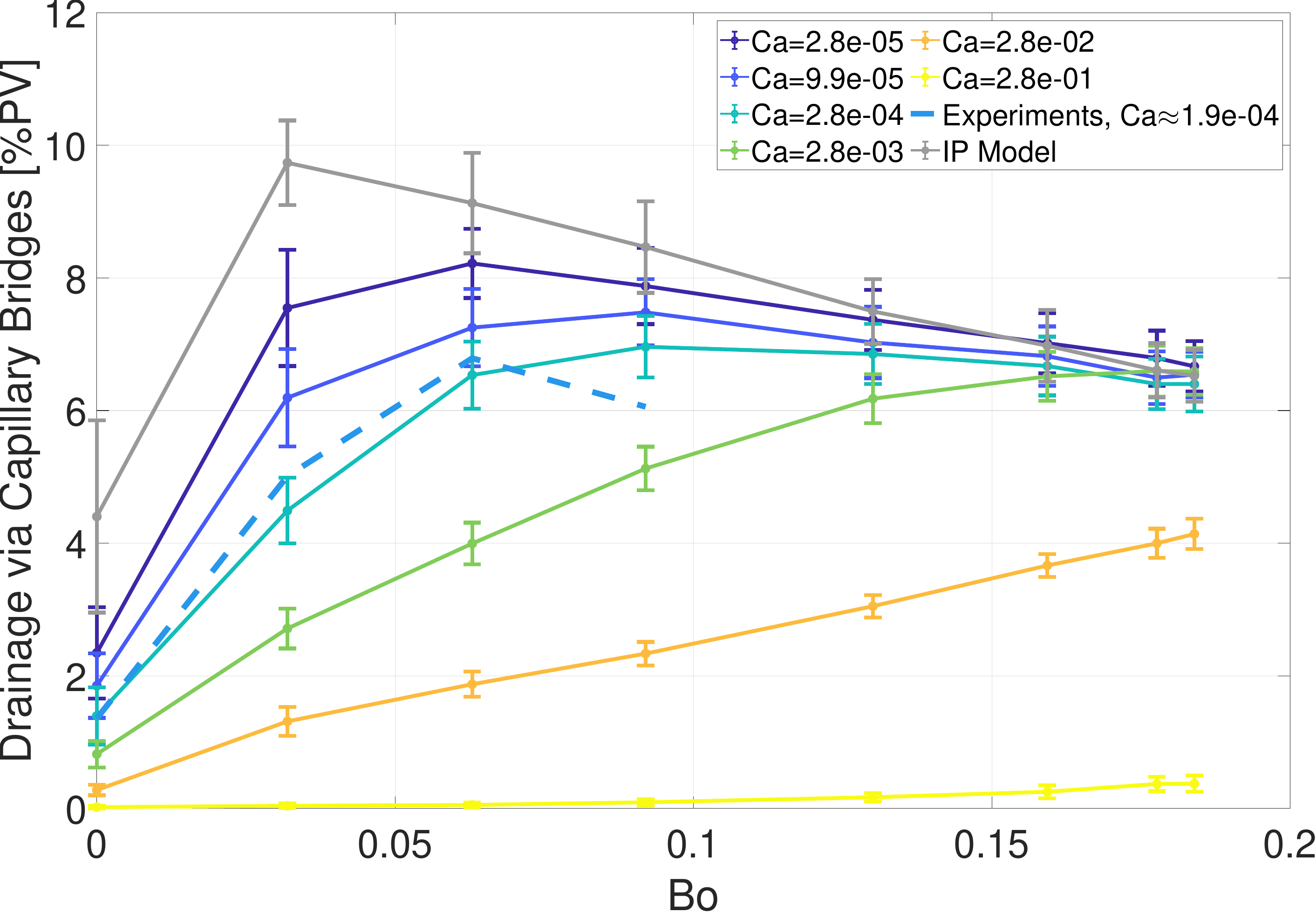}
    \caption{Total corner flow drainage obtained at breakthrough, measured as pore volumes (PV) vs. $Bo$. Here, all flow conditions tested in this work, quasi-static results from \citet{reis2023simplified} and experiments from \citet{moura2019connectivity} are represented.}
    \label{fig:SatRecFilm}
\end{figure}

The total amount of liquid mobilized from the pore networks through the secondary drainage mechanism, represented as a fraction of the pore volume (PV), is presented in Fig. \ref{fig:SatRecFilm}. In this graph, continuous lines represent numerical results, and the dashed line represents experiments performed in \citet{moura2019connectivity}. The line colors from dark blue to yellow indicate the capillary number of the displacement, in ascending order. The gray line is associated with numerical results obtained with the modified invasion-percolation (IP) model proposed by \citet{reis2023simplified}, in which the effects of viscous forces are not taken into consideration. It can be verified here that, for the same Bond number, increasing the drainage flow rate can reduce significantly the corner flow, as observed before in Fig. \ref{fig:InvasionMode}. This effect is related to the different time scales associated with the primary and secondary drainage mechanisms. An increase in $Ca$ produces a faster movement of the drainage invasion front, but the gravity-driven drainage from the clusters behind the front remains slow. In this way, clusters connected to the front by paths of capillary bridges are allowed less time to drain before eventually becoming disconnected from the fast-moving front. For low and moderate $Bo$, even the results obtained with a very low imposed drainage rate, $Ca\approx3\times10^{-5}$, differ considerably from the IP results, highlighting the importance of representing viscous effects in the new PNM proposed in this study. For high Bond numbers, however, connected clusters tend to be drained closer to the front, as seen in Fig.\ref{fig:InvasionTime}(g), reducing the effect of $Ca$ in the secondary drainage mechanism. Consequently, as the influence of gravitational forces increases, both the invasion-percolation and dynamic models lead to similar predictions of corner-flow drainage, given that $Ca\leq 3\times 10^{-3}$ .

It is also noticeable in Fig. \ref{fig:SatRecFilm} that, for low capillary numbers, the effect of gravitational forces on the amount of liquid drained through corners is non-monotonic. This tendency was also verified in the experiments conducted by \citet{moura2019connectivity}, as seen in the graph's dashed and continuous medium blue lines. We attribute this effect to two main reasons. During the stable displacement regime, increasing the Bond number leads to flatter invasion fronts and a smaller fraction of liquid remaining in clusters in the unsaturated zone \cite{birovljev1991gravity,maaloy2021burst}. Then, as previously pointed out in \citet{reis2023simplified}, although gravitational effects drive the drainage through corners in this region, at high $Bo$ there is less liquid to be drained, reducing the importance of the secondary drainage mechanism. On top of that, we noticed a strong correlation between gravitational effects and the probability of snap-off of capillary bridges, as presented in Fig. \ref{fig:Snap-off}. For all evaluated drainage flow rates, there is a significant increase in the fraction of capillary bridges formed behind the invasion front that reaches the snap-off capillary pressure and break, as $Bo$ increases. With this, the connectivity between liquid clusters and the drainage front is more likely to be interrupted, reducing corner flow. More details on how viscous and capillary forces affect the connectivity between clusters and invasion front are presented in the next section.

\begin{figure}[ht!]
    \centering
    \includegraphics[width=0.75\textwidth]{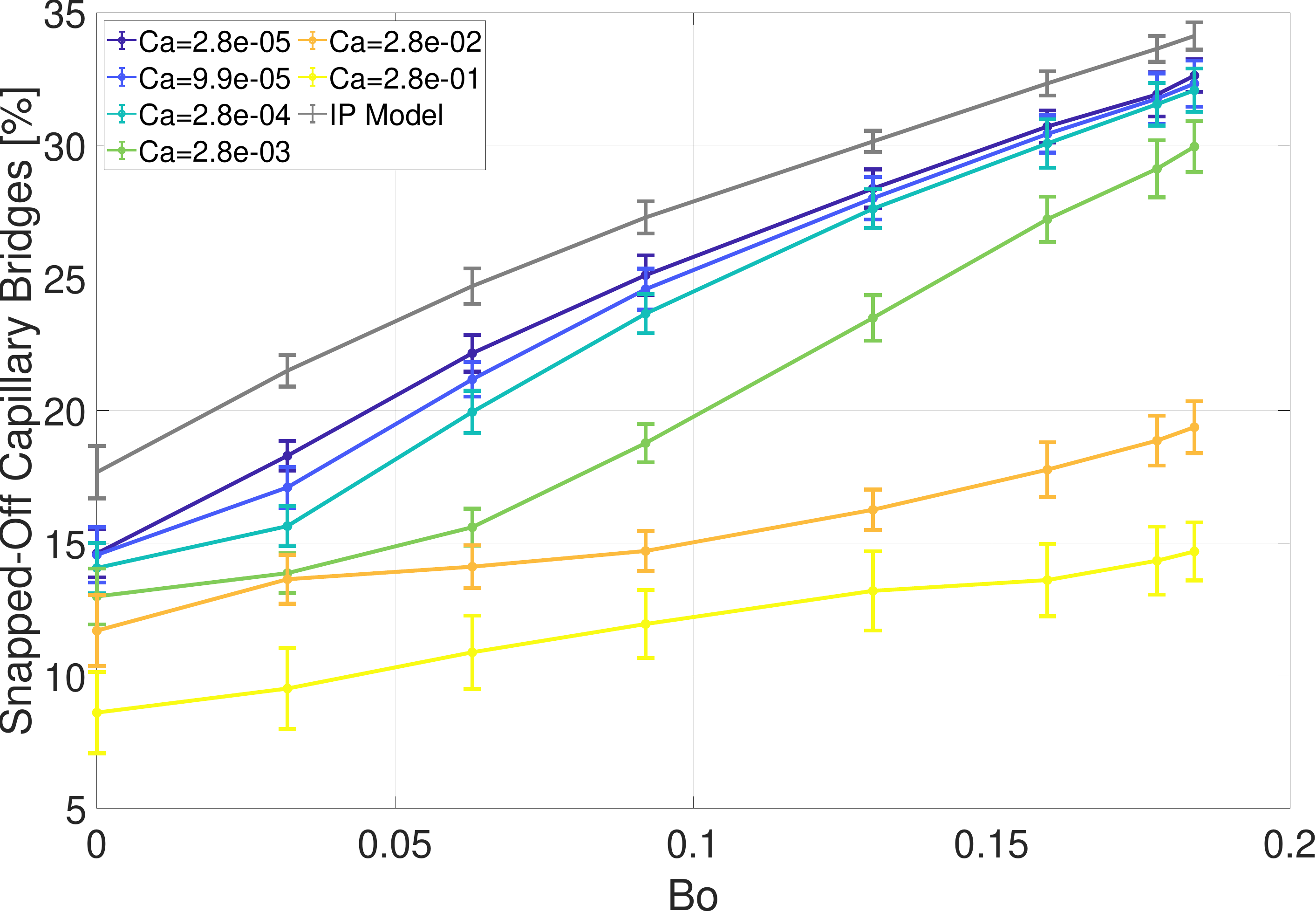}
    \caption{Fraction of capillary bridges that suffered snap-off during drainage under different flow conditions.}
    \label{fig:Snap-off}
\end{figure}

\subsection{Active Zone for corner flow}
\label{sec:AZ}

As introduced by \citet{moura2019connectivity}, the active zone for corner flow denotes the region trailing the invasion front where corner-flow drainage events are more likely to happen. Here, we measure the width of the area containing clusters connected to the invasion front by paths of capillary bridges as a surrogate for the active zone for corner flow. During drainage, this area undergoes large fluctuations, as the liquid long-range connectivity often relies on single capillary bridges that snap off, and on clusters that are drained \cite{moura2019connectivity, reis2023simplified}. Still, for the stable drainage regime, these fluctuations tend to stabilize around a mean, leading to a well-defined finite region where secondary drainage may occur.

Figure \ref{fig:ActiveZone} shows the extent of the liquid-connected zone above the front under different drainage conditions. Unlike Figs. \ref{fig:InvasionMode} and \ref{fig:InvasionTime}, here we show in color the pores and throats filled with liquid, while the black background represents the gas-filled and solid portions of the porous media. 
In the images, the main defending cluster is represented in light blue, the pores belonging to the invasion front are highlighted in yellow, and the clusters connected to the front by capillary bridges are shown in pink. Clusters that are disconnected from the front, and therefore trapped, are shown in green. For all nine flow conditions represented in Fig. \ref{fig:ActiveZone}, the depicted snapshots correspond to a time $t=0.75t_{bt}$, where $t_{bt}$ is the breakthrough time. The comparison between the cases represented in Fig. \ref{fig:ActiveZone} suggests that variations in the relevance of both gravitational and viscous forces during the flow significantly affect the liquid connectivity in the region trailing the invasion front. In fact, their effects on the active zone extent seem much more pronounced than those on the front width, which are well described in the literature \cite{birovljev1991gravity,auradou1999competition,meheust2002interface,lovoll2005competition,ayaz2020gravitational,maaloy2021burst,toussaint2012two,vincent2022stable}. 

As previously reported in both experimental \cite{moura2019connectivity} and numerical \cite{reis2023simplified} slow drainage studies, increasing the Bond number leads to a compaction of the active zone, with connected clusters becoming progressively limited to the vicinity of the drainage front. For the cases with $Ca\approx3\times 10^{-4}$, this reduction of the active zone with $Bo$ explains why corner-flow drainage events could take place way past the front passage with low $Bo$, as seen in Fig. \ref{fig:InvasionTime}(a), but not with higher $Bo$, as in Fig. \ref{fig:InvasionTime}(g). For the cases with $Ca\approx3\times 10^{-3}$, we also notice the formation of limited regions where clusters are connected to the front by capillary bridges, which decrease with rising gravitational forces. Interestingly, as seen in Fig. \ref{fig:SatRecFilm}, the amount of liquid drained through corners with $Ca\approx3\times 10^{-3}$ increases monotonically with $Bo$. These results suggest that extensive liquid connectivity in the unsaturated region is not linked to increased secondary drainage. In reality,  the flow at low $Bo$ and high $Ca$ accentuates the difference between the characteristic drainage time scales associated with the movement of the front and the corner-dominated flow behind it. Therefore, the conditions leading to the formation of extensive networks of clusters connected by capillary bridges are the same leading to reduced drainage efficiencies.

\begin{figure}[ht!]
     \centering
     
     \begin{subfigure}[t]{0.32\textwidth}
         \centering
         \includegraphics[width=0.95\textwidth]{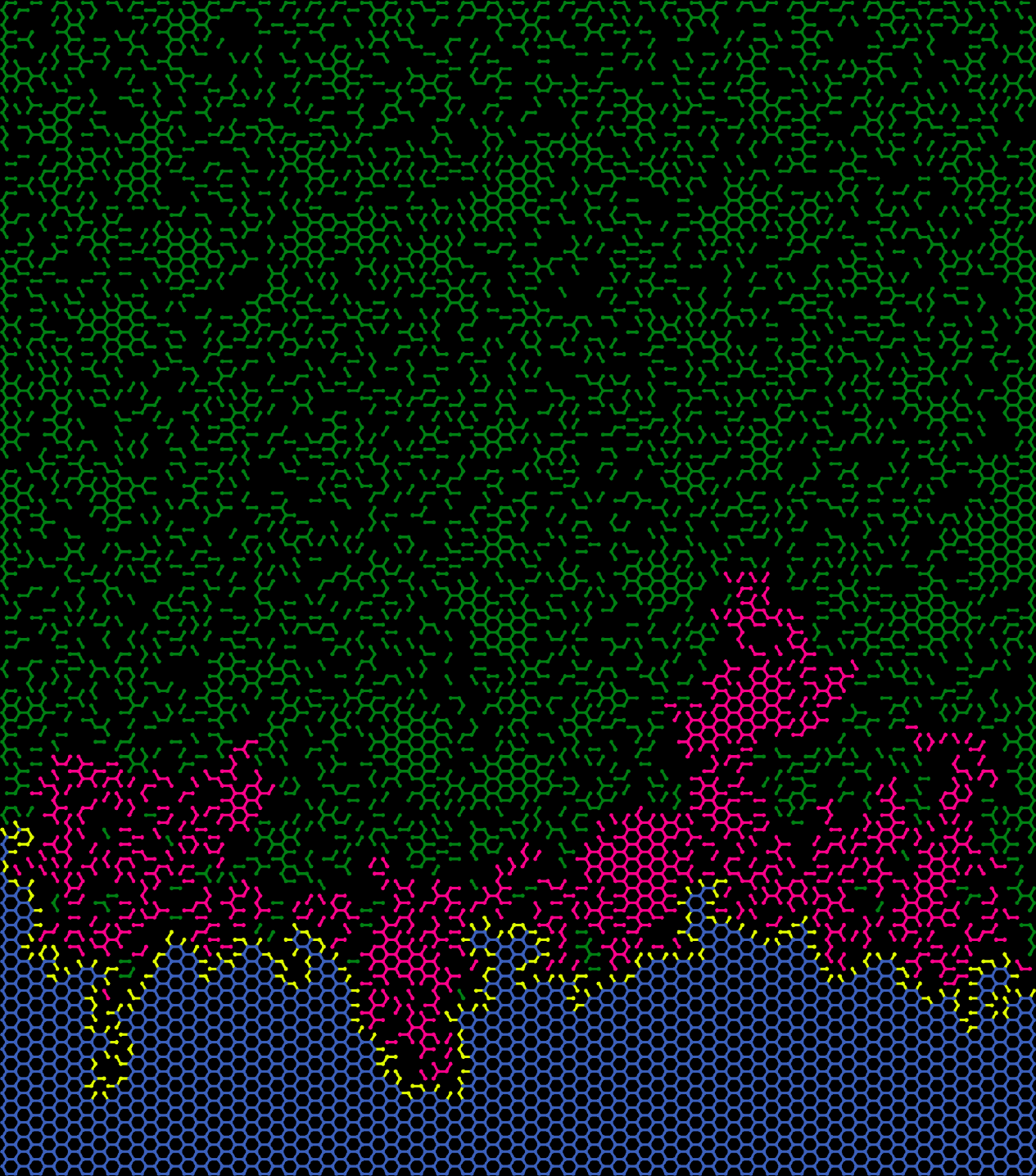}
         \caption{$Bo\approx 6 \times 10^{-2}$ $Ca\approx3\times10^{-4}$}
         \label{fig:az_20_0.01}
     \end{subfigure}
     \hfill
     \begin{subfigure}[t]{0.32\textwidth}
         \centering
         \includegraphics[width=0.95\textwidth]{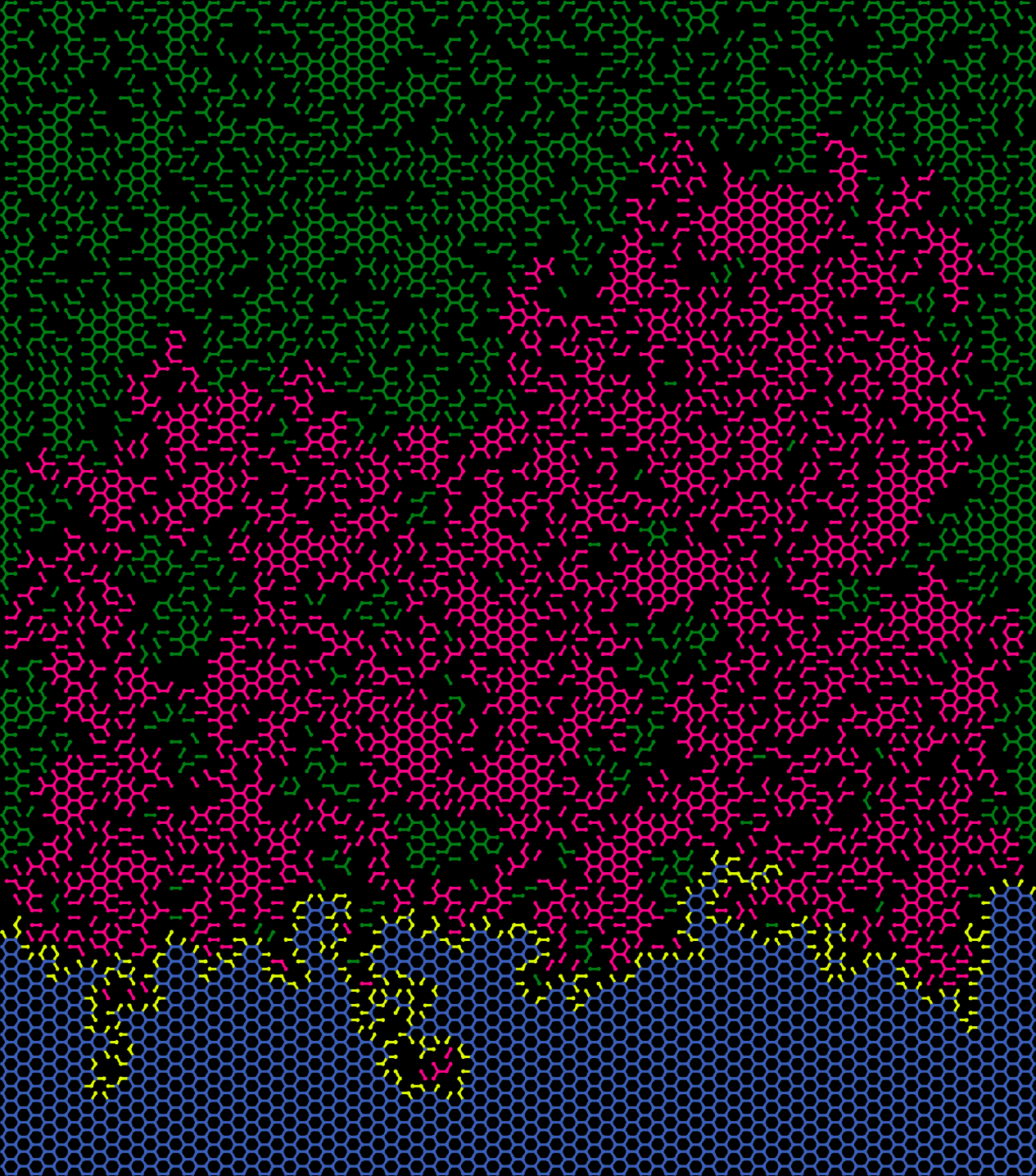}
         \caption{$Bo\approx 6 \times 10^{-2}$ $Ca\approx3\times10^{-3}$}
         \label{fig:az_20_0.1}
     \end{subfigure}
     \hfill
     \begin{subfigure}[t]{0.32\textwidth}
         \centering
         \includegraphics[width=0.95\textwidth]{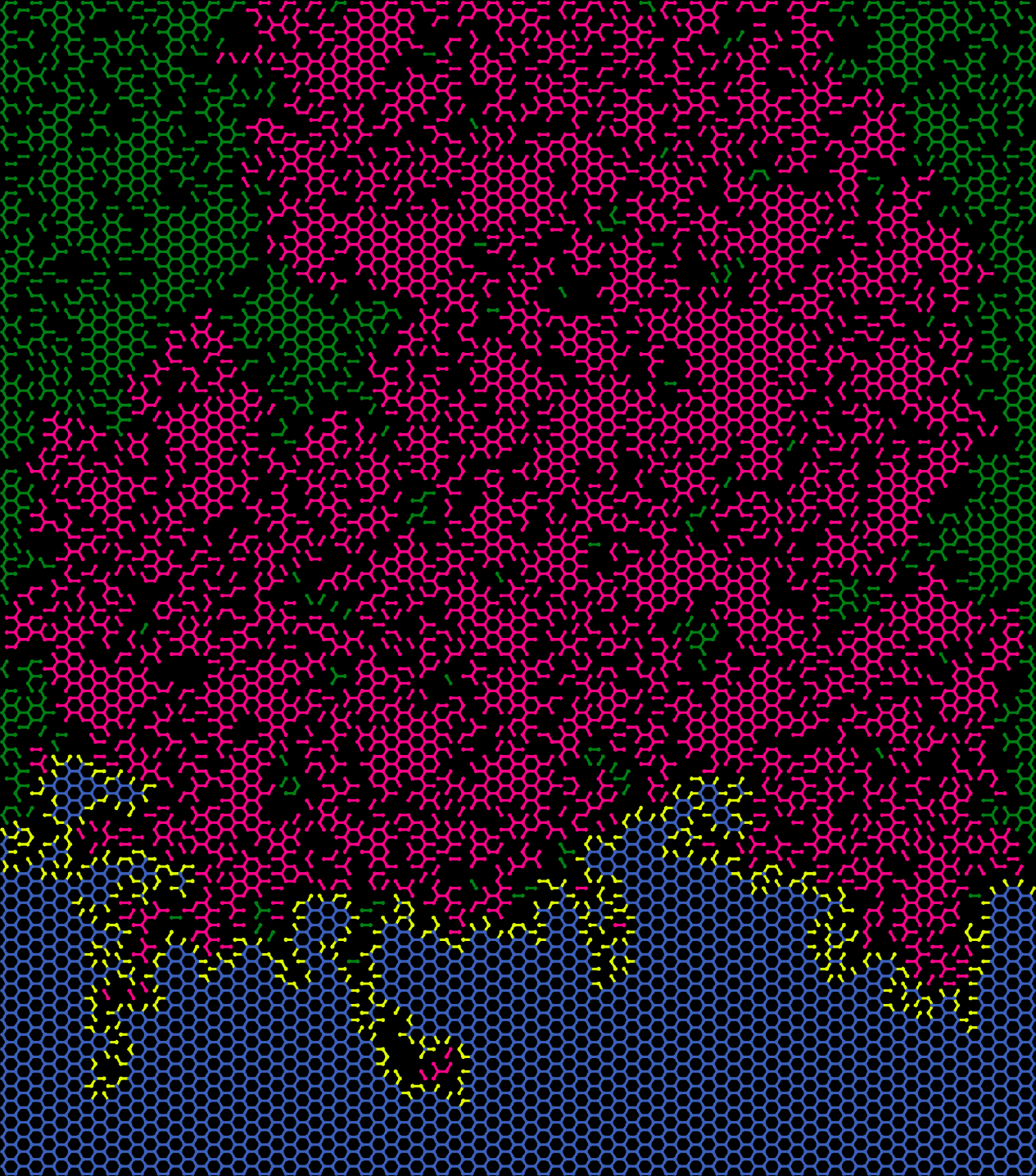}
         \caption{$Bo\approx 6 \times 10^{-2}$ $Ca\approx3\times10^{-2}$}
         \label{fig:az_20_1}
     \end{subfigure}
     \hfill
     \begin{subfigure}[t]{0.32\textwidth}
         \centering
         \includegraphics[width=0.95\textwidth]{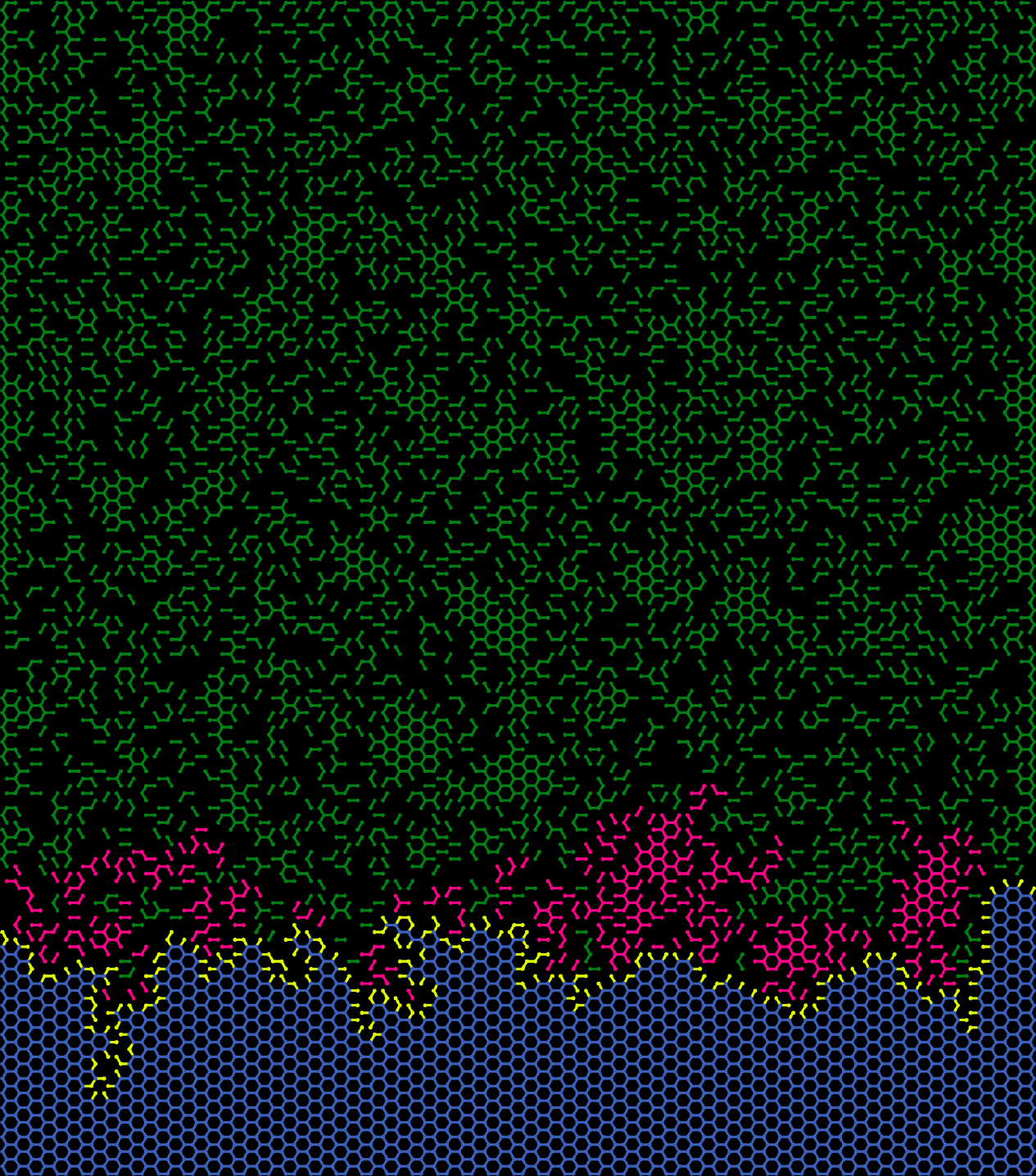}
         \caption{$Bo\approx9\times 10^{-2}$ $Ca\approx3\times10^{-4}$}
         \label{fig:az_30_0.01}
     \end{subfigure}
     \hfill
     \begin{subfigure}[t]{0.32\textwidth}
         \centering
         \includegraphics[width=0.95\textwidth]{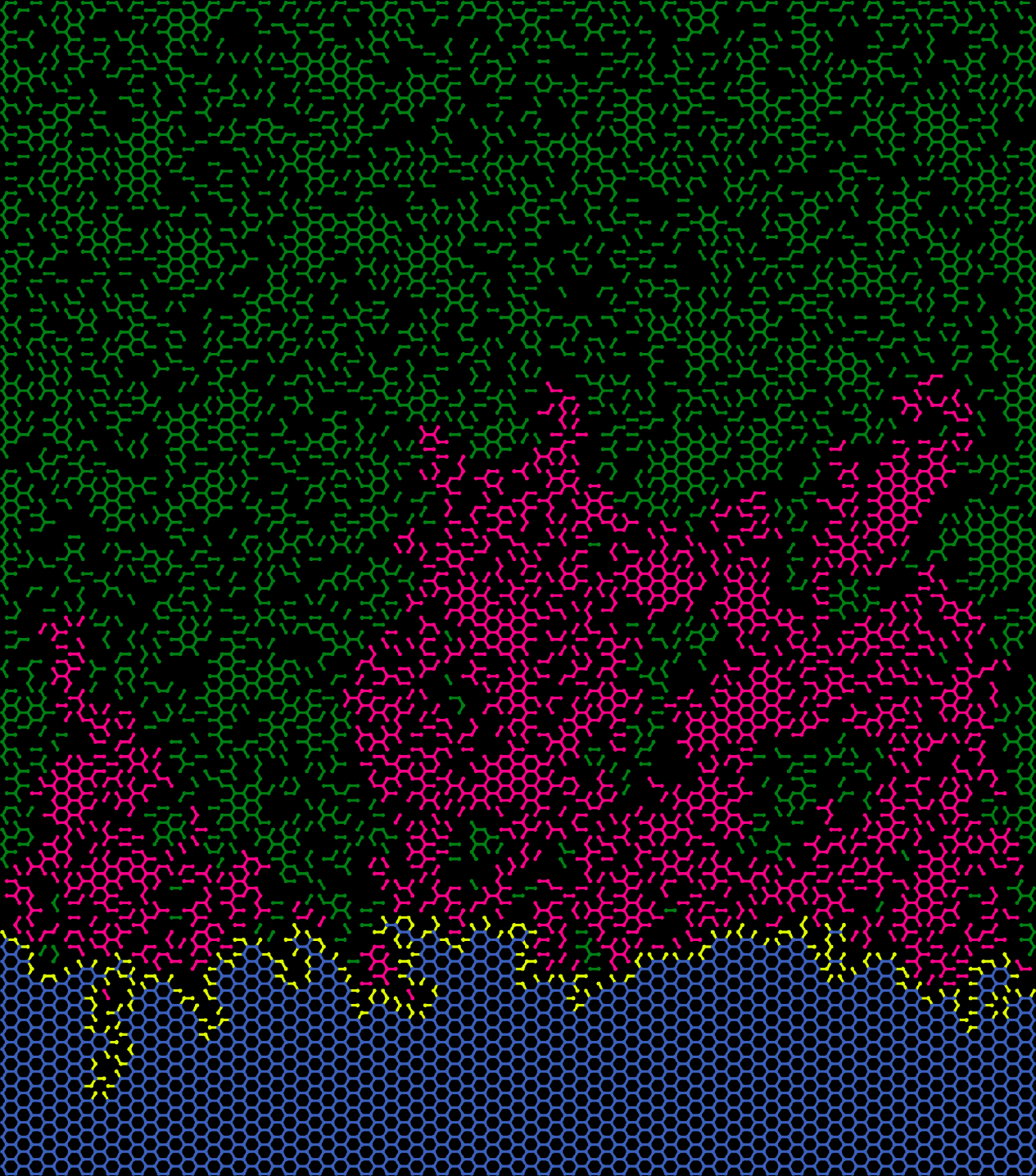}
         \caption{$Bo\approx9\times 10^{-2}$ $Ca\approx3\times10^{-3}$}
         \label{fig:az_30_0.1}
     \end{subfigure}
     \hfill
     \begin{subfigure}[t]{0.32\textwidth}
         \centering
         \includegraphics[width=0.95\textwidth]{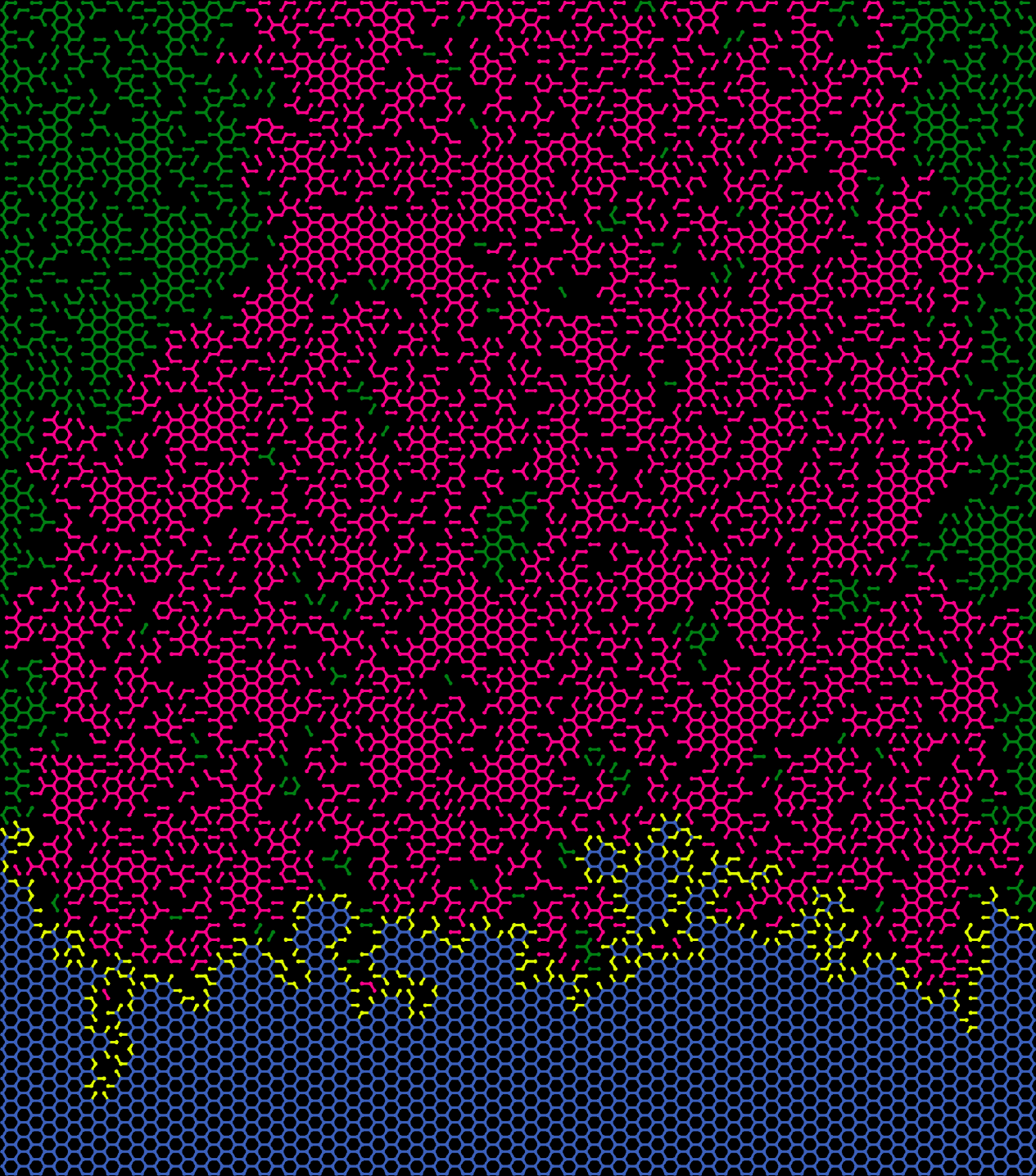}
         \caption{$Bo\approx9\times 10^{-2}$ $Ca\approx3\times10^{-2}$}
         \label{fig:az_30_1}
     \end{subfigure}
     \hfill
     \begin{subfigure}[t]{0.32\textwidth}
         \centering
         \includegraphics[width=0.95\textwidth]{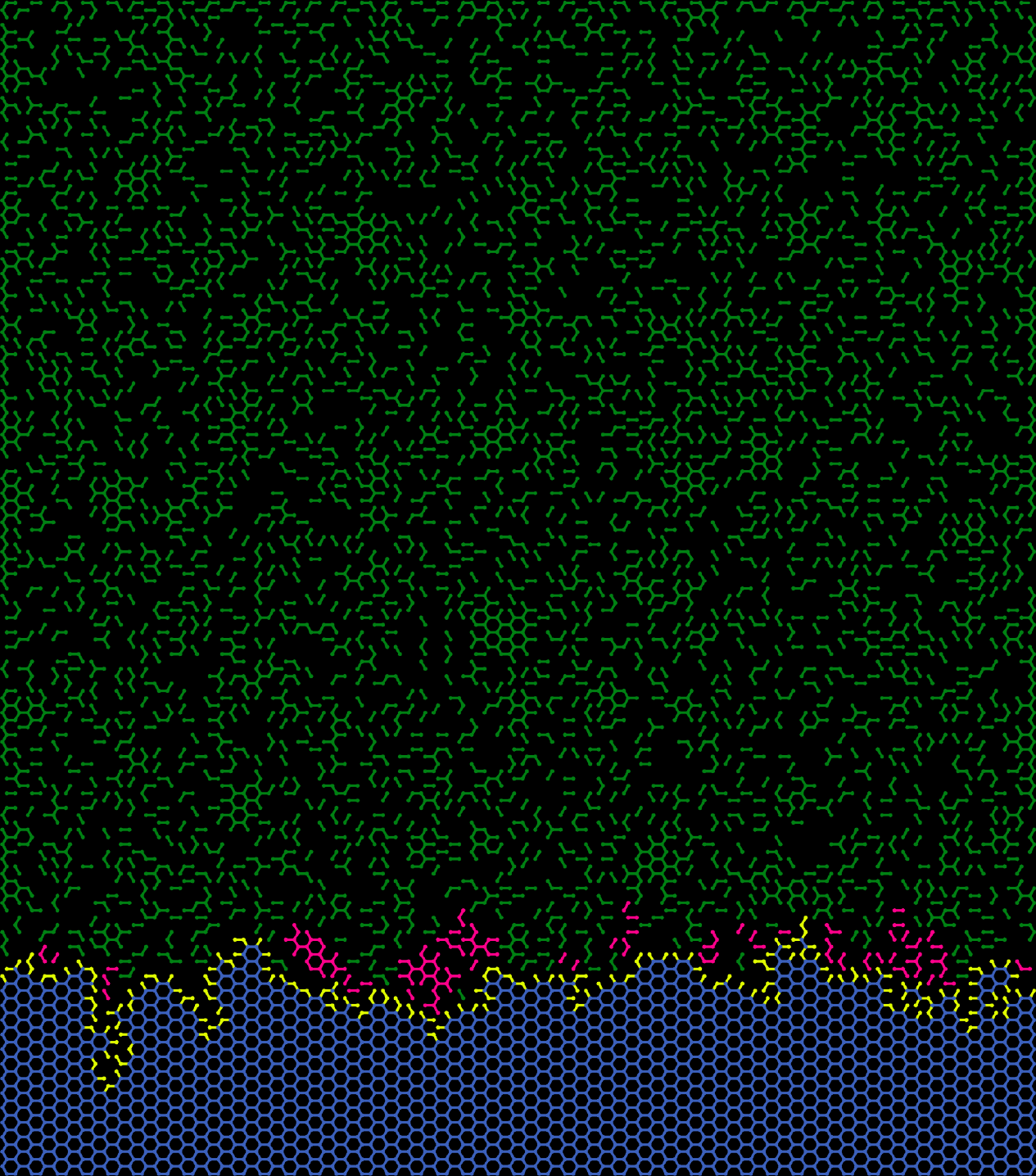}
         \caption{$Bo\approx1.6\times 10^{-1}$ $Ca\approx3\times10^{-4}$}
         \label{fig:az_60_0.01}
     \end{subfigure}
     \hfill
     \begin{subfigure}[t]{0.32\textwidth}
         \centering
         \includegraphics[width=0.95\textwidth]{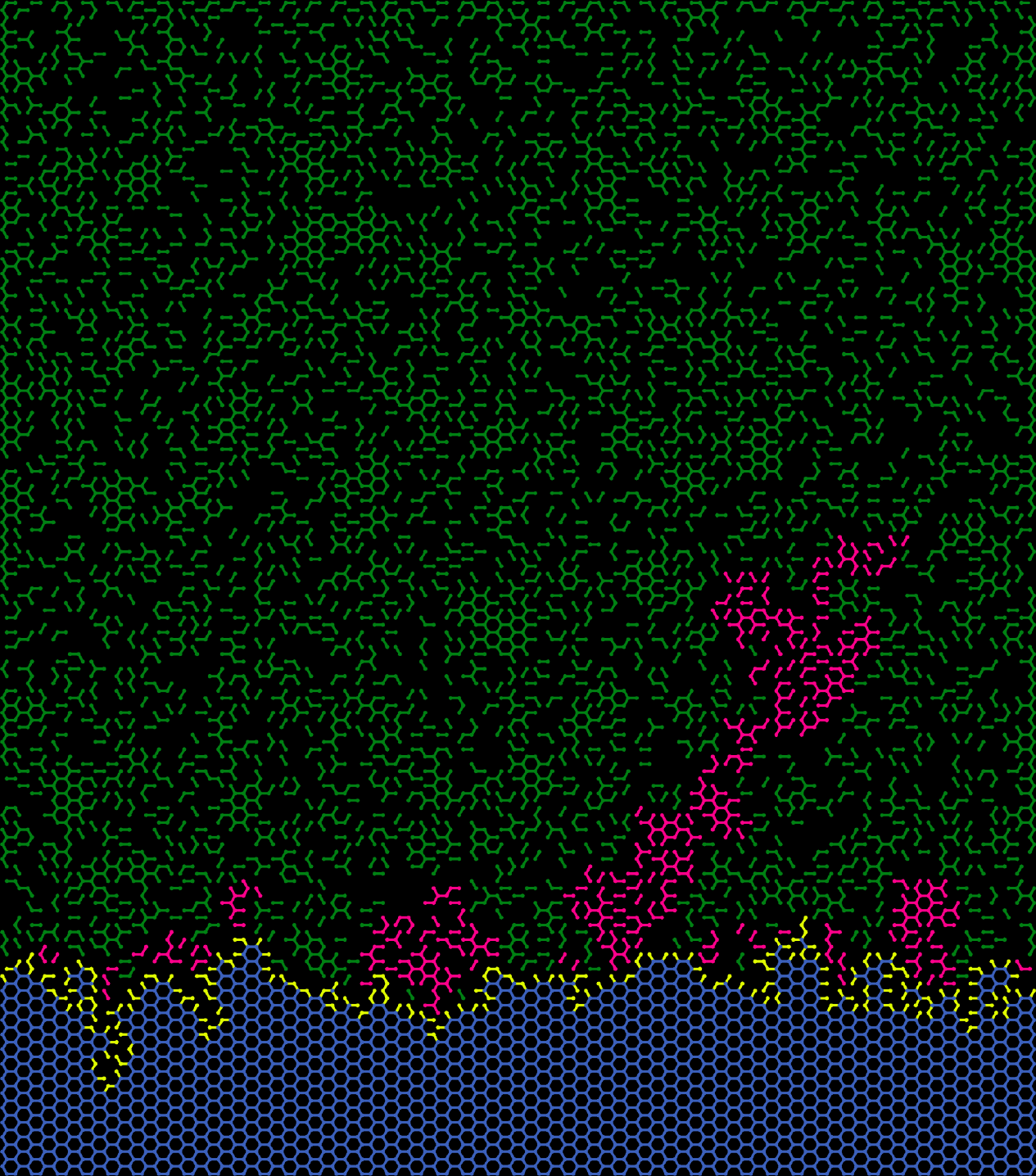}
         \caption{$Bo\approx1.6\times 10^{-1}$ $Ca\approx3\times10^{-3}$}
         \label{fig:az_60_0.1}
     \end{subfigure}
     \hfill
     \begin{subfigure}[t]{0.32\textwidth}
         \centering
         \includegraphics[width=0.95\textwidth]{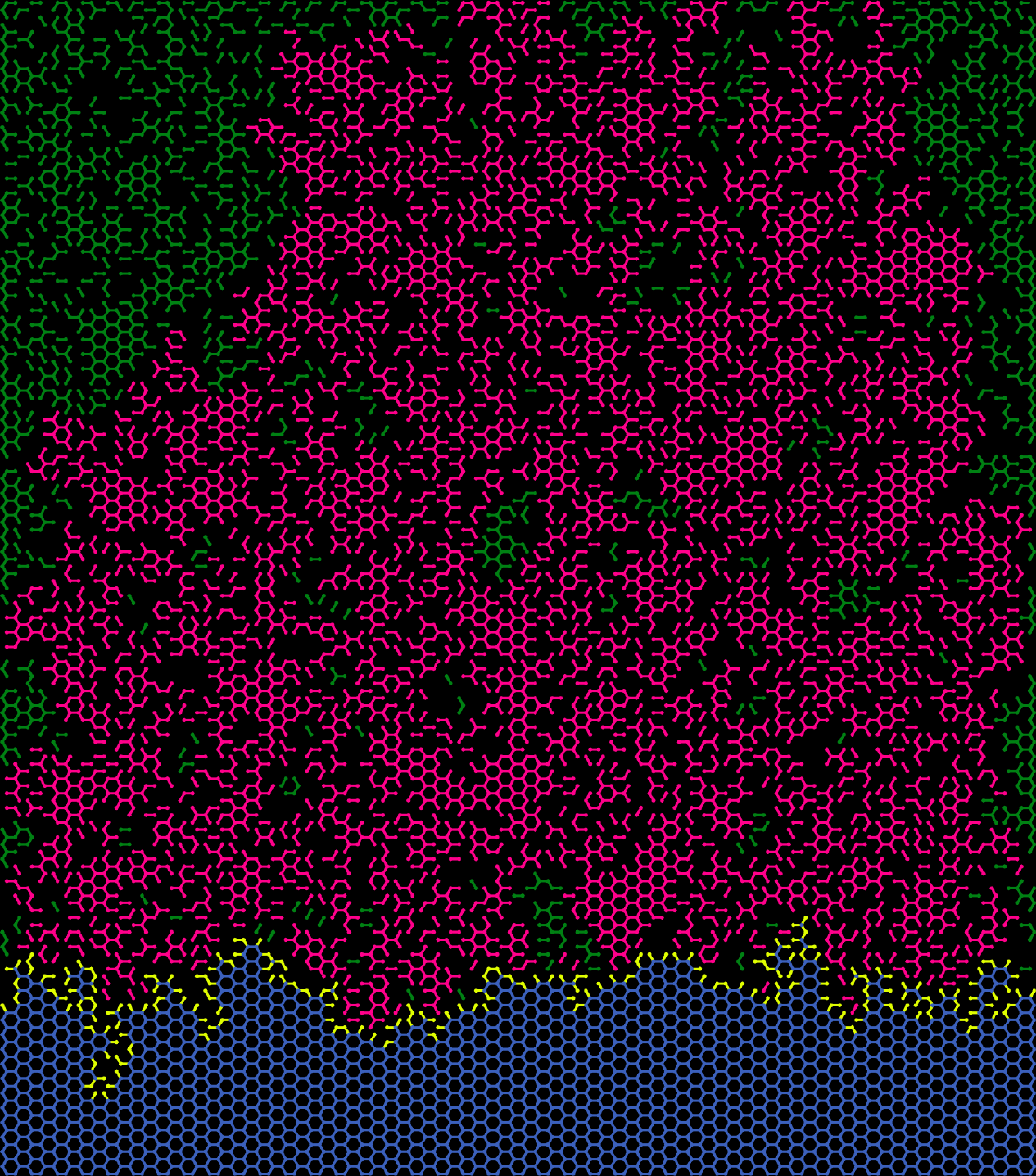}
         \caption{$Bo\approx1.6\times 10^{-1}$ $Ca\approx3\times10^{-2}$}
         \label{fig:az_60_1}
     \end{subfigure}
     
        \caption{Representation of the main defending cluster (light blue), invasion front (yellow), clusters connected to the invasion front by capillary bridges and liquid rings (pink), and clusters that lost connectivity to the front (green). From (a) to (i), different flow conditions are represented, at a time $t=0.75 t_{bt}$, where $t_{bt}$ is the breakthrough time.}
        \label{fig:ActiveZone}
\end{figure}

For the drainage cases with $Ca\approx3\times 10^{-2}$, shown in Figs. \ref{fig:ActiveZone}(c), (f), and (i), we observe that the liquid-connected zone above the front spans the entire simulated porous-medium domain. These results, therefore, cannot indicate whether the connected region would be reduced as $Bo$ increases, as seen for slower flows. The analysis of these images, however, clarifies the difference in the behavior between the curves representing $Ca < 3\times 10^{-2}$ and $Ca\geq 3\times 10^{-2}$ in Fig. \ref{fig:SatRecFilm}. For $Ca < 3 \times 10^{-2}$, the liquid amount drained through corners either decreases or stabilizes, as $Bo$ goes from intermediate to high values. As briefly discussed in Sec. \ref{sec:sat_mech}, although gravity is the main driving force for flow in the unsaturated region, its positive effects on drainage can be counterbalanced by the reduced liquid connectivity span seen in Fig. \ref{fig:ActiveZone}. On the other hand, for $Ca\geq 3 \times 10^{-2}$, Fig. \ref{fig:SatRecFilm} points to a linear increase in corner-flow drainage with $Bo$. With a seemingly constant width of the connected liquid region for all evaluated Bond numbers, this counterbalancing effect vanishes, and the gravitational influence becomes predominantly positive on the secondary drainage mechanism. 

As a way to quantify the active zone width for all investigated stable drainage cases, the mean distance between the pores in the active zone to the pores in the invasion front is presented in Fig. \ref{fig:AZ_FF}(a). For each pore in the connected region, this distance is measured as the minimum Euclidean distance to any point on the front. Then, the measured values for all connected pores and all times $t\geq0.75t_{bt}$ are averaged.
For $Ca<3\times 10^{-4}$, we can observe that the average distance from the clusters connected by capillary bridges to the front is very small, encompassing a region approximately 10 pores wide. For $Ca\approx 3 \times 10^{-3}$, as previously suggested in Figs. \ref{fig:ActiveZone}(b), (e), and (h), the effects of gravitational forces on the active zone are prominent, with the average distance from connected pores to the front decreasing threefold within the range of $Bo$ in which the displacement is stable. Finally, for $Ca\approx 3\times 10^{-2}$, we verify that the measured distances are nearly constant with $Bo$ and consistent with an active zone that spans the entire length of the pore networks used in this study.

\begin{figure}[ht!]

    \centering
     \begin{subfigure}[t]{0.65\textwidth}
         \centering
         \includegraphics[width=0.95\textwidth]{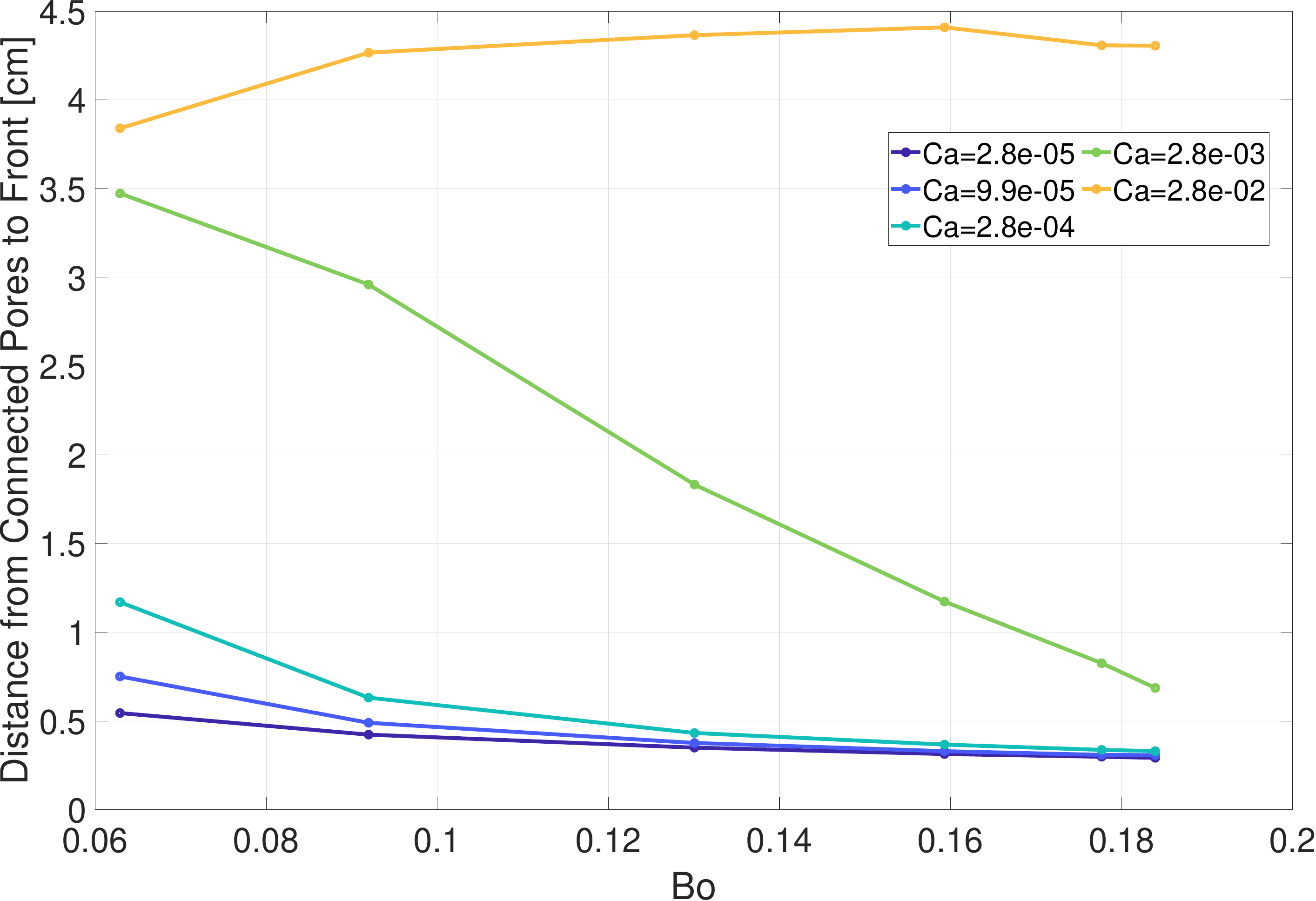}
         \caption{}
         \label{fig:AZNodes}
     \end{subfigure}
     \hfill
     \begin{subfigure}[t]{0.65\textwidth}
         \centering
         \includegraphics[width=0.95\textwidth]{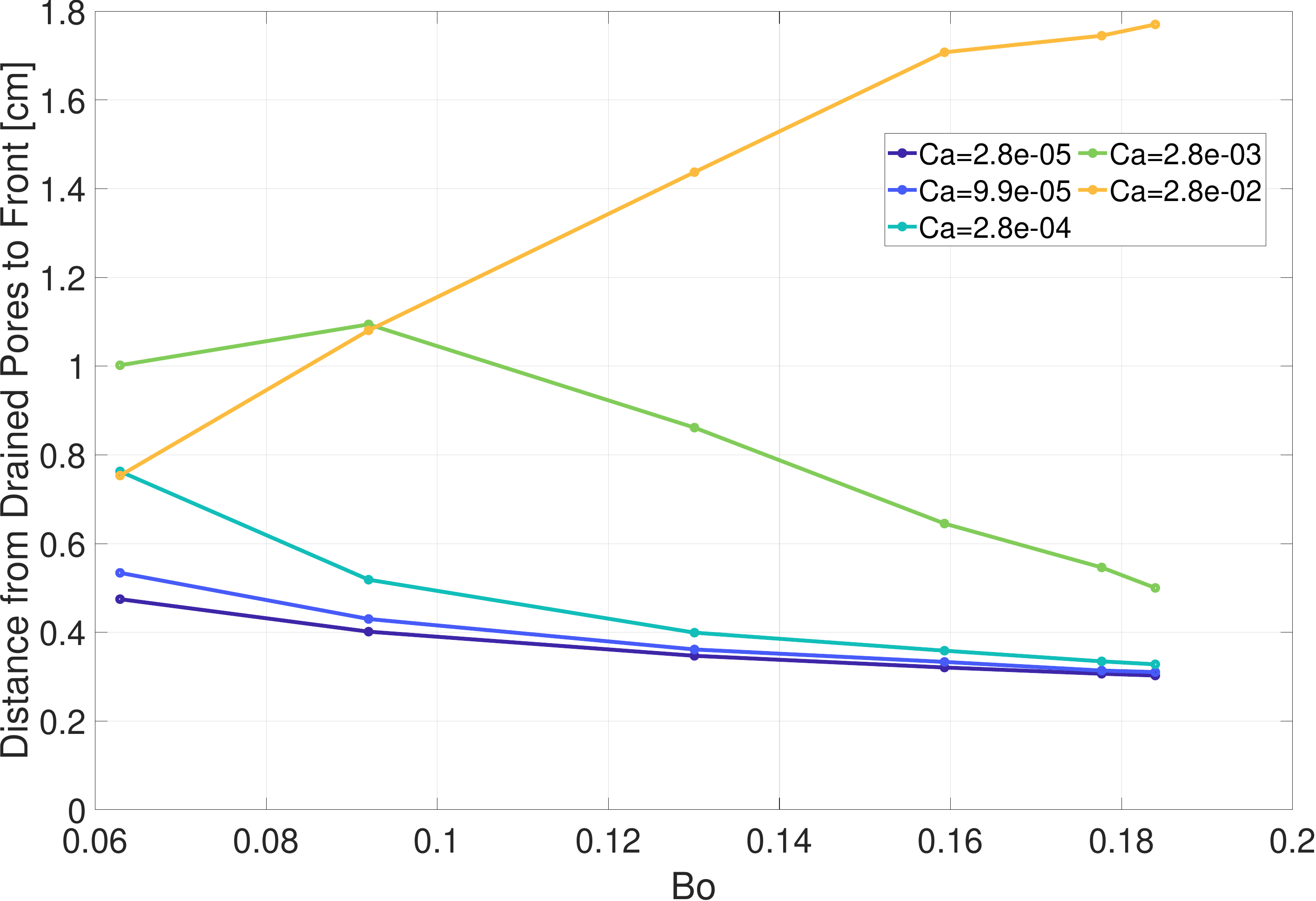}
         \caption{}
         \label{fig:FFNodes}
     \end{subfigure}
     \hfill
    \caption{ (a) Average Euclidean distance from pores belonging to clusters connected to the front by capillary bridges. (b) Average Euclidean distance from pores effectively drained due to corner flow to the front.}
    \label{fig:AZ_FF}
\end{figure}

Additionally, the average distance between pores drained via corner flow and the front was measured and presented in Fig. \ref{fig:AZ_FF}(b). This information, along with the results in Fig. \ref{fig:AZ_FF}(a), can indicate whether drained pores from connected clusters tend to be evenly distributed within the connected region, or concentrate at a certain distance from the front. At low Bond numbers, drainage is more likely to occur in a subset of the connected-region pores closer to the invasion front, especially at higher $Ca$. At $Bo\approx 6 \times 10^{-2}$, for instance, the average distance from secondary drainage events to the front is the same for $Ca\approx3\times 10^{-2}$ and $Ca\approx3\times 10^{-4}$, even though the total connected region is much narrower for the slower flow. 

Congruently, as $Bo$ increases, the probability of drainage events in pores from the connected region farther from the front becomes higher. This effect is clear in the $Ca\approx 3 \times 10^{-2}$ curve in Fig. \ref{fig:AZ_FF}(b). For this case, the width of the liquid-connected region above the front is constant (given the pore-network size limitation), and we can observe that the average distance from secondary drainage events to the front is extended with gravitational forces. For $Ca< 3 \times 10^{-2}$, the distance from drainage events to the front decreases with $Bo$, as the whole connected region also narrows.

\section{Discussion}
\label{sec:disc}

The results obtained with the dynamic pore-network model proposed in this study suggest that variations in viscous and gravitational forces acting during drainage substantially impact liquid connectivity and flow in the unsaturated zone. Similar to the effects observed on the front width under the stable displacement regime, the width of the region where clusters can still be drained by corner flow increases with the capillary number and decreases with the Bond number. When it comes to drainage efficiency, increasing $Ca$ reduces not only the amount of liquid mobilized by the primary drainage mechanism -- as the displacement gradually shifts from stable to viscous fingering -- but also hinders the secondary drainage mechanism. Even though the liquid content and connectivity in the unsaturated zone increase with $Ca$, the different time scales related to the movement of the front and the drainage through liquid rings and capillary bridges impede substantial liquid mobilization from the clusters. The influence of $Bo$ on the total drainage from the unsaturated zone to the front can also be split into contrasting effects. As $Bo$ increases, the gravity-driven corner flow from the clusters to the front is favored. Still, the formation of smaller liquid clusters and more frequent snap-off of capillary bridges reduce the permeability of the corner-flow network, potentially reducing drainage. 

Besides the novelty of the results, the model proposed in this study is in itself relevant, as it successfully follows an unconventional approach to representing corner flow in granular porous media. The identification of liquid-connectivity networks decoupled from the pore-throat networks \cite{bryant2003wetting,vorhauer2015drying,chen2017control,chen_2018,moura2019connectivity,kharaghani2021three,reis2023simplified} can provide a more physically-realistic image of the phase distribution among the grains during drainage. On top of identifying liquid-connected paths formed by liquid rings and capillary bridges, the adopted dual-network approach, along with the study of bridge shapes and conductances, provides a suitable way to quantify corner-flow drainage. In contrast, previous works in the literature either tracked the liquid-connected structures in the unsaturated region in quasi-static models \cite{vorhauer2015drying,kharaghani2021three,reis2023simplified}, or calculated the viscous flow through predictable linear chains of capillary bridges \cite{chen2017control,chen_2018}.

As for this study's potential applications, the connectivity and flow of liquid in granular media during drainage can be linked, for instance, to water retention in soils. Understanding such processes is relevant for environmental and agricultural applications, as it controls the plant water availability and biological functioning of soils \cite{hoogland2016drainage}. Therefore, further studies with the proposed model could focus on obtaining macro water-flow properties in soil for large continuous models. Additionally, understanding water flow and distribution in granular media can have important implications for the transport and mixing of solutes in soils. Although the results presented in Sec.\ref{sec:res} focused on flow, identifying the span and form of liquid-connected structures in unsaturated soil can be instrumental to understanding how far and how fast nutrients or contaminants can travel within this type of porous medium.

Nevertheless, some simplifying assumptions of the presented model and analyses should be taken into consideration prior to their application. A first important simplification is the adoption of a 2D hexagonal lattice to represent the network of pore bodies and throats. This geometry retains some important aspects of 3D granular porous media, e.g. the coordination number \cite{dong2009pore}, and we expect that the type of connectivity established by bridges and liquid rings defined in this work to be valid also in 3D \cite{bryant2003wetting,Scheel_2008}. Still, the results presented in Sec. \ref{sec:res} could vary in actual 3D porous media, as the liquid phase in the unsaturated region tends to be more connected than in the 2D case \cite{wilkinson1984percolation}.

Another source of uncertainty in the results could be the simplified shape used to calculate the conductance of the model's pore throats. While the conductances of the capillary bridges were based on DNS of Stokes flow through their shapes, the throats were approximated by converging-diverging circular tubes. This choice allowed us to have analytical solutions for their conductances and capillary pressures, but do not represent the throat's shapes to a high level of accuracy. On top of that, the conductances of the capillary bridges also incorporate some simplifications: we considered that neither the flow through their structures nor the presence of neighboring bridges would lead to distortions in their shapes. In any case, such simplifications are typically adopted in PNMs and should not significantly impact the level of predictability expected for this modeling approach.

As a final remark, we considered in this work that liquid in the unsaturated region could only be connected and flow through the main network or pore bodies and throats -- as inside liquid clusters -- or by the network of capillary bridges and liquid rings. Different studies in the literature point out the additional connectivity provided by crevices on rough grains \cite{bryant2003wetting} and adsorbed films covering the entire grain surface \cite{zhang2024anomalous}. Yet, such structures are expected to have thicknesses -- hence permeabilities -- much lower than the liquid flow paths incorporated in our model. As our focus here was to quantify the extra drainage allowed by corner flow within the time scale of the drainage front advancement, we assumed that including the flow through crevices and adsorbed films would not lead to significant deviations in the presented results.

\section{Conclusion}
\label{sec:conc}

In this work, we proposed a fully-implicit dynamic pore-network model for drainage in granular media, when the wetting phase is a liquid and the non-wetting phase is an inviscid and weightless gas. In the model, dual-lattice networks for the flow through pores connected by throats and the flow through capillary bridges and liquid rings were adopted. With this approach, we could efficiently reproduce and distinguish drainage events associated with the primary and secondary drainage mechanisms \cite{hoogland2016drainage,moura2019connectivity,reis2023simplified}. To appropriately quantify the liquid mobilization through the secondary drainage mechanism, a dedicated analysis of the capillary bridges' shapes and capacity to accommodate flow was conducted and used as input for the pore-network model.

Using the resulting model, we investigated drainage in quasi-2D granular porous media under a broad range of capillary and Bond numbers. In particular, the widths of the regions where clusters are still connected to the receding liquid front, as well as the amount of liquid drained from them, were calculated and examined in Sec. \ref{sec:res}. The results suggested that liquid connectivity and flow in the unsaturated region strongly depend on drainage conditions, challenging the assumption that the wetting phase is always fully connected by corner flow.

This study's findings and new modeling approach can be relevant to natural and engineered processes in the subsurface, such as the water cycle in soils and gas storage in saturated geological formations. Therefore, future developments in the presented line of work should include expanding the model to accommodate irregular and 3D networks, aiming at a more fitting representation of real granular media. In addition, conducting a more rigorous analysis of the flow and interface displacement through the networks of pores and throats found in granular media -- as done here for the capillary bridges using DNS -- could improve the accuracy of the results. It could also be interesting to incorporate in the model the possibility of flow through crevices and adsorbed films on the grains' surface, to investigate if these structures would significantly impact drainage under particular flow conditions. In a similar direction, the effects of evaporation in the unsaturated region could also be considered, especially in the limit of slow flows, when liquid removal by drainage and drying potentially have similar characteristic time scales.

\section{Acknowledgements}

We acknowledge the support of the University of Oslo, the Njord Centre, and the Research Council of Norway through the PoreLab Center of Excellence (project number 262644), the Researcher Projects for Young Talents FlowConn (project number 324555) and M4 (project number 325819), and the CNRS/University of Oslo IRP D-FFRACT.

\appendix

\section{Scaling the Stokes equations for flow through liquid bridges}
\label{sec:scaling_stokes}
We introduce the scaled variables
\begin{align}
    \tilde{\v x} = \v x / d, \quad
    \tilde{\v u} = \v u / U, \quad \text{and} \quad
    \tilde{p} = p' / P ,
\end{align}
where we recall that $p' = p - F x$ and $F = \Delta p /\ell$ is the imposed
macroscopic pressure gradient.
Plugging these definitions into the Stokes eqs.\ \eqref{eq:stokes}, we obtain
\begin{align}
    \frac{\mu U}{Fd^2} \tilde{\grad{}}^2 \tilde{\v u} = \frac{P}{Fd} \tilde{\grad{}}\tilde{p} - \hat{\v x},
    \quad \tilde{\grad{}} \cdot \tilde{\v u} = 0,
\end{align}
where $\tilde{\grad{}}$ is the gradient operator in the scaled spatial coordinate $\tilde {\v x}$.
Choosing the scalings
\begin{align}
    U &= \frac{Fd^2}{\mu}, \label{eq:Uscal}\\
    P &= Fd,
\end{align}
means that the scaled velocity field $\tilde{\v u}$ resulting from solving the scaled Stokes equations,
\begin{align}
    \tilde{\grad{}}^2 \tilde{\v u} = \tilde{\grad{}}\tilde{p} - \hat{\v x},
    \quad \tilde{\grad{}} \cdot \tilde{\v u} = 0,
    \label{eq:scaled_stokes}
\end{align}
can be related to the physical field $\v u$ by \eqref{eq:Uscal}:
\begin{align}
    \v u (\v x) = \frac{F d^2}{\mu} \tilde {\v u} \left( \frac{\v x}{d} \right).
    \label{eq:scalingU}
\end{align}
To quantify the conductance we compute the cross-sectional flux (for arbitrary $x$):
\begin{align}
    \Phi &= \int_{\Omega_x} \hat{\v x} \cdot \v u (\v x ) \, \diff^2 \v x 
    = \frac{F d^4}{\mu} \int_{\tilde \Omega_x} \hat{\v x} \cdot \tilde{\v u} (\tilde{\v x} ) \, \diff^2 \tilde{\v x} 
    = \frac{F d^4}{\mu} k_0.
\end{align}
Here, we have identified the dimensionless shape factor $k_0$.
Note that if the integral is taken over half the cross section, like the computational domain above, a factor 2 must be included here.
The conductance $\Gamma$ used for pore-network modelling is implicitly defined by:
\begin{align}
    \Phi = \Gamma \Delta P
\end{align}
where $\Delta p = F \ell$ is the total bead-bead pressure difference.
This yields
\begin{align}
    \Gamma = \frac{\Phi}{F \ell} = \frac{k_0 d^4}{\mu \ell}.
\end{align}
Note that since the shape factor $k_0$ only depends on the shape of the bridge, it may only depend on the length scales involved: the bead diameter $d$, the bead-bead separation $\ell$, and the mean curvature $\kappa$. (Note that we here assume that we can neglect any hysteresis effects that could make $k_0$ multi-valued or history dependent.) 
From these dimensional quantities we can construct exactly two independent dimensionless quantities, such that the Buckingham \textpi{} theorem \cite{buckingham1914physically} implies:
\begin{align}
    k_0 = k_0\left( \kappa d, \frac{\ell}{d} \right).
\end{align}

\section{Validation of bridge shapes}
\label{sec:validation_bridge_shape}
\begin{figure}[htb]
    \centering
    \begin{subfigure}[t]{0.45\textwidth}
        \centering
        \includegraphics[width=0.99\textwidth,trim=0.3cm 0.3cm 7.7cm 0.3cm,clip]{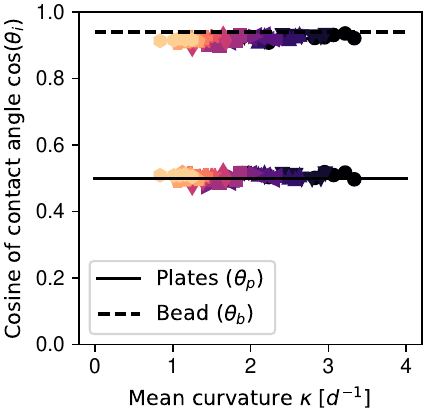}
        \caption{}
        \label{fig:bridge_shapes_cos}
    \end{subfigure}
    \begin{subfigure}[t]{0.45\textwidth}
        \centering
        \includegraphics[width=0.99\textwidth,trim=7.7cm 0.3cm 0.3cm 0.3cm,clip]{validation_shapes.pdf}
        \caption{}
        \label{fig:bridge_shapes_sc}
    \end{subfigure}
    \caption{Validation of bridge shapes. The symbols correspond to the same simulations as plotted in Fig.\ \ref{fig:shape_factor_kappa}, and were obtained by the fitting procedure described in Sec.\ \ref{sec:validation_bridge_shape}.
    (a) Cosine of contact angles at the bead and at the plates versus mean curvature. The solid (dashed) line represents the imposed contact angle at the plates (beads).
    (b) Distance $s_c$ from the $z$ axis to the center of the fitted circle as a function of the radius of principal curvature. The solid line corresponds to Eq.\ \eqref{eq:sc_exact} while the dashed line corresponds to the leading-order approximation \eqref{eq:s_approx}.}
    \label{fig:bridge_shapes}
\end{figure}
We verify the shapes obtained by Surface Evolver by fitting a circle to the shape to the air-liquid interface that crosses the inlet plane (i.e.\ the plane that crosses the bead center; see Fig.\ \ref{fig:inlet_plane_figure} below). 
The effective contact angles can be then be read off by measuring the intersection of the fitted circle and the top plane or the bead.
As shown in the left panel of Fig.\ \ref{fig:bridge_shapes}, the measured angles are in good agreement with the imposed angle.
As a second test, we can check the exact relation \eqref{eq:sc_exact} against the data. 
As shown in the right panel of Fig.\ \ref{fig:bridge_shapes}, the agreement with the numerical data is excellent.
Furthermore, Eq.\ \eqref{eq:s_approx} is shown to yield a very good approximation for the data range considered, establishing confidence in the modelling approximations used in \ref{sec:theoretical_modelling_bridge_g}.

\section{Finite element implementation of  flow through liquid bridges}
\label{sec:fem_bridge}
We want to solve the dimensionless Stokes Eqs.\ \eqref{eq:stokes} (with $\mu=F=1$) with the boundary conditions \eqref{eq:stokes_bcs} using the finite element method.
The main challenge is to implement the free-slip condition on the air-water interface $\Gamma$, which we opt to enforce weakly using Nitsche's method \cite{nitsche1971variationsprinzip,freund1995weakly,urquiza2014weak,gjerde2022nitsche}.
The corresponding variational problem can be expressed as follows.
Find $(\v u, p) \in \mathcal{V} \cross \mathcal{Q}$ such that for all $(\v v, q) \in \mathcal{V} \cross \mathcal{Q}$:
\begin{multline}
     \int_\Omega \left[ 2 \v D \v u : \v D \v v   - p \div \v v - q \div \v u  \right] \diff V  \\
    - \int_\Gamma \left[ ((( 2 \v D \v u - p \v I) \cdot \hat{\v n}) \cdot \hat{\v n} )(\v v \cdot \hat{\v n}) + ((( 2 \v D \v v - q \v I) \cdot \hat{\v n} ) \cdot \hat{\v n} ) (\v u \cdot \hat{\v n}) \right] \diff S \\
    + \beta \int_\Gamma h^{-1} (\v u \cdot \hat{\v n} )(\v v \cdot \hat{\v n} ) \diff S 
    = \int_\Omega v_x \diff V .
    \label{eq:variational_stokes}
\end{multline}
Here, the velocity subspace $\mathcal{V}$ encodes the Dirichlet conditions $u_y = 0 $ at $y = 0$; $\v u = \v 0 $ at $z = \pm 1/2$ and $\sqrt{x^2 + y^2 + z^2} = 1/2$; and $ u_y = u_z = 0 $ at $ x = 0, \ell/2$. 
The pressure subspace $\mathcal{Q}$ includes the Dirichlet conditions $p = 0$ at $x=0$ and $x=\ell/2$.
For the discretized (into tetrahedral cells) domain $\Omega$, we used the Taylor--Hood combination
$\mathcal P_2(\Omega)^3$ for $\mathcal{V}$ and $\mathcal P_1(\Omega)$ for $\mathcal{Q}$.
For $h$, we used the globally minimal mesh size $h_{\rm min}$ measured as the smallest distance between any two vertices of a cell.
As the penalization parameter we heuristically found $\beta=100$ to yield good results, but we verified that the results were not sensitive to this choice.
The variational form \eqref{eq:variational_stokes} with the boundary conditions above was discretized and solved numerically using the FEniCS framework \cite{logg2012automated}.
The resulting linear system was solved using a built-in direct LU solver.
The implementation is available at the Git repository \url{https://github.com/gautelinga/liquid_bridge_conductance}.

\section{Theoretical modelling of bridge conductance}
\label{sec:theoretical_modelling_bridge_g}
Here we motivate the functional form for the shape factor $k_0$ used for modelling liquid bridge conductance $\Gamma$.

We consider a bead of radius $R = d/2$ centered at the origin.
At an in-plane distance $s$ from the bead center, under (or over) the bead, the thickness $h(s)$ of the liquid corner is given by
\begin{align}
    h = R - \sqrt{R^2 - s^2} \simeq \frac{s^2}{d},
\end{align}
where the last expression follows from Taylor expansion to the leading order in $s/R$.
This approximation is quite good even for moderate $s/R$, e.g.\ for a typical maximum value of $s$ where the flow is most constrained, $s = 0.6 R$, the approximation gives $h \simeq 0.18 R$ compared to the exact $0.2 R$.

The lubrication approximation for the flux per area in the corner yields
\begin{align}
    \v q = - \frac{h^3}{12 \mu} \grad p \simeq - \frac{s^6}{12 \mu d^3} \grad p.
    \label{eq:qvec_lub}
\end{align}
We now assume that the flow is azimuthal in the region under the bead where the flow resistance is largest.
Eq.\ \eqref{eq:qvec_lub} then becomes
\begin{align}
    q_\theta  \simeq - \frac{s^5}{12 \mu d^3} \partial_\theta p.
\end{align}
We linearize the pressure in the azimuthal angle $\theta$ such that
\begin{align}
    \partial_\theta p \simeq - \frac{\Delta p}{2\varphi},
    \label{eq:qtheta_lub}
\end{align}
where $\Delta p$ is the total pressure drop (bead-bead) and $\varphi(\ell)$ is an angle over which the pressure drop occurs, which depends weakly on the bead-bead separation $\ell$.
We model the latter angle by the linear relation
\begin{align}
    \varphi \sim \frac{\ell}{d},
\end{align}
meaning that the further the beads are apart, the larger will be the angle over which the pressure drop occurs.
The total flux $\Phi$ is given by radially integrating Eq.\ \eqref{eq:qtheta_lub} to some outer radius $s$ (at an arbitrary $\theta$).
We skip all numerical prefactors for convenience:
\begin{align}
    \Phi &
    \sim \int_0^s q_\theta(s') \diff s' 
    \sim \frac{\Delta p}{\mu d^2 \ell} 
     \int_0^s s'^5 \diff s' 
    \sim \frac{\Delta p \cdot s^6}{\mu d^2 \ell} .
    \label{eq:flux_lub}
\end{align}
The remaining task is to relate the outer radius $s$ to the mean curvature $\kappa$.
As an intermediate quantity, we will use the largest principal curvature, $1/r$, of the air-liquid interface at the inlet plane, which is sketched in Fig.\ \ref{fig:inlet_plane_figure}.
\begin{figure}
    \centering
    \includegraphics[width=0.6\linewidth]{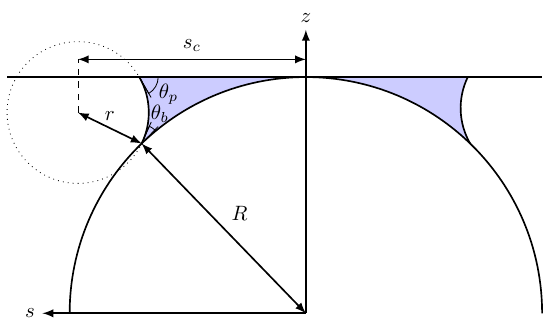}
    \caption{Top half of the inlet plane geometry, with the wetting corner indicated in light blue color.}
    \label{fig:inlet_plane_figure}
\end{figure}
Assuming the interface to have a partially circular shape, $r$ is the radius of the corresponding circle. 
We denote $s_c$ as the in-plane distance from an axis $s$ passing through the bead center, to the center of the circle (see Fig.\ \ref{fig:inlet_plane_figure}).

By considering the inlet geometry shown in Fig.\ \ref{fig:inlet_plane_figure}, we find an exact relation between $s_c$ and $r$:
\begin{align}
    s_c = \sqrt{(\cos \theta_b + \cos \theta_p) d r + r^2 \sin^2 \theta_p}, 
    \label{eq:sc_exact}
\end{align}
where $\theta_b, \theta_p$ are the contact angles between the air-liquid interface and the bead and the plates, respectively.
For small radii $r$, we may approximate
\begin{align}
    s \simeq s_c \simeq \sqrt{(\cos \theta_b + \cos \theta_p) d r}
    \sim \sqrt{dr}.
    \label{eq:s_approx}
\end{align}
Inserting the latter expression into Eq.\ \eqref{eq:flux_lub}, we obtain
\begin{align}
    \Phi
    &\sim \frac{\Delta p d r^3}{\mu \ell}.
    \label{eq:flux_lub2}
\end{align}
The last step is to relate $r$ to the mean curvature $\kappa$.
Rather than modelling the smallest principal curvature $1/R_\parallel$, where $R_\parallel \sim s$ and using the formula
\begin{equation}
    \kappa = \frac{1}{2}\left(\frac{1}{r} - \frac{1}{R_\parallel} \right),
    \label{eq:kappa_expr}
\end{equation}
we found that our numerical data was remarkably well fitted by the simple linear expression
\begin{align}
    \kappa = \frac{a}{r} -  \frac{\tilde\kappa_0}{d},
    \label{eq:kappa_lin}
\end{align}
with the dimensionless coefficients $a = 0.342$ and $\tilde \kappa_0 = 0.312$ (which we expect to depend only on $\theta_b, \theta_p$).
\begin{figure}
    \centering
    \includegraphics[width=0.65\textwidth]{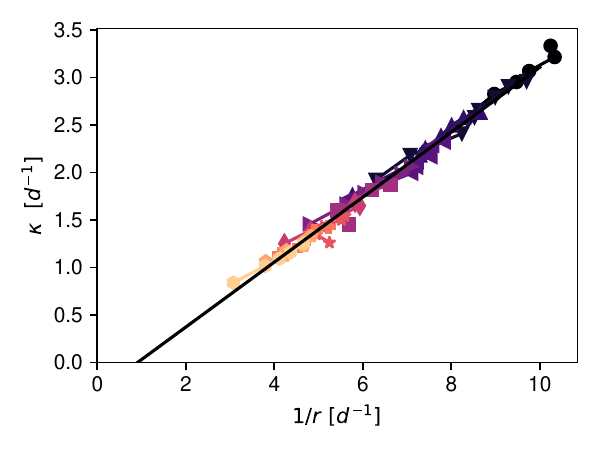}
    \caption{Mean curvature of the air-liquid interface versus the principle curvature at the inlet plane.
    A linear fit is superimposed at the numerical data. 
    The symbols correspond to the same simulations as plotted in Fig.\ \ref{fig:shape_factor_kappa}.}
    \label{fig:kappa_vs_rinv}
\end{figure}
The fit is shown in Fig.\ \ref{fig:kappa_vs_rinv}.
Note that \eqref{eq:kappa_expr} and \eqref{eq:kappa_lin} together suggest that $1/R_\parallel$ and $1/r$ are linearly related.
Inserting \eqref{eq:kappa_lin} into \eqref{eq:flux_lub2} yields
\begin{align}
    \Phi
    & \sim \frac{\Delta p d^4 }{\mu \ell} \left( \tilde\kappa + \tilde\kappa_0 \right)^{-3},
\end{align}
where $\tilde \kappa = \kappa d$ is the nondimensionalized mean curvature.
From this expression we can extract the conductance $\Gamma = \Phi / \Delta p$ and thereby, upon comparison with \eqref{eq:g_bridge_gen}, the shape factor $k_0$:
\begin{align}
    k_0 
    \sim \left( \tilde\kappa + \tilde\kappa_0 \right)^{-3}.
\end{align}
This motivates the expression
\begin{align}
    k_0(\tilde \kappa) = \left( \frac{A}{\tilde\kappa + B} \right)^3,
    \label{eq:k0cubic}
\end{align}
used to fit, with $A, B$ as fitting parameters, our numerical data for liquid bridge conductances.

\bibliography{Biblio}
\end{document}